\documentclass[bibyear]{aa}
\usepackage[varg]{txfonts}
\usepackage{graphicx}
\usepackage{epstopdf}
\usepackage{longtable}
\usepackage{lscape}
\newcommand{\e}{et al.\ }
\newcommand{\ha}{H$\alpha$}
\newcommand{\hii}{H$_{\rm II}$}
\begin{document}

\title{The Gaia-ESO Survey: dynamics of ionized and neutral gas in the
Lagoon nebula (M8)
\thanks{
Based on observations collected with the FLAMES spectrograph at VLT/UT2
telescope (Paranal Observatory, ESO, Chile), for the Gaia-ESO Large
Public Survey (program 188.B-3002).
Tables~\ref{table-sky} and~\ref{table-sodium} are only available in
electronic form at the CDS via anonymous ftp to cdsarc.u-strasbg.fr
(130.79.128.5) or via http://cdsarc.u-strasbg.fr/viz-bin/qcat?J/A+A/
}
}

\date{Received date / Accepted date}

\author{F. Damiani\inst{1},
R. Bonito\inst{1,2},
L. Prisinzano\inst{1},
T. Zwitter\inst{3},
A. Bayo\inst{4},
V. Kalari\inst{5},
F.~M. Jim\'{e}nez-Esteban\inst{6},
M.~T. Costado\inst{7},
P. Jofr\'e\inst{8,9},
S. Randich\inst{10},
E. Flaccomio\inst{1},
A.~C. Lanzafame\inst{11},
C. Lardo\inst{12},
L. Morbidelli\inst{10},
\and
S. Zaggia\inst{13}
}
\institute{INAF - Osservatorio Astronomico di Palermo G.S.Vaiana,
Piazza del Parlamento 1, I-90134 Palermo,
Italy \\
\email{damiani@astropa.inaf.it}
\and
Dipartimento di Fisica e Chimica, Universit\`a di Palermo,
Piazza del Parlamento 1, 90134, Palermo, Italy
\and
Faculty of Mathematics and Physics, University of Ljubljana, Jadranska
19, 1000, Ljubljana, Slovenia
\and
Instituto de F\'isica y Astronomi\'ia, Universidad de Valparai\'iso,
Chile
\and
Departamento de Astronom\'{\i}a, Universidad de Chile, Casilla 36-D
Santiago, Chile  
\and
Departmento de Astrof\'{\i}sica, Centro de Astrobiolog\'{\i}a
(INTA-CSIC), ESAC Campus, Camino Bajo del Castillo s/n, E-28692
Villanueva de la Ca\~{n}ada, Madrid, Spain  
\and
Instituto de Astrof\'{i}sica de Andaluc\'{i}a-CSIC, Apdo. 3004, 18080
Granada, Spain
\and
Institute of Astronomy, University of Cambridge, Madingley Road,
Cambridge CB3 0HA, United Kingdom
\and
N\'ucleo de Astronom\'{i}a, Facultad de
Ingenier\'{i}a, Universidad Diego Portales, Av. Ej\'ercito 441,
Santiago, Chile
\and
INAF - Osservatorio Astrofisico di Arcetri, Largo E. Fermi 5, 50125,
Florence, Italy
\and
Dipartimento di Fisica e Astronomia, Sezione Astrofisica, Universit\`{a}
di Catania, via S. Sofia 78, 95123, Catania, Italy
\and
Laboratoire d'astrophysique, Ecole Polytechnique F\'ed\'erale de
Lausanne (EPFL), Observatoire de Sauverny, CH-1290 Versoix, Switzerland
\and
INAF - Padova Observatory, Vicolo dell'Osservatorio 5, 35122 Padova,
Italy
}

\abstract
{}
{We present a spectroscopic study of the dynamics of the ionized and
neutral gas throughout the Lagoon nebula (M8), using VLT/FLAMES data from the
Gaia-ESO Survey. The new data permit to explore the physical connections
between the nebular gas and the stellar population of the associated
star cluster NGC6530.}
{We characterize through spectral fitting emission lines of \ha, [N II]
and [S II] doublets, [O III], and absorption lines of sodium D doublet,
using data from the FLAMES/Giraffe and UVES spectrographs, on more than
1000 sightlines towards the entire face of the Lagoon nebula. Gas
temperatures are derived from line-width comparisons, densities from the
[S II] doublet ratio, and ionization parameter from \ha/[N II] ratio.
Although doubly-peaked emission profiles are rarely found, line
asymmetries often imply multiple velocity components along the same line
of sight. This is especially true for the sodium absorption, and for the
[O III] lines.}
{Spatial maps for density and ionization are derived, and compared to
other known properties of the nebula and of its massive stars 9~Sgr,
Herschel~36 and HD~165052 which are confirmed to provide most of the
ionizing flux. The detailed
velocity fields across the nebula show several expanding shells,
related to the cluster NGC6530,
the O stars 9~Sgr and Herschel~36, and the massive protostar M8East-IR.
The origins of kinematical expansion and ionization of the NGC6530 shell 
appear to be different.
We are able to put constrains on
the line-of-sight (relative or absolute) distances between some of these
objects and the molecular cloud. The data show that
the large obscuring band running through the middle of
the nebula is being compressed by both sides, which might explain its
enhanced density.
We also find an unexplained large-scale velocity gradient across the
entire nebula.
At larger distances, the transition from ionized to neutral gas is
studied using the sodium lines.
}
{}

\keywords{ISM: individual objects: (Lagoon nebula)
-- ISM: general -- HII regions
}

\titlerunning{Gas dynamics in Lagoon nebula}
\authorrunning{Damiani et al.}

\maketitle

\section{Introduction}
\label{intro}

\begin{figure*}
\resizebox{\hsize}{!}{
\includegraphics[width=17.5cm]{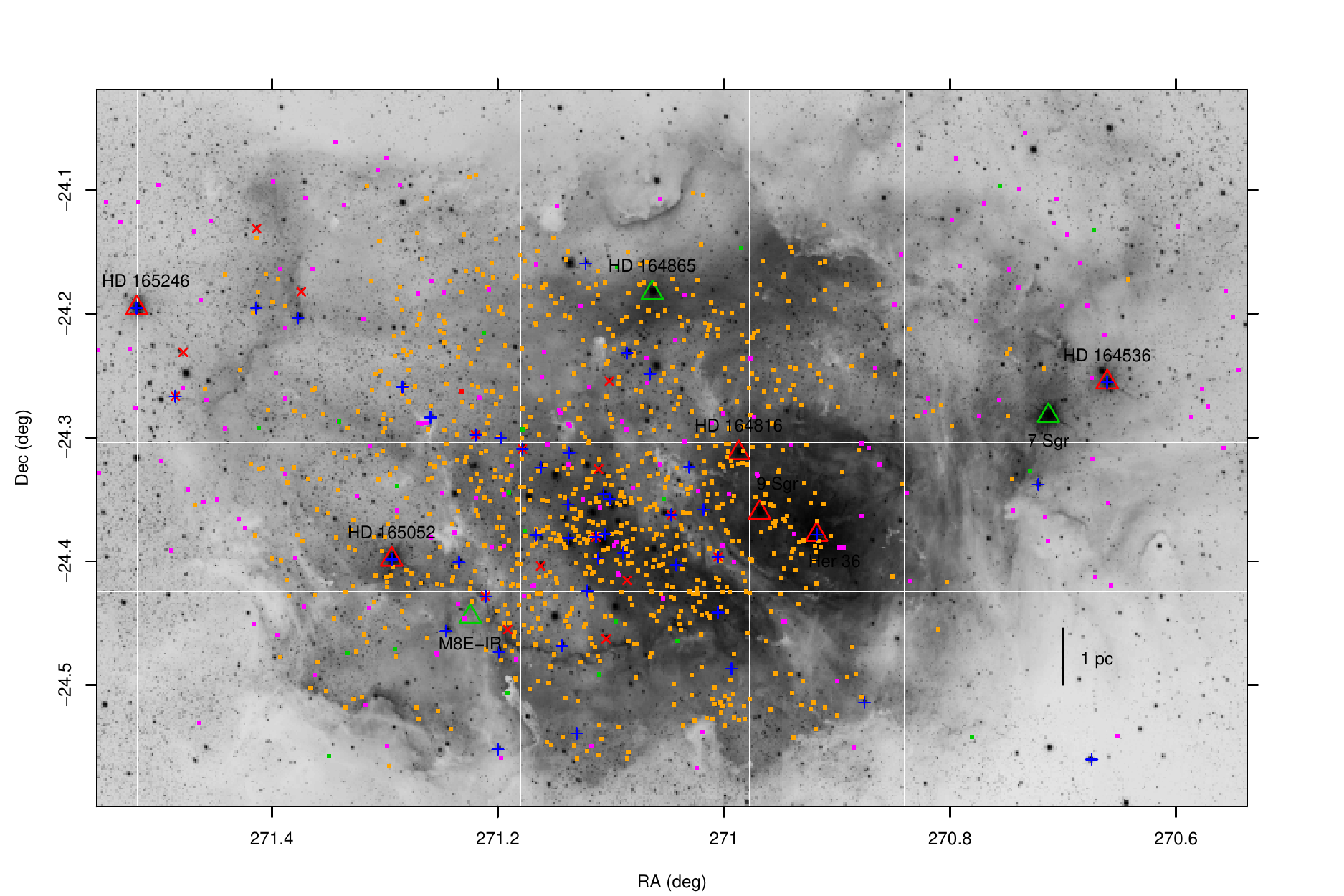}}
\caption{VPHAS$+$ image of the Lagoon nebula (size $55.8^{\prime} \times
34.8^{\prime}$, corresponding to $20.36 \times 12.7$~pc),
with superimposed all Gaia-ESO target positions used in this work.
Giraffe targets are indicated with
magenta (sky fibres) or orange (faint stars)
{filled squares.
}
UVES sky fibres are indicated with red (setup 580) or green (setup 520)
{filled squares. Only one UVES-580 sky fibre position is not
coincident with (and its symbol not hidden by) an UVES-520 sky fibre position.
}
UVES stars are
indicated with red crosses (setup 580) or blue `$+$' signs (setup 520).
Red triangles indicate the O stars
HD~165246, HD~165052, HD~164816, 9~Sgr, Herschel~36, and HD~164536.
Green triangles indicate the massive objects M8E-IR, HD~164865,
and 7~Sgr. North is up and East to the left.
\label{vphas-fibers}}
\end{figure*}

The Lagoon nebula (M8, NGC6523) is one of the brightest \hii\ regions in
the solar neighborhood, and has been the subject of many observational
studies (e.g., Lada \e 1976 in CO and optical lines, Tothill \e 2002 in
CO and sub-mm, Takeuchi \e 2010 in CO).
It harbors the young cluster NGC6530, only a few Myrs
old, and also intensively studied especially in recent years
at optical, infrared and X-ray wavelengths
(e.g.\ Walker 1957, van den Ancker \e 1997, Sung \e 2000,
Damiani \e 2004, 2006, Prisinzano \e 2005, 2007,
Kumar and Anandarao 2010, Povich \e 2013).
The \hii\ region is illuminated by many
massive stars of O and B spectral types,
the hottest one being 9~Sgr (HD~164794, type O4V((f))z); a
few other late-O/B type stars are also found in the region. The
optically brightest part of the Lagoon nebula is the so-called Hourglass
nebula, which surrounds and partially oscures the O7:V star Herschel~36
(Woodward \e 1986); stars in the Hourglass are thought to be younger
than the NGC6530 cluster. Also noteworthy is the presence, to the East of
NGC6530, of the embedded massive
protostar M8E-IR (Wright \e 1977, Simon et al.\ 1984,
Henning and G\"urtler 1986),
indicating that star formation in the region has also taken place recently.
The most recent determination of the distance of the NGC6530 cluster
(and by inference of the \hii\ region as well) is 1250~pc (Prisinzano \e
2005). The properties of the whole region were reviewed by Tothill \e (2008).

The Lagoon nebula and its stellar population show a well-defined spatial
organization. The brightest nebular region (Hourglass) does not lie near
the geometrical center of the whole nebula, but several arcmin ($\sim
2$~pc) to the West. Closer to the nebula center lies instead the bulk of
low-mass cluster stars (Damiani \e 2004), cospatial with the B stars.
The most massive member 9~Sgr is also offset with respect to the B
stars, but appears not directly related to any localized bright
nebulosity, leading Lada \e (1976) to suggest that it actually lies
several parsecs in front of the nebula, not within it. All around the
central region, many bright-rimmed dark clouds are found,
{
which being located along the outer border of the nebula suggest
}
the ``blister" nature of the \hii\ region. Behind one of them
to the South-East of the main cluster, the mentioned young, massive star
M8E-IR is found. This and other low-mass stars in the same neighborhood
show indications of being younger than the main NGC6530 cluster (Damiani
\e 2004).
The bright Hourglass region has received much more
attention than the outer nebula parts, which remain relatively little studied.

\begin{figure*}
\sidecaption
\includegraphics[width=12.0cm,angle=0]{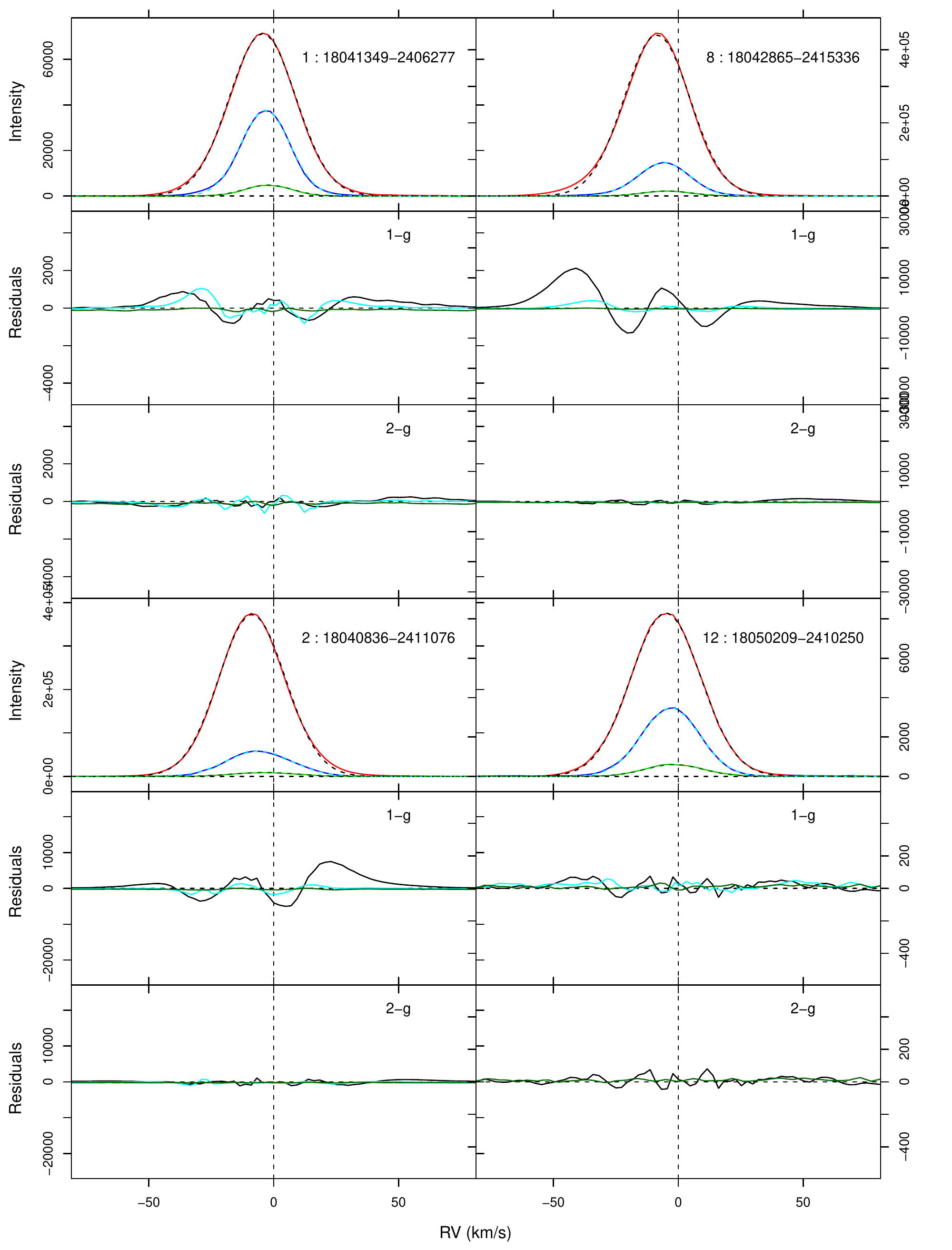}  
\caption{Four examples of nebular line profiles of \ha\ (red), [N II] 6584
(blue), and [S II] 6731 (green) lines, and their best-fit models using
both one- and two-Gaussian models (black, cyan and dark-green dashed
lines, for \ha,
[N II] and [S II] respectively). The ordinate scale is the same for all
lines.  Only pure-sky spectra from Giraffe are shown.
Each panel labeled with sky position shows the observed spectrum with 
superimposed the one-Gaussian best-fit model for each line. Below it,
the fit residuals (labeled `1-g') for the one-Gaussian model are shown,
with an ordinate scale enlarged 15 times; still below, the fit residuals
for the two-Gaussian model (labeled `2-g') are shown on the same scale.
\label{atlas-fits-1}}
\end{figure*}

\begin{figure*}
\sidecaption
\includegraphics[width=12.0cm,angle=0]{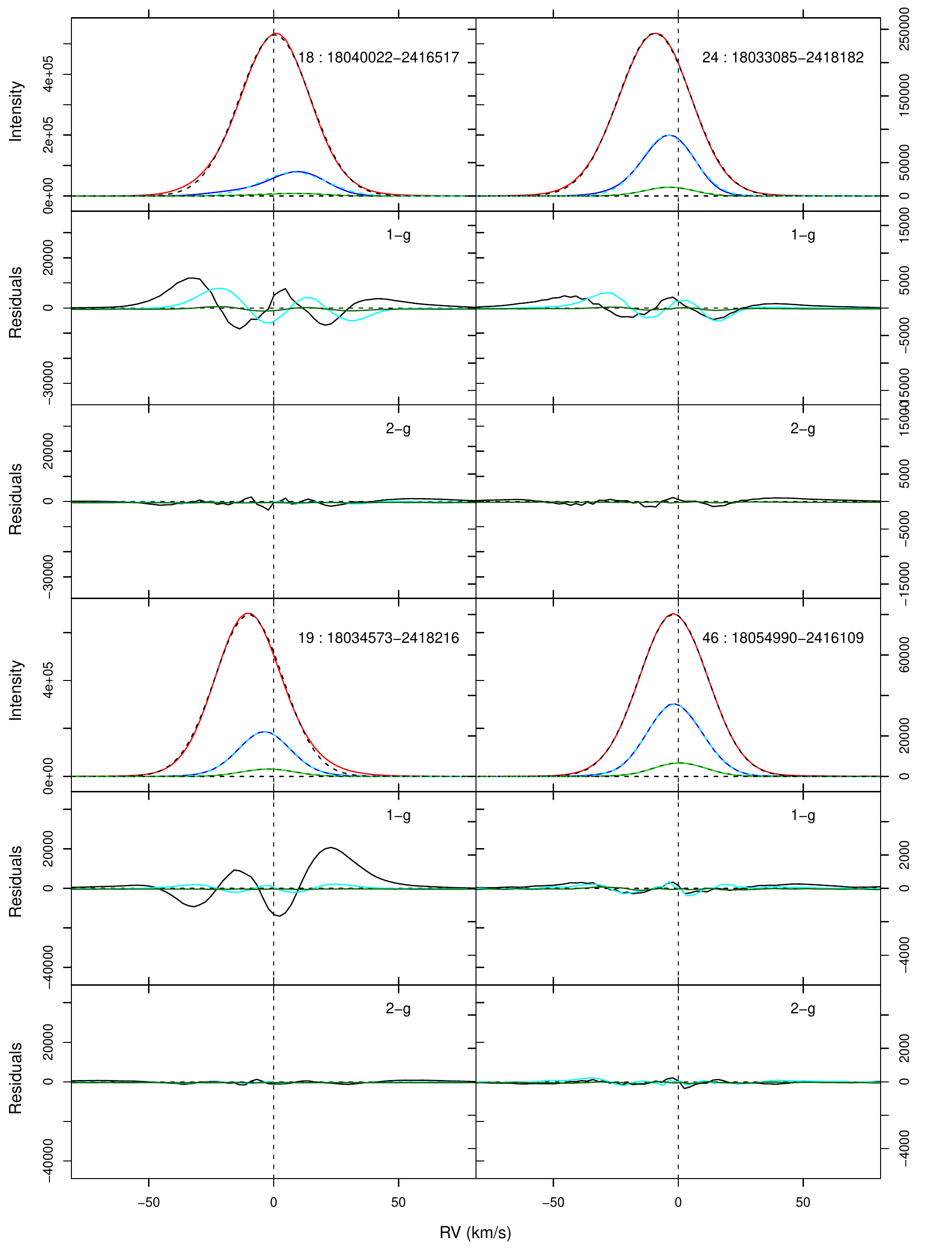}  
\caption{Additional examples of observed line profiles and their
best-fit models, as in Fig.~\ref{atlas-fits-1},
showing large differences between the peak radial velocities
of \ha\ and [N II], and occasional asymmetrical residuals.
\label{atlas-fits-2}}
\end{figure*}

\begin{figure*}
\sidecaption
\includegraphics[width=12.0cm,angle=0]{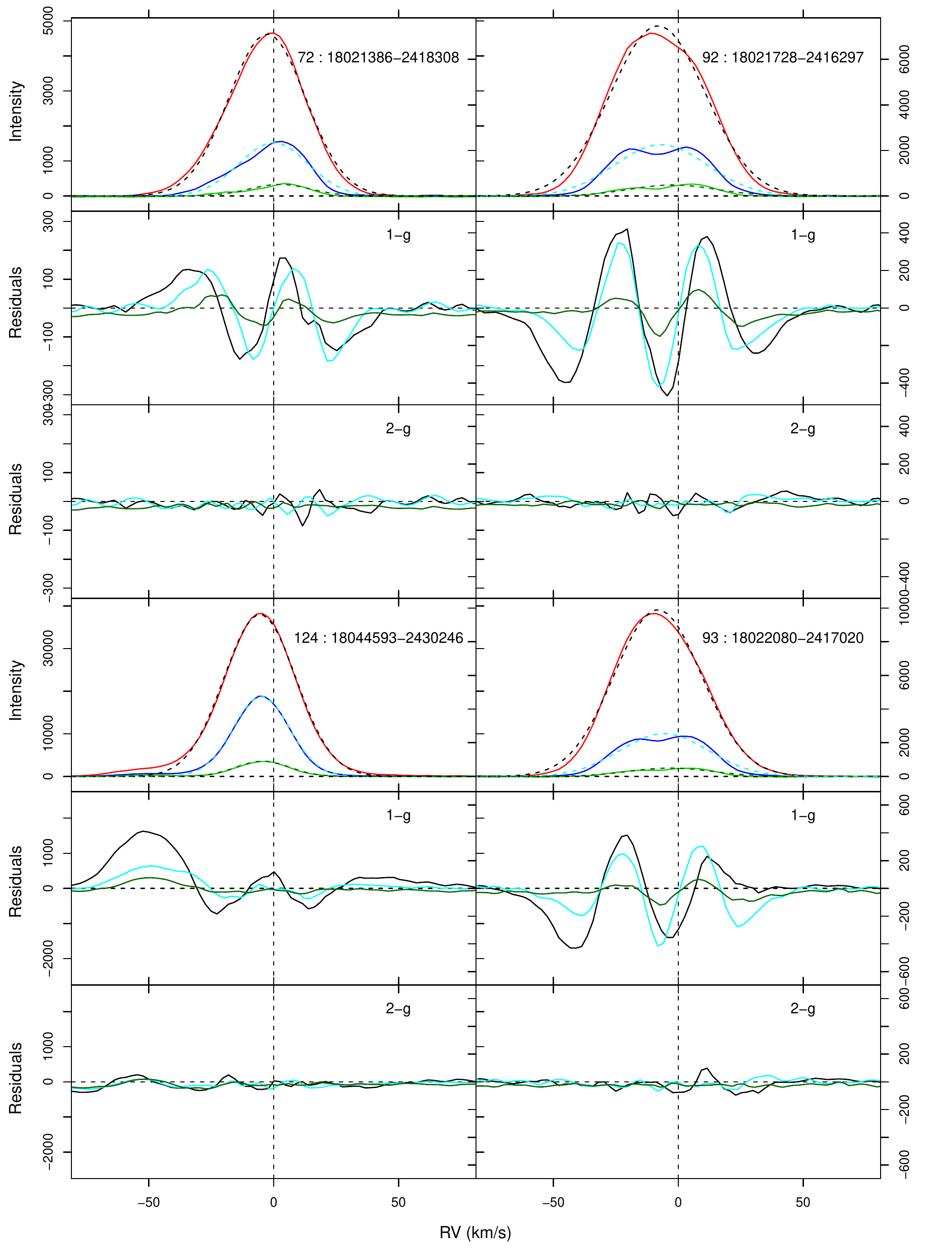}  
\caption{Additional examples of observed line profiles and their
best-fit models, as in Fig.~\ref{atlas-fits-1}.
Here we show line profiles with the most evident distortions, even
showing double peaks in [N II] (right panels).
\label{atlas-fits-3}}
\end{figure*}

In this work we study the kinematics of the ionized and neutral gas
giving rise to the strong optical emission lines (\ha, [N II], [S II],
[O III]), and sodium D absorption lines, across the whole nebula, using
new spectroscopic data from the Gaia-ESO Survey (Gilmore \e 2012, Randich
\e 2013).  In Section~\ref{data} we
describe the observations, while in Section~\ref{results} we present
our results. In Section~\ref{discuss} we
discuss the main implications of our work for the structure and dynamics of the
Lagoon nebula, including issues that deserve further research.

\section{Observational data}
\label{data}

The observations of the Lagoon nebula were obtained as
part of the Gaia-ESO Survey (internal release {\em iDR4}), targeting
Milky Way stars and clusters across a wide range of parameters with the
ESO VLT/FLAMES multi-fibre spectrograph (Pasquini \e 2002).
Twenty-seven observing blocks (OBs) were devoted to NGC6530 as part of
the Survey program, down to a limiting stellar magnitude $V=19$.
Observations were spread over 17 nights, in September 2012 and
June-September 2013.
Fifteen
fibres per OB are targeted at star-free sky positions, interspersed
between stellar position, with the intent of obtaining purely nebular
spectra needed to correct the stellar spectra through subtraction.
Spectrographs used were Giraffe (setup HR15N, $R \sim 19000$, wavelength
range 6444-6818\AA) and UVES (setup 580, $R \sim 47000$, range
4768-6830\AA; and setup 520, $R \sim 47000$, range 4180-6212\AA).
The number of sky spectra obtained with Giraffe is 647 (on 197
individual sky positions), 4 spectra with UVES setup 580 (3 positions),
and 25 spectra with UVES setup 520 (21 positions).
In addition, being so bright, the nebula dominates clearly the
stellar spectra of faint stars, and the strongest nebular lines could be
studied with good results whenever the ratio between \ha\ peak and
stellar continuum is larger than 100 (henceforth ``faint stars"), so
that the \ha\ line cannot be dominated by the star, even in strongly
accreting T~Tauri stars.
The same method was successfully used by us in a previous work on the
emission of the Carina nebula using Gaia-ESO data (Damiani \e 2016).
The number of faint star spectra in the NGC6530 Gaia-ESO dataset is 980
(892 individual stars, with some having been observed more than once).
Therefore, only considering Giraffe spectra, we have nebular spectra for
1089 positions across the Lagoon nebula. The Giraffe HR15N wavelength range
includes several strong nebular lines: \ha, the neighboring [N II] lines
at 6548, 6584\AA, and the two [S II] lines at 6716, 6731\AA.
Also the He~I line at 6678\AA\ falls in the range, but is found to be too
weak in our spectra for a proper study.
Spectra for the same sky position were coadded to improve the S/N ratio.
All the studied nebular lines are orders of magnitude stronger than
atmospheric lines, as estimated from other Gaia-ESO spectra in clusters
free of nebulosity.

The UVES sky spectra are very few as mentioned, and add little information
despite the much wider wavelength range; they are therefore not considered
further.  On the other hand, the UVES spectra of
stellar targets in NGC6530 are more numerous (16 stars using setup 580,
and 44 using setup 520, mostly of early type), and were used to study two
important lines: the [O III] line at 5007\AA, clearly detected in most
spectra, and the Na~I D1, D2 absorption lines at 5895.92, 5889.95\AA.
The latter enable us to perform a comparative study of the ionized and
the neutral gas in the region.
Like the ionic lines, also these sodium absorption lines are much
stronger than typical atmospheric sodium absorption.
An \ha\ image of the Lagoon nebula from the VPHAS$+$ survey (Drew \e 2014)
is shown in Figure~\ref{vphas-fibers}, together with all sky positions
considered here, and positions of several of the most massive stars.

Since the exposure times (20-50 min) were determined by the
requirements dictated by
the faintest stars, the signal-to-noise ratio (S/N) in the main nebular lines
is usually very high (but lines remain unsaturated).
The high S/N and the large number of nebular positions
comprised in our dataset make it one of the richest datasets ever available
for the study being performed.

\section{Results}
\label{results}

\subsection{\ha, [N II] and [S II] lines from Giraffe data}
\label{giraffe}

\begin{figure*}
\includegraphics[width=9cm]{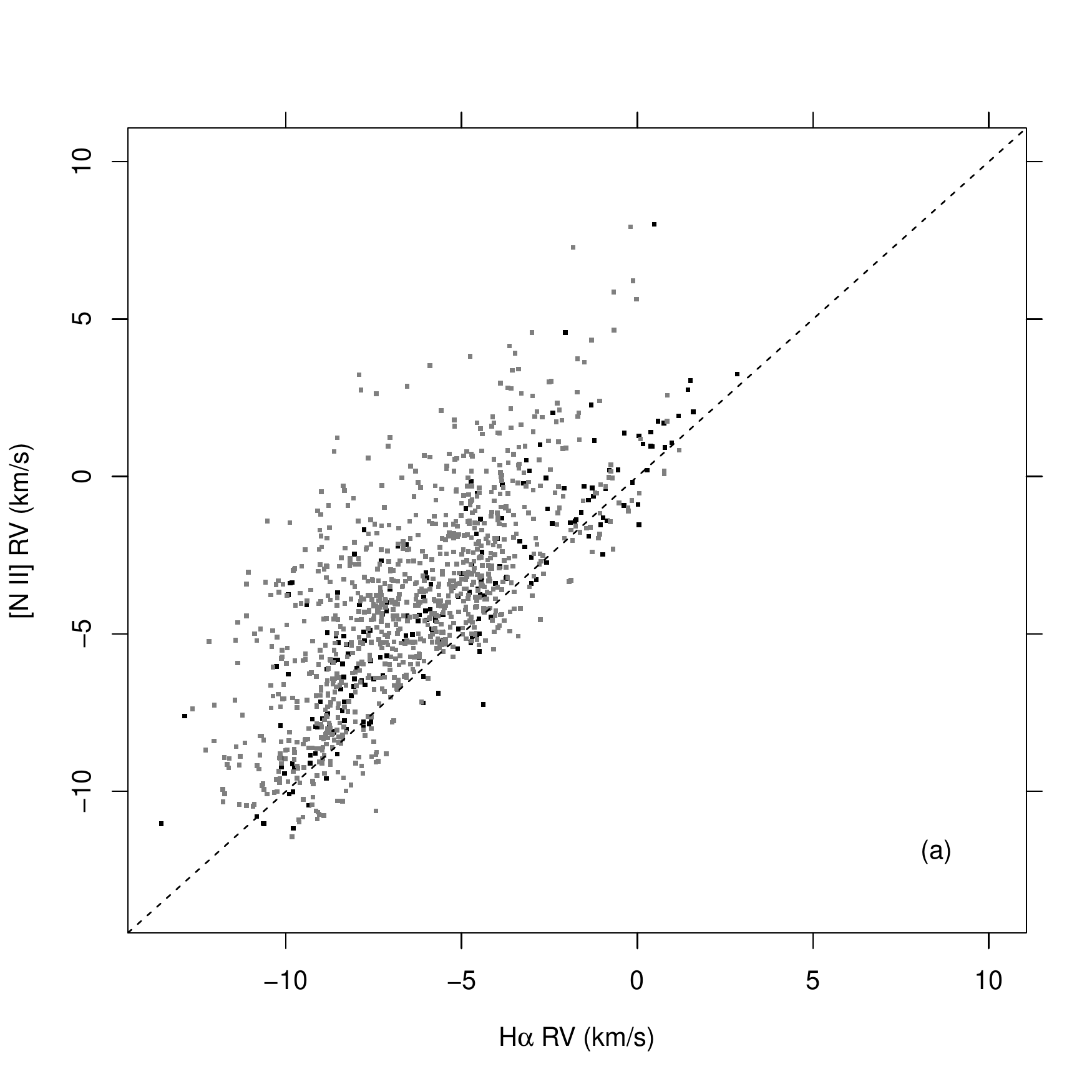}
\includegraphics[angle=90,width=9cm]{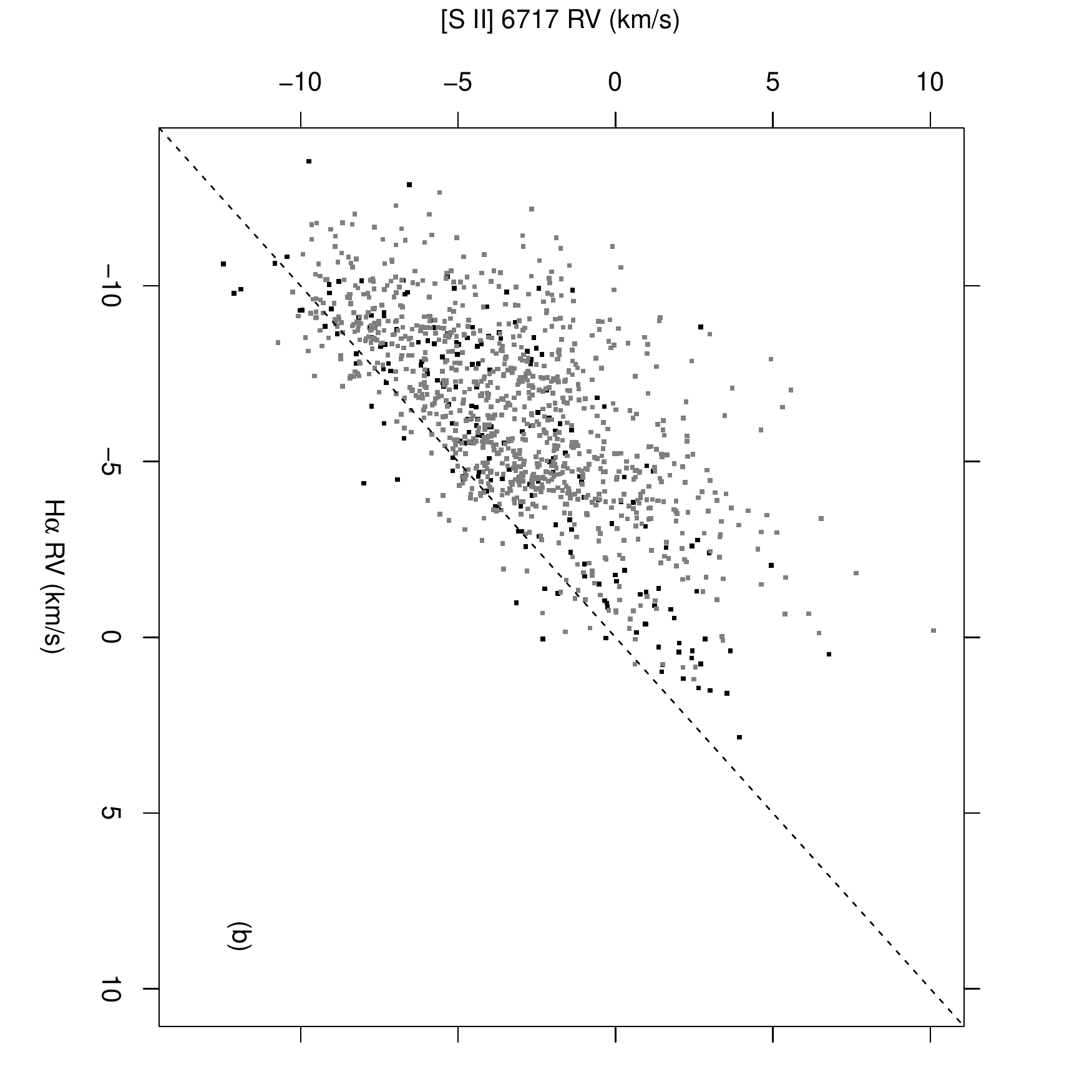} \\
\includegraphics[angle=0,width=9cm]{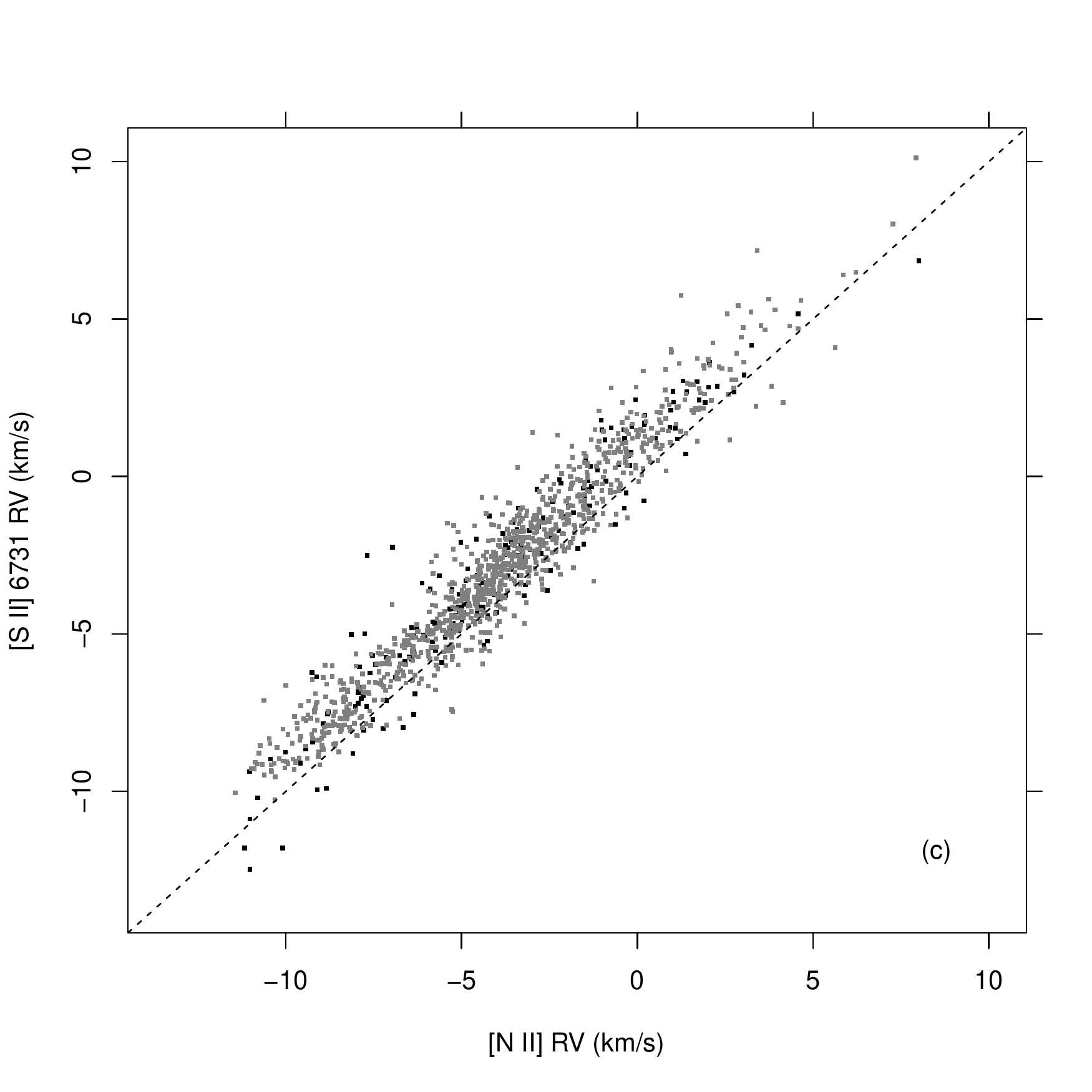}
\includegraphics[angle=90,width=9cm]{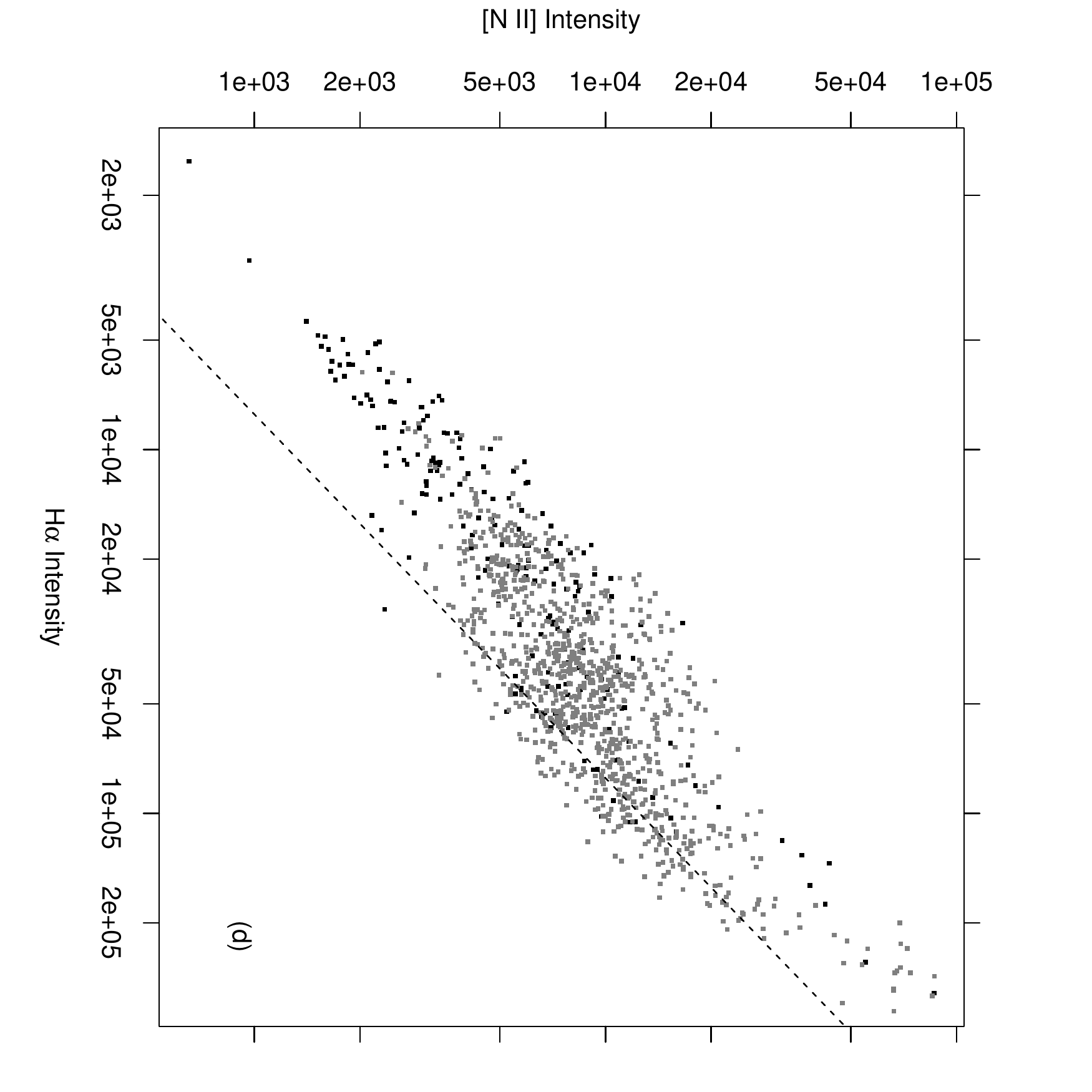}
\caption{
$a$: Comparison between 1-g best-fit RVs from \ha\ and [N II] 6584
lines. Black dots indicate pure-sky fibres while gray dots indicate
faint stars. The dashed line indicates identity.
%
$b$: Same comparison as in panel $a$, between \ha\ and [S II] 6731.
%
$c$: Same comparison as in panel $a$, between [N II] 6584 and [S II] 6731.
%
$d$: Comparison between intensities of \ha\ and [N II] 6584, as
resulting from 1-g models. Symbols as in panel $a$.
The dashed line represents a ratio of 1/8.
\label{ha-n2-s2}}
\end{figure*}

In order to study the dynamics of the nebular gas the optical emission
lines were analyzed and modelled with analytical functions, starting with
lines from Giraffe spectra.
Figures~\ref{atlas-fits-1} to \ref{atlas-fits-3} show several examples
of the observed lines of \ha, [N II] 6584\AA, and [S II] 6731\AA\
(panels labeled with coordinates), and the residuals after subtraction of
single-Gaussian (panels labeled with `1-g') or double-Gaussian
(labeled `2-g') best-fit models.  In nearly all cases
the optical lines are found single-peaked: among all 197 pure-sky fibres
only two cases of doubly-peaked lines are found
(Fig.~\ref{atlas-fits-3}, right panels), and only in the [N II] and [S II]
lines while not in \ha. In general, lines are found to be nearly
Gaussian in shape, suggesting a single kinematical component along each
line of sight for each emitting ion, as the representative examples of
Figs.~\ref{atlas-fits-1} and~\ref{atlas-fits-2} show.
However, while a single Gaussian (henceforth ``1-g" model) accounts well
for the bulk of the
emission in each line, examination of the fitting residuals (panels
`1-g') reveals that the detailed line shape is systematically
misreproduced by a simple Gaussian function, by a small but significant
amount in terms of the available S/N (note that the ordinate scale in
each of the 1-g residuals panels is 1/15 that of the panel above it).
The systematic pattern of the 1-g
model residuals is suggestive of at least two
{
unresolved
}
velocity components along the same sightline; only in rare cases the 1-g
residuals show no such pattern (Fig.~\ref{atlas-fits-1}, spectrum \#12).
In order to model the line profiles in detail, double-Gaussian (``2-g") models
were therefore attempted, whose residuals are also shown in
Figs.~\ref{atlas-fits-1}-\ref{atlas-fits-3} (panels `2-g'): these show
in nearly all cases no systematic patterns, indicating that two Gaussian
components provide a sufficient characterization of emission lines in the
Lagoon nebula. It should be remarked that each emission line (\ha, [N
II] 6584 and [S II] 6717, 6731\AA) was fitted independently; as
the first three panels of Fig.~\ref{atlas-fits-2} show particularly
well, the \ha\ emission peak may lie at velocities significantly different
than the peaks of [N II] and [S II] lines.
All velocities here are heliocentric.
Results from our 1-g best fits are reported in Table~\ref{table-sky}.

While the 2-g models can be considered ``better" than the 1-g ones
because of the smaller residuals, the corresponding best-fit parameters
must be treated with caution. The ``blue" and ``red" best-fit radial
velocities (RVs) of each modelled line are in most cases only a few km/s from
one another, i.e.\ much closer together than the line widths ($\sigma$)
themselves: this makes the relative intensities of the two Gaussian
components much more uncertain than their total value, which is instead
robustly modelled by 1-g fits (this is especially true of low S/N
spectra).
{
%
}
Therefore, the choice of considering one set of model
parameters or the other will depend on the specific problem. We find
that diagnostics involving two lines simultaneously (e.g.,
line-intensity ratios, or linewidths comparisons) are derived much more
robustly from 1-g fitting parameters; on the other hand, the description
of the gas dynamics considering RVs alone benefits also from the 2-g
model velocities, whenever S/N is sufficient. In this respect, we remark
that the smaller thermal widths of the [N II] and [S II] lines permit
often a more accurate derivation of components' RVs, despite these lines
being weaker than \ha.

{
In several cases among those shown in Figures~\ref{atlas-fits-1}
to~\ref{atlas-fits-3}, residuals from 1-g fits show a blue component.
This might be indicative of an approaching ionized layer,
blueshifted with respect to the bulk of ionized gas, and reminiscent of
the layer in the outer part of the Orion nebula, known as the Veil (see
the reviews by O' Dell 2001, O' Dell et al. 2008).
More detailed indications on the dynamics of the neutral gas probably
associated with this layer in the M8 nebula are derived from the sodium
absorption lines in Section~\ref{uves} below.
}

Some general properties
{of the ionized gas
}
may be derived from the best-fit parameters.
Figure~\ref{ha-n2-s2}$a$ shows a comparison between RVs derived from 1-g
fits to \ha\ and [N II] 6584\AA\ lines.
The same pattern is shown by measurements from pure-sky fibres (black
dots) and from faint stars (gray), confirming that the usage of
faint-star spectra introduces no biases in the derived nebular
properties.  The datapoints scatter is not
caused by errors ($<1$ km/s as a rule) but is real: in many cases the [N
II] RV is less negative than the \ha\ RV.
{
}
The velocity of the low-mass
stars in the NGC6530 cluster is $RV=0.5 \pm 0.2$ km/s (Prisinzano
\e 2007), which is assumed to coincide with the center-of-mass velocity
$RV_{cm}$ of all cluster stars.
The Figure then shows than the bulk of ionized gas
emitting in these lines has negative velocities (approaching us) from
the standpoint of the cluster center of mass. This is unlike the case of
the Carina nebula, where the double emission line peaks bracket usually
the center-of-mass RV, indicating expansion towards both the near and
far sides (Damiani \e 2016). In NGC6530, the ionized gas seems instead to expand
predominantly towards our side (but more details will be studied in
Sections~\ref{maps} and~\ref{pos-vel} below).
Figure~\ref{ha-n2-s2}$a$ however shows that along some sightlines the [N II]
moves away from us, while hydrogen moves towards us: clearly, the
different lines arise from dynamically distinct gas layers, a feature
which cannot be understood from narrow-band images, which mix emission
from all velocity layers (e.g.\ Tothill \e 2008, fig.4).
The comparison between \ha\ and [S II] RVs in Figure~\ref{ha-n2-s2}$b$ 
shows a pattern very similar to Figure~\ref{ha-n2-s2}$a$. This suggests
that the [N II] and [S II] lines originate from gas layers which are
more closely related mutually than with \ha. This is confirmed by the
good correlation shown in Figure~\ref{ha-n2-s2}$c$, between the [N II] and
[S II] RVs, the small systematic shifts being attributable to the
uncertainty in the adopted line wavelengths\footnote{Wavelengths adopted
here are 6583.43\AA\ for [N II], and 6716.44, 6730.815\AA\ for [S II],
the latter two from Kaufman and Martin (1993).}.
This figure also permits to estimate empirically the maximum errors in the
best-fit RVs from the scatter of datapoints in the correlation,
on the order of $\leq 1$~km/s. We remark that the [S II] and [N II] lines are
the weakest being studied, the \ha\ line being much stronger; therefore,
the scatter of datapoints in both Figures~\ref{ha-n2-s2}$a$ and~$b$ is
certainly dominated by real effects.

{
The above results are reminiscent of those found in the Orion nebula,
a well studied blister \hii\ region, where ionized gas flows away from the
ionization front, and ionized hydrogen is found at more negative speeds
with respect to [N II] and [S II], the ionization level gradually
increasing as the gas acquires larger and larger speeds in a champagne-flow
geometry (e.g., Balick \e 1974, O' Dell \e 1993).
The analogy with the Orion nebula, although very interesting, cannot
however be pushed too far, since there are also important differences
between it and M8: this latter is a much larger region, with more than
one ionizing O-type star, and is probably also a more evolved blister,
where at least the most massive star 9~Sgr has excavated a larger cavity
in the parent cloud compared to that excavated by the Orion most massive
member, $\theta^1$~Ori~C. We will examine in more detail the relative
geometry of 9~Sgr and the M8 nebula in Section~\ref{9sgr}.
}

%

\begin{figure}
\resizebox{\hsize}{!}{
\includegraphics[angle=90]{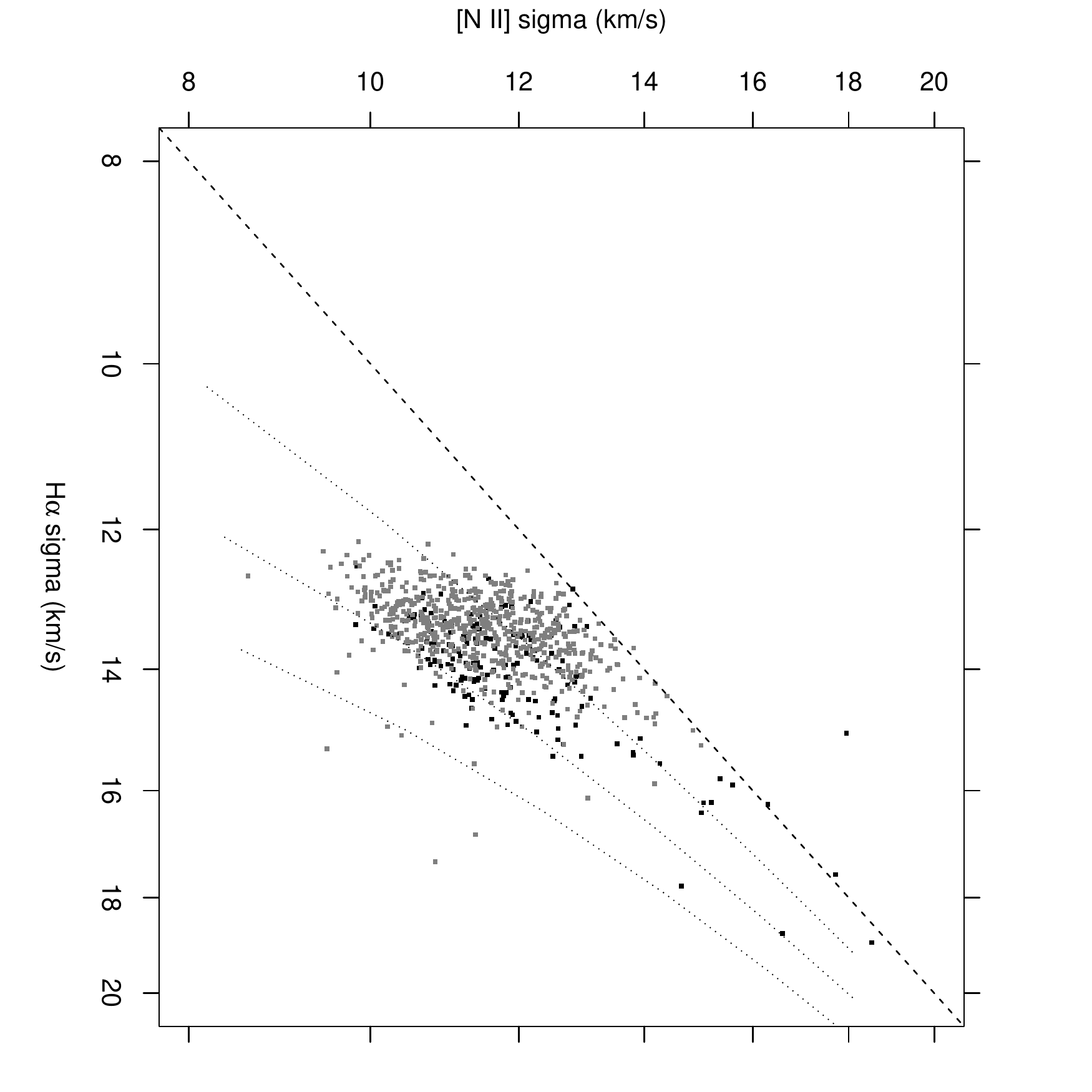}}
\caption{Comparison between line widths $\sigma$ of \ha\ and [N II] 6584, as
resulting from 1-g models. Symbols as in Fig.~\ref{ha-n2-s2}.
The dotted lines indicate loci of thermally broadened lines for
$T=5000$, 10000, and 15000~K, respectively from left to right.
The dashed line represents identity.
\label{ha-n2-sigma}}
\end{figure}

A comparison between the 1-g model intensities of \ha\ and [N II] lines
is shown in Figure~\ref{ha-n2-s2}$d$. The intensity ratio is
significantly non-uniform; this, under the typical conditions found in
\hii\ regions, suggests significant differences in the ionization
parameter across the region (e.g., Viironen \e 2007), which will be
studied in Section~\ref{maps}.
{
In very general terms, in regions with high Lyman-continuum flux
ionization will be highest, and hydrogen lines dominate over [N II]
lines; in the same region, the diagram suggests that the highest
densities and largest surfact brightnesses are also found.
In Section~\ref{maps} we will examine in much better detail how these
quantities depend on position across the nebula.
}
Figure~\ref{ha-n2-sigma}
{
shows}
the best-fit linewidths $\sigma$ of \ha\ and [N II],
{
whose comparison provides}
a measure of temperature (since
{turbulent and instrumental broadenings
}
are the same for
the two lines). To avoid mixing unrelated gaseous layers, we only show
datapoints having maximum absolute RV differences of 3~km/s between the
two lines. In the Figure, dotted lines indicate loci for
{fixed
}
temperatures of 5000, 10000, and 15000~K,
{
and a range of combined turbulent+instrumental broadening (between [8-18]~km/s).
}
Most datapoints lie between 5000-10000~K, however with considerable spread.

%

\begin{figure}
\resizebox{\hsize}{!}{
\includegraphics{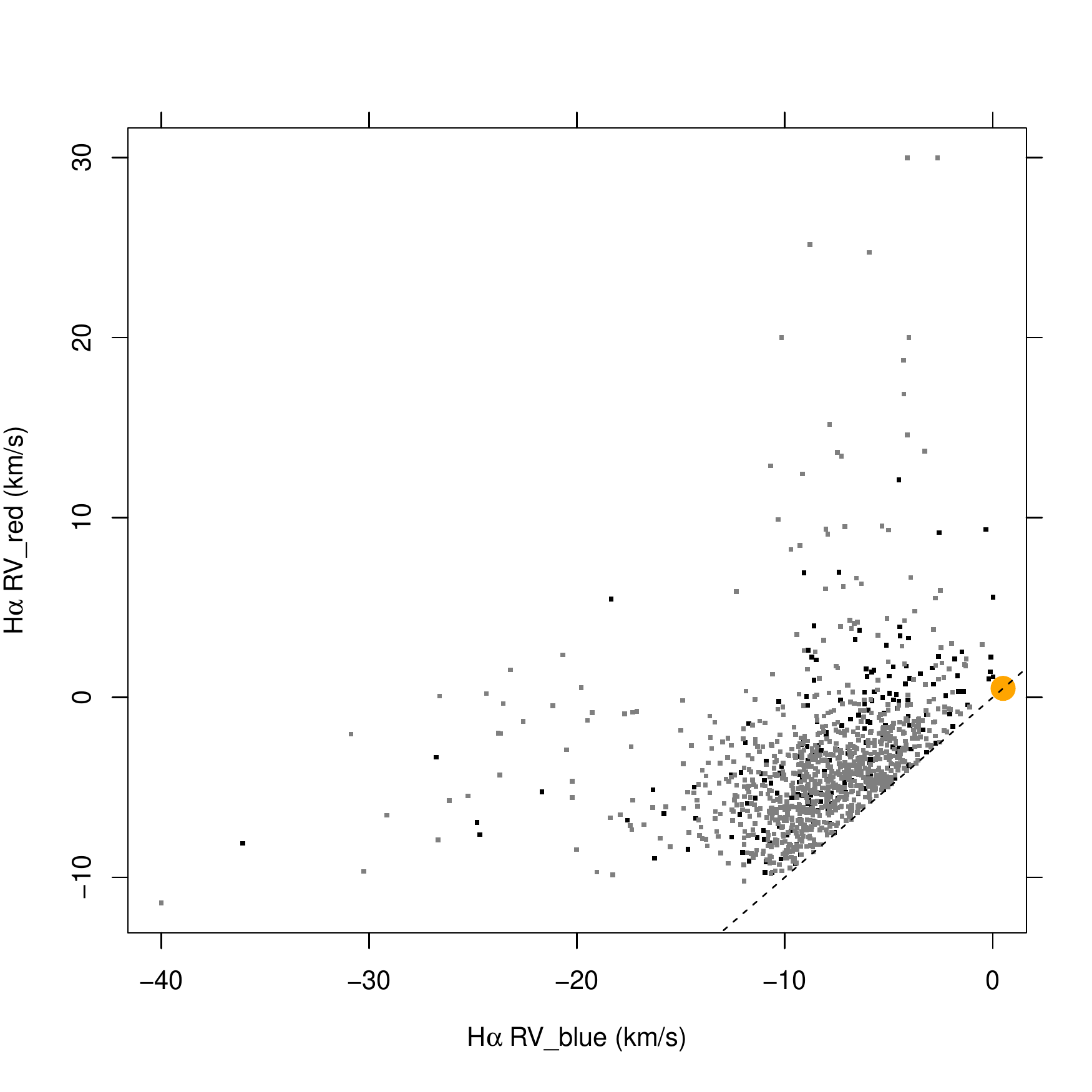}}
\caption{Comparison between RV of ``blue" and ``red" components as
resulting from 2-g model fits of the \ha\ line.
Symbols as in Fig.~\ref{ha-n2-s2}. The orange circle indicates the RV of
the late-type stars in NGC6530, from Prisinzano \e (2007).
\label{ha-rv2}}
\end{figure}

\begin{figure}
\resizebox{\hsize}{!}{
\includegraphics{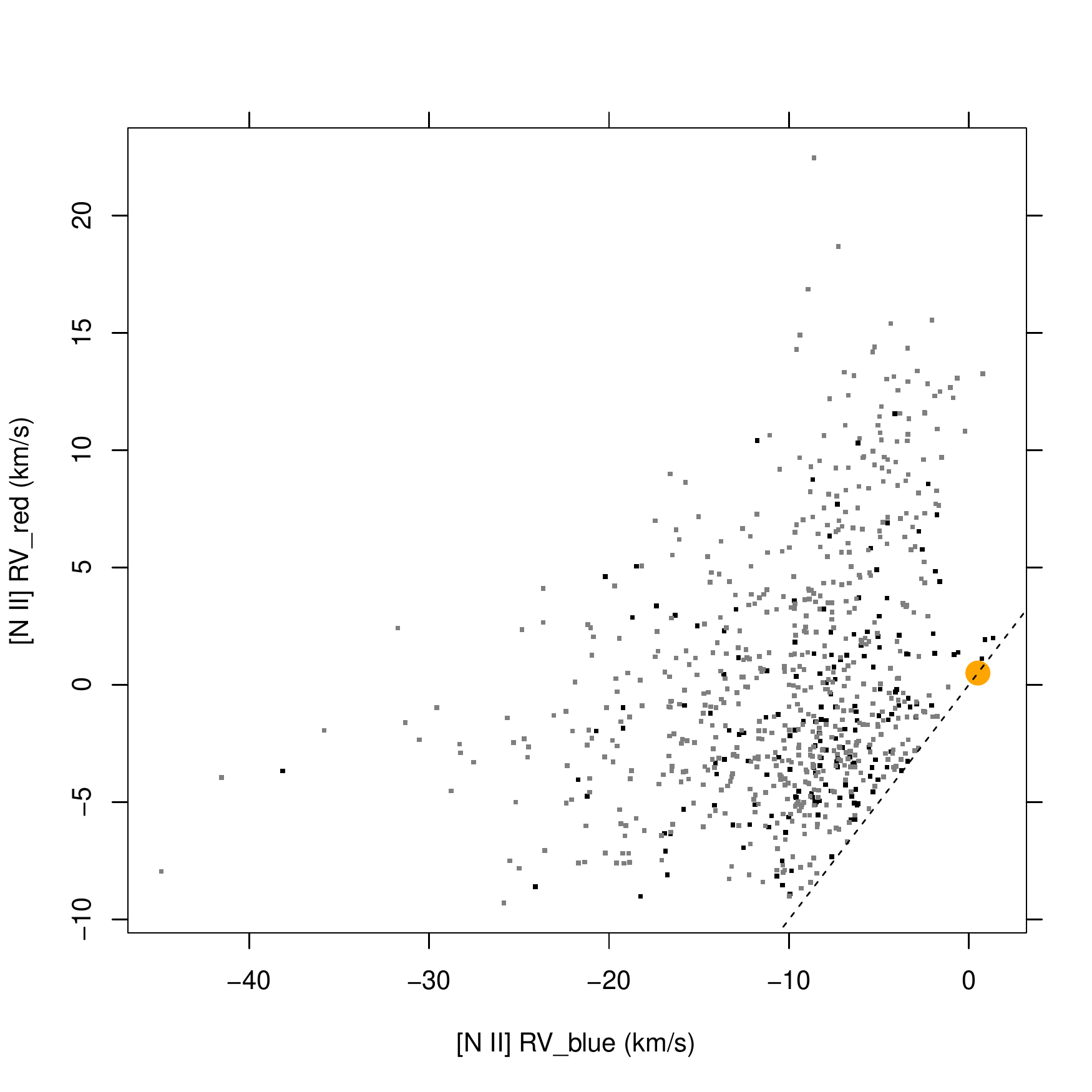}}
\caption{Same as Fig.~\ref{ha-rv2}, for the [N II] line.
\label{n2-rv2}}
\end{figure}

We then examine the details of line profiles, as provided by our 2-g
best-fit models. In Figure~\ref{ha-rv2} the RV of the ``blue" component
is compared with that of the respective ``red" component in the same
spectrum; the orange point indicates the cluster $RV_{cm}$. Also these
more detailed RVs show motions predominantly towards us (as seen from
center of mass); only a small number of spectra show blue and red RV
components lying on opposite sides of $RV_{cm}$ (i.e.,
$RV_{red}>0.5$~km/s and $RV_{blue}<0.5$~km/s), as it would be expected
for an expansion originated from center of mass velocity.
In the large majority of cases, the two components, both approaching us,
have RVs differing by only few km/s, whose physical origin is not
completely clear. One possibility is that the two components are
actually an oversimplification of reality, and that they only represent
the approximate RV range found in a rapidly decelerating layer emitting
\ha. This accounts well for the significant correlation found in
Fig.~\ref{ha-rv2} between $RV_{red}$ and $RV_{blue}$ in the \ha\
line\footnote{Alternatively, this might be an instrumental effect
arising from the non-gaussianity of the line-spread-function, a
systematic effect that becomes observable at the highest signal levels,
see Damiani \e (2016, Appendix).}.
{
The median intensity of the blue component is only slightly higher than that of
the red component (1.3 to 1.5 times, from pure-sky and faint-star fibres
respectively), thus backscattering from dust is unlikely to account for the
bulk of the red component.
}

\begin{figure}
\resizebox{\hsize}{!}{
\includegraphics[angle=90]{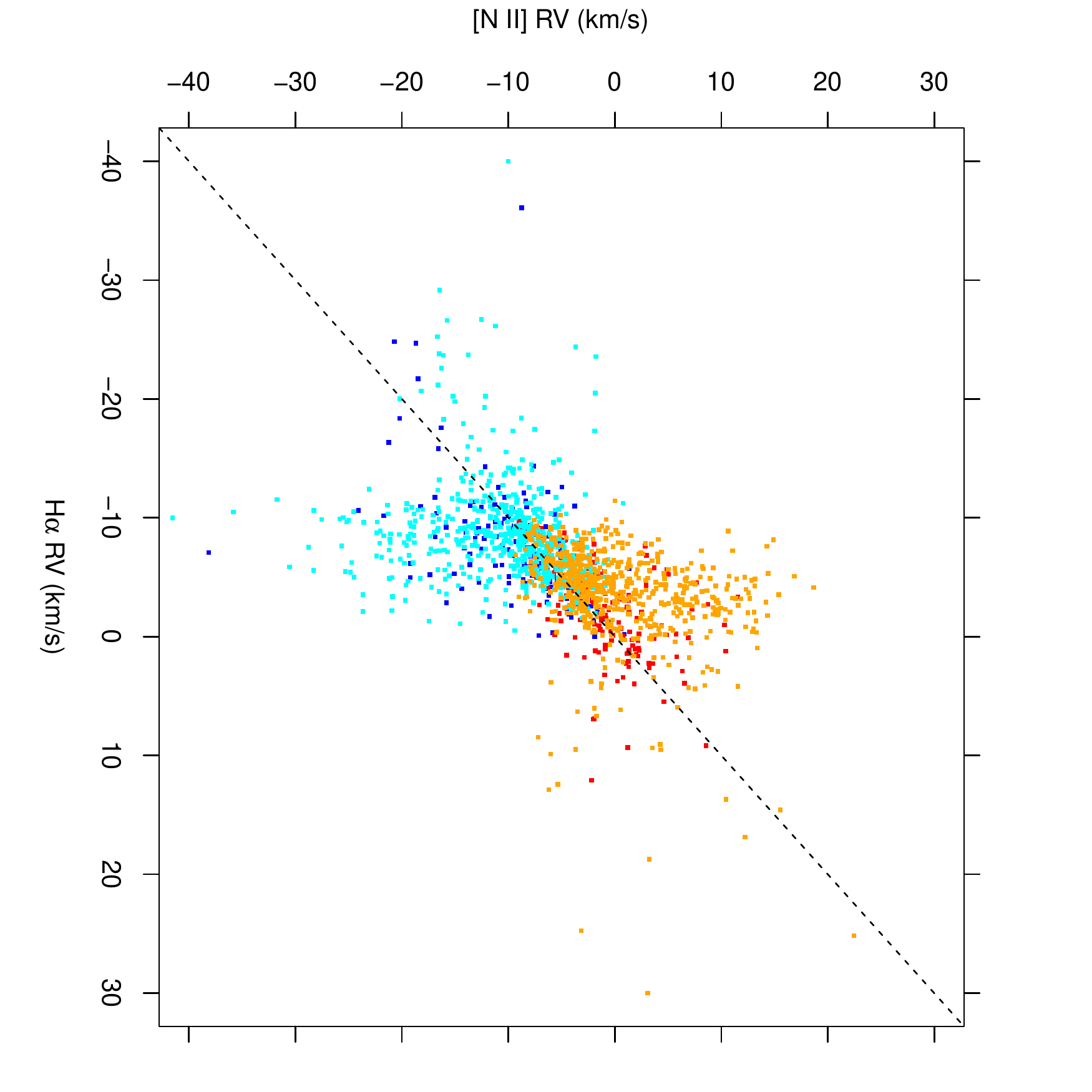}}
\caption{Comparison between RVs of \ha\ and [N II] 2-g fits.
{
The results of each individual spectrum are plotted twice, once for the red
\ha\ and [N II] components, and again for the blue components of the
same lines.
}
Blue/red
dots refer to ``blue/red" components from sky fibres, while cyan/orange
dots refer to the same components from faint stars.
\label{ha-n2-rv2}}
\end{figure}

\begin{figure}
\resizebox{\hsize}{!}{
\includegraphics[angle=90]{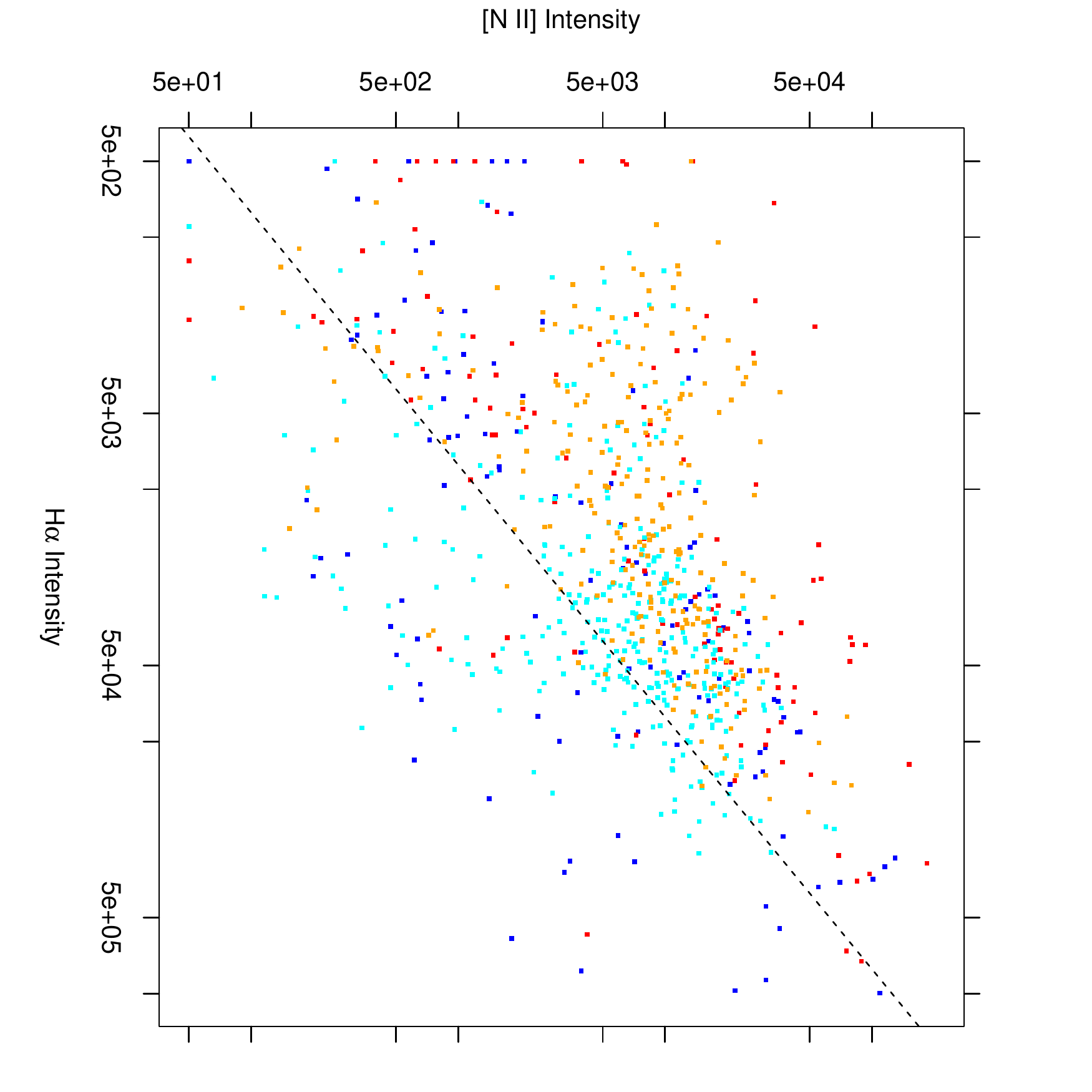}}
\caption{Comparison between intensities of \ha\ and [N II] (log scales),
from 2-g fits.  Symbols as in Fig.~\ref{ha-n2-rv2}.
\label{ha-n2-norm2}}
\end{figure}

Still different is the picture derived from the corresponding diagram
involving the [N II] line (Figure~\ref{n2-rv2}). Motion receding from
us is much more frequently found, and the correlation between $RV_{red}$
and $RV_{blue}$ is much less tight, if existing at all. The blue and red
components of [N II] are therefore in many cases indicative of
dynamically distinct gas layers, whose spatial characteristics will be
examined in detail in Section~\ref{maps}.
It is interesting to compare the RVs for the \ha\ and [N II] lines, as
given by the 2-g fits, analogously to the 1-g RV comparison of
Fig.~\ref{ha-n2-s2}$a$. This is done in Figure~\ref{ha-n2-rv2}: here we see
that, unlike the 1-g fit RVs, there is in most cases a fairly good match
between RVs of the two lines, apart from a minority of strong outliers.
In order to reconcile this agreement with the disagreement of 1-g RVs
shown by Fig.~\ref{ha-n2-s2}$a$, one may expect that the relative
intensities of the blue and red components are different between \ha\
and [N II], with the red component being dominant in [N II] over \ha, and
viceversa for the blue component.
This expectation is confirmed by the diagram of
Figure~\ref{ha-n2-norm2}, where the intensities of the different lines
are compared (for each blue/red component separately). Only components
with an absolute RV difference less than 3~km/s were plotted.
It is clear that red components tend to have a larger [N II]/\ha\
intensity ratio than blue components, and therefore lower ionization, as
mentioned above. Recalling the above result that the two components are
diagnosing different layers from gas moving in the same direction, we
infer that the gas moving faster (larger negative velocities: blue
component) is more ionized than the slower red component.


\subsection{[O III] and sodium D lines from UVES data}
\label{uves}

\begin{figure}
\resizebox{\hsize}{!}{
\includegraphics[angle=90]{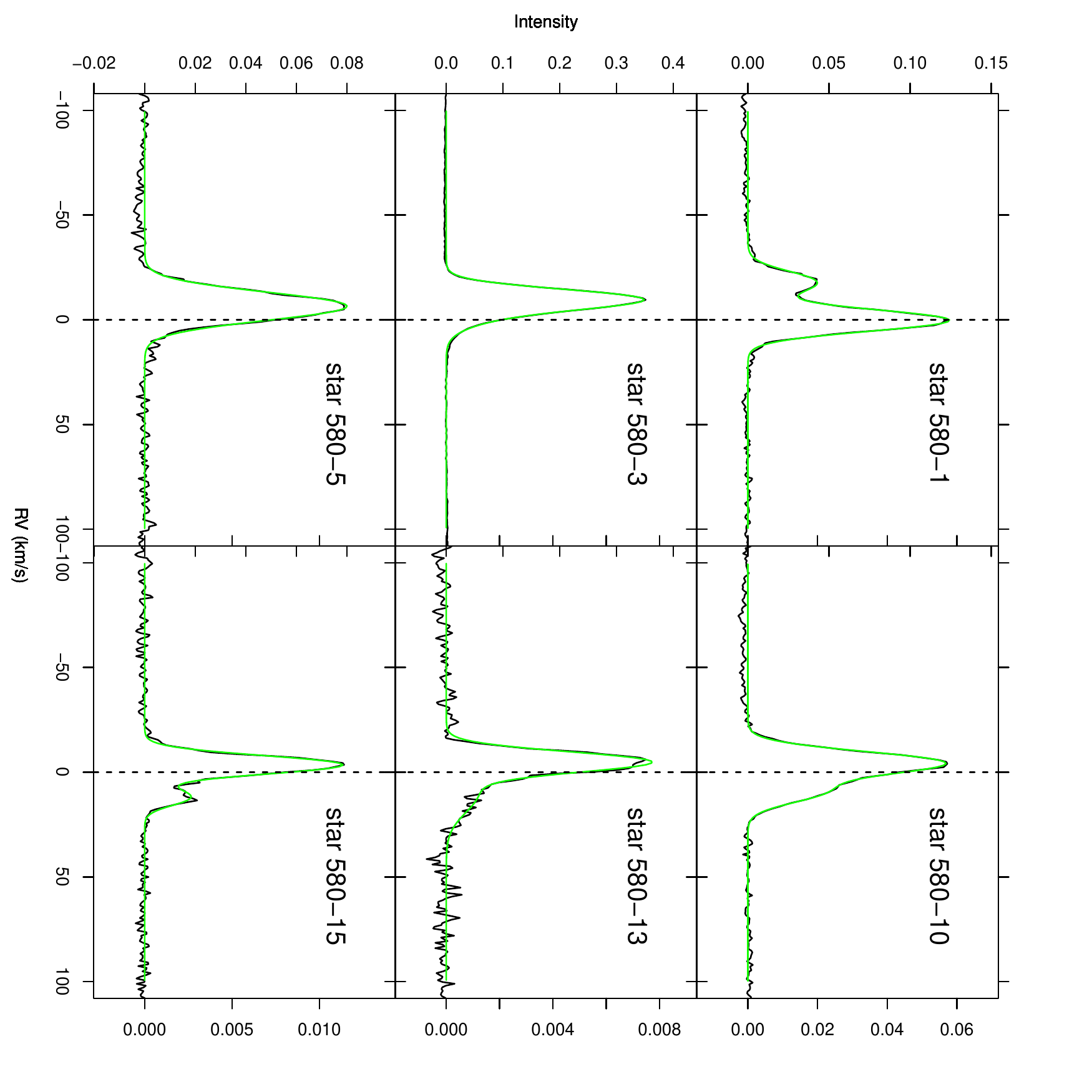}}
\caption{Line profiles (black) and 2-g best-fit models (green) for the [O III]
5007 line, from UVES stellar spectra.
\label{oxygen-1}}
\end{figure}

\begin{figure}
\resizebox{\hsize}{!}{
\includegraphics[angle=90]{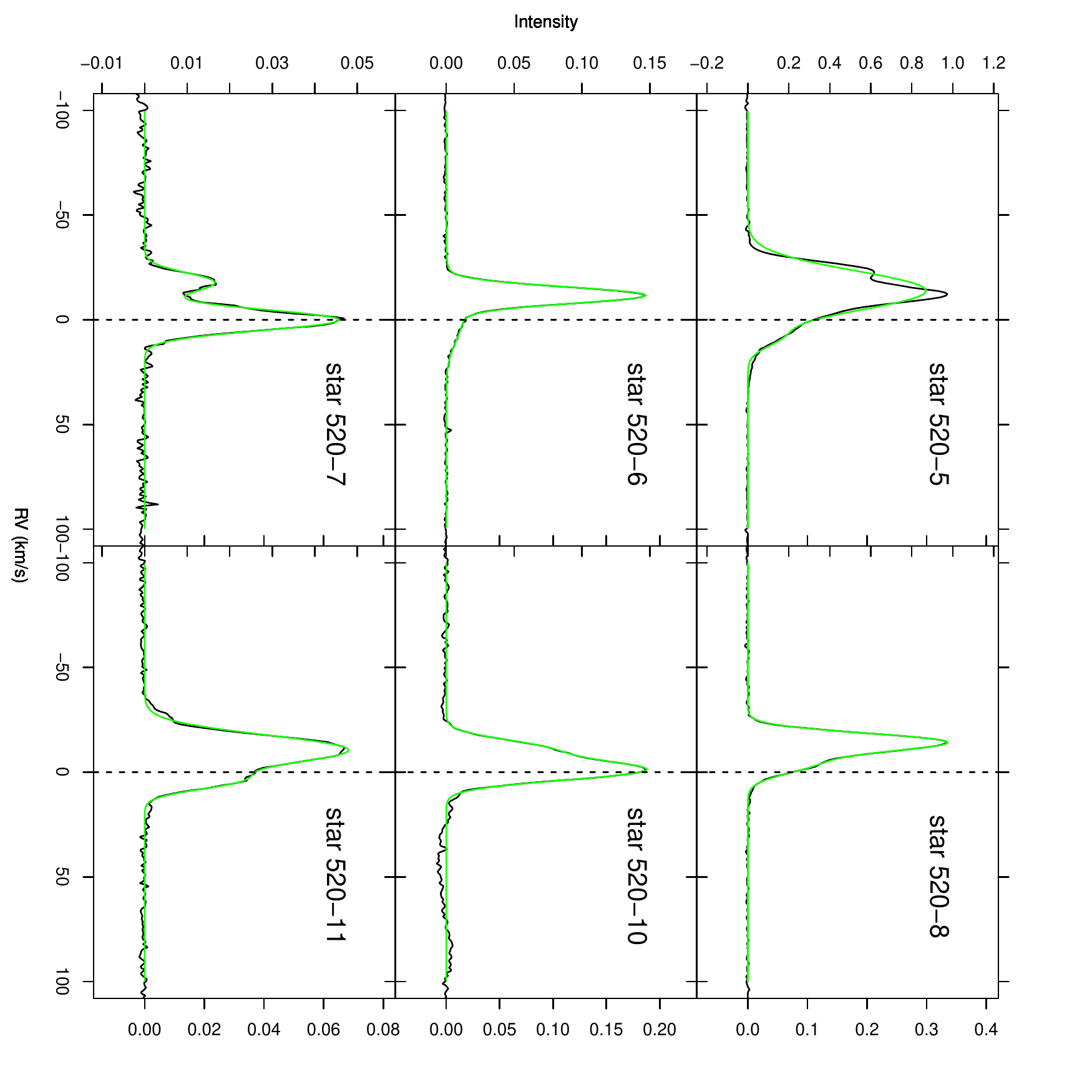}}
\caption{Additional [O III] line examples, as in Fig.~\ref{oxygen-1}. 
Star 520-5 is the O star Her~36, the only case in which the two-Gaussian
profile fails to match well the observed profile.
\label{oxygen-2}}
\end{figure}


A representative selection of line profiles of the [O III] 5007\AA\
line\footnote{The exact wavelength used here is 5006.843\AA, from NIST
(http://physics.nist.gov/asd, Kramida \e 2015).} from UVES
spectra of several stars is shown in Figures~\ref{oxygen-1} 
and~\ref{oxygen-2}. Although all spectra are from stars and not sky
fibres, the nebular line is clearly evident; no stellar spectrum
subtraction was performed (apart from a constant continuum level) for
any of the spectra shown.
This line is frequently the second strongest nebular line in our
spectra, after \ha.
In the [O III] line doubly-peaked profiles are much
more frequently found (as already reported by Elliot and Meaburn 1975)
than in lower-ionization lines like those of [N II].
Even if no double peaks are found, line distortions are evidently
present in most cases, and require 2-g fits. In only one case (labeled
star 520-5, upper left panel in Fig.~\ref{oxygen-2}) two Gaussians
provide a bad fit (as shown), and three components are suggested by the
line profile: this is the O7 star Herschel~36 in the Hourglass nebula,
having also the strongest [O III] line among our sample.
Shown with green lines in Figures~\ref{oxygen-1}-\ref{oxygen-2} are our
2-g model fits, which reproduce the [O III] profiles very well, except
for Herschel~36.

\begin{figure*}
\sidecaption
\includegraphics[width=12.5cm]{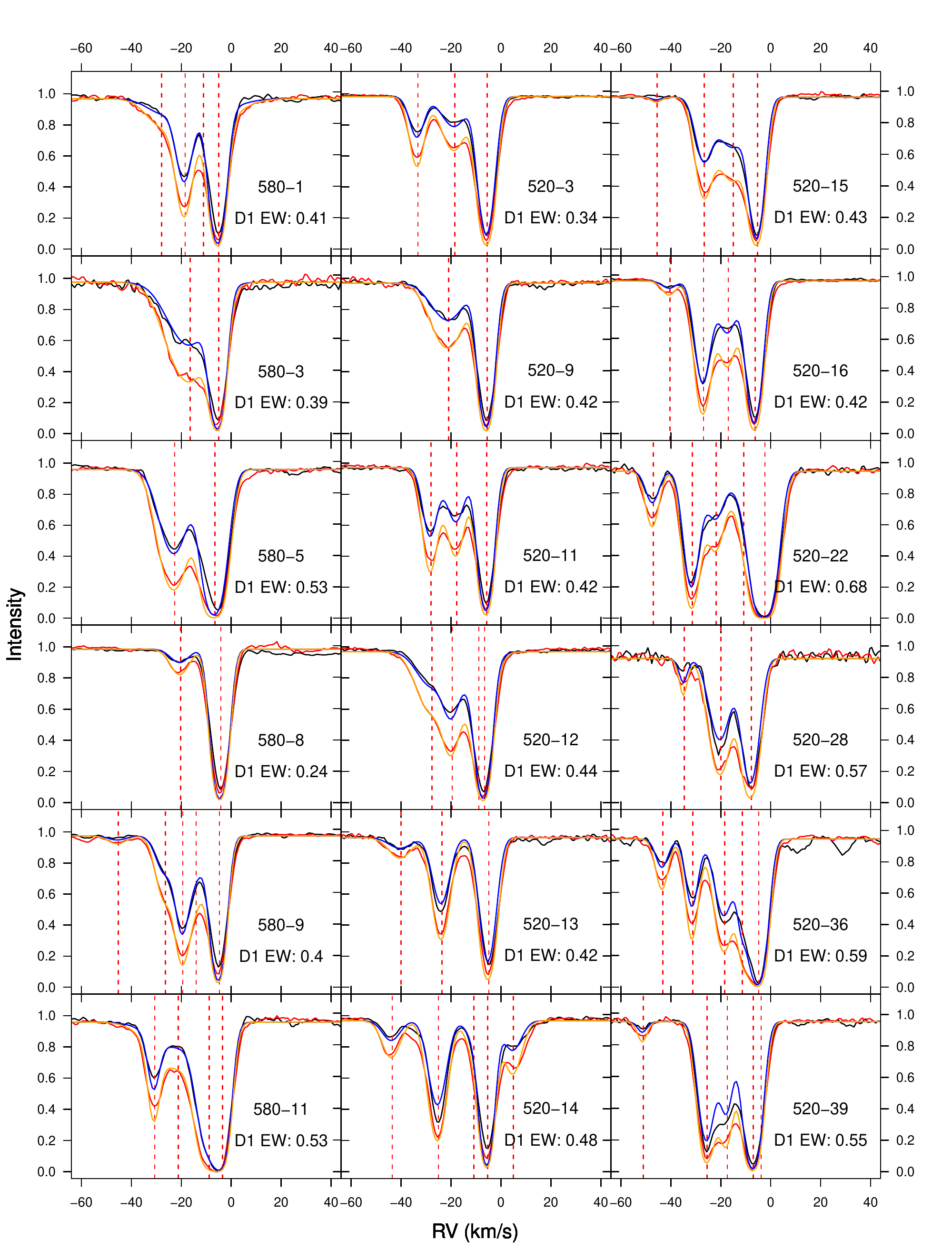}  
\caption{Profiles of interstellar sodium  D1 (black) and D2 (red) lines
from UVES spectra (both setups 580 and 520, as labeled).
The stellar line contributions,
when present, were removed by division with suitable template spectra.
The orange curves are multiple-Gaussian best-fit models to the D2
profiles; the blue curves are D1 profile predictions based on the D2
models, and assuming pure absorption (see text).
Vertical red dashed lines indicate the best-fit RVs of model components.
Total equivalent widths for the D1 lines are also indicated (in Angstroms).
\label{sodium-1}}
\end{figure*}

The UVES spectra of stars in NGC6530 also enable us to study the neutral
gas along the line of sight to these stars, using the sodium D doublet.
The selection of UVES targets in the Gaia-ESO Survey is such as to
maximize the probability of their cluster membership (Bragaglia \e, in
preparation); therefore, most (not all)
UVES spectra are likely to sample the entire column of neutral sodium
between us and NGC6530. A wide selection of sodium absorption line
profiles for these stars is presented in Figure~\ref{sodium-1}.
For not-too-hot stars, the stellar sodium lines were also evident in the
spectra: these were divided out, using suitable template spectra chosen
among the UVES-POP library (Bagnulo \e 2003), such that Fig.~\ref{sodium-1}
shows only non-stellar absorption components.
Both doublet lines are shown, the D1 line in black, and the stronger D2
line in red. The occasional features at $\sim 10$ and ~30~km/s in the D1
line (only) are telluric. The intensity ratios of the two lines are
closely related, being originated from the same (ground) level, and
having oscillator strengths of $f_{D2}=0.6405$ (D2) and $f_{D1}=0.3199$ (D1).
The low ionization energy of sodium (5.139~eV) implies that these lines
must originate from layers more distant from the OB stars
(nearer to us) than the ionic lines studied in Section~\ref{giraffe}.
In basically all cases, more than one absorbing layer, each with a
distinct RV, is needed to model the sodium absorption profiles.
The most complex line profiles is modelled with five Gaussians.

The fitting function was chosen as follows: we rewrite eq.(1) from Hobbs
(1974) using $\lambda$ instead of $\nu$ as (for a single component)

\begin{equation}
N \frac{\pi e^2}{m_e c} \frac{\lambda^2}{c} f = N \int \alpha_{\lambda}
\; d \Delta \lambda = \int (- \ln r_{\lambda}) \; d \lambda
\label{eq1}
\end{equation}

and the term containing the residual intensity $r_{\lambda}$, for a
combination of absorbing layers $i$, as
\begin{equation}
\int (- \ln r_{\lambda}) \; d \lambda = \sum_i N_i \int \frac{1}{\sqrt{2 \pi}
\sigma_i} \exp{\left(
-\frac{1}{2}\frac{(\lambda-\lambda_i)^2}{\sigma_i^2}\right)} \; d \lambda
\label{eq2}
\end{equation}

where $\lambda_i = \lambda_0 +v_i/c$ is the central wavelength of absorbing
component $i$, at velocity $v_i$, and $\lambda_0$ is the line rest
wavelength. $\sigma_i$ is the component intrinsic width in wavelength units.
The explicit form for $r_{\lambda}$ is therefore:

\begin{equation}
r_{\lambda} = \exp \left(-
\sum_i \frac{N_i}{\sqrt{2 \pi} \sigma_i} \exp{\left(
-\frac{1}{2}\frac{(\lambda-\lambda_i)^2}{\sigma_i^2}\right)}
\right)
\label{eq3}
\end{equation}

Since the lines from cold neutral sodium may be very narrow,
and saturated absorption profiles may become highly non-Gaussian,
the instrumental resolution of UVES was separately introduced as a fixed
Gaussian broadening through convolution, and the final functional form used to
fit the observed, normalized line profiles is:

\begin{equation}
r_{\lambda}^{obs} = r_{\lambda} \otimes \int \frac{1}{\sqrt{2 \pi}
\sigma_{UVES}} \exp{\left(
-\frac{1}{2}\frac{\Delta \lambda^2}{\sigma_{UVES}^2}\right)} \; d \Delta\lambda
\label{eq4}
\end{equation}

where $\otimes$ is the convolution operator and $\sigma_{UVES}$ ($\sim
2.7$~km/s) is the UVES instrumental linewidth.
The chosen function for $r_{\lambda}$ corresponds to absorption from
cold gas, whose emissivity is approximated as zero.
Using this function, non-linear fits were attempted to the D2
line profiles (the strongest of the doublet) with two to five
components; the optimal number of fitting components was determined by
inspecting the results visually. The chosen best-fit models of D2 lines
are superimposed to the observed profiles in Fig.~\ref{sodium-1} as
orange curves: these match so well the observed profiles as to result often
indistinguishable from them.  Velocities of the individual best-fit components
are indicated in the Figure with vertical red dashed lines.
They range from $\sim -50$~km/s to $\sim +10$~km/s; however,
positive-velocity components are found only in two cases, the most
evident example being that of star 520-14 in the Figure.
As it should be clear from eq.\ref{eq1}, the profiles of the D1 line
can be obtained from those of the D2 line (eqs.\ref{eq3} and \ref{eq4})
by multiplying the intensity
of each component $N_i$ by the constant factor $f_{D1}/f_{D2}$.
Model profiles of the D1 line obtained in this way are also shown in
Fig.~\ref{sodium-1} as blue lines. In most cases, they match well the
observed D1 profiles (black), providing a test of the goodness of
the adopted best-fit models. Best-fit parameter values are reported in
Table~\ref{table-sodium}. Intensity values $N_i>1000$ correspond to saturated
components and are highly uncertain.

\begin{figure}
\resizebox{\hsize}{!}{
\includegraphics{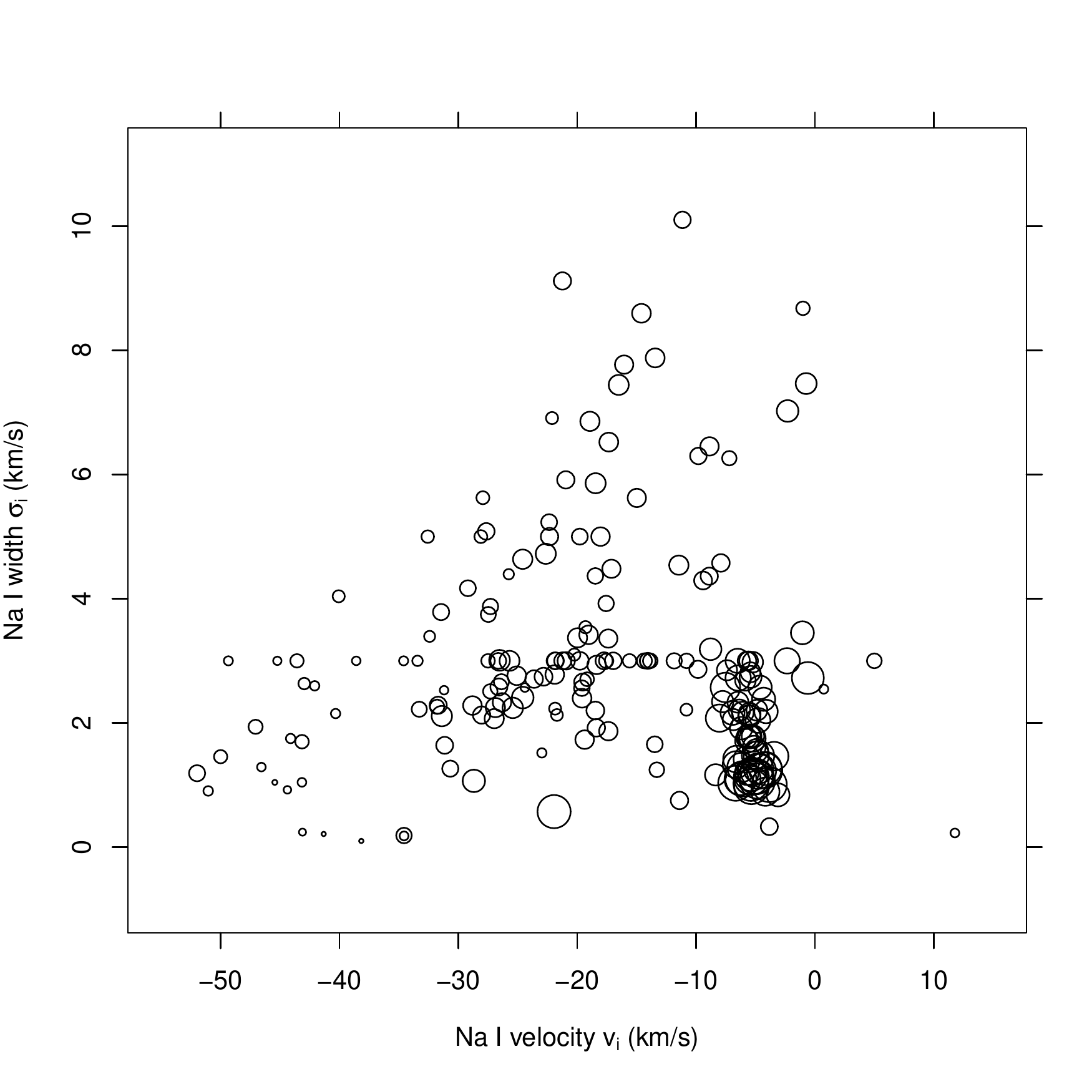}}
\caption{Sodium absorption linewidths $\sigma_i$ vs.\ radial velocities
$v_i$, for all fitted components.
Circle size is proportional to component intensity.
\label{na-rv-sigma}}
\end{figure}

However, a number of cases are found where the observed D1 profile and
its model do not match (see e.g.\ stars 520-13, 520-14, 520-28, 520-36,
520-39), to a significant degree for the given S/N. The discrepancy
occurs always near velocities of $\sim -30$ to $-20$~km/s. It is always
in the same sense, with the observed D1 profile being deeper and closer
to the respective D2 profile than the model would predict: this is
suggestive of that particular component saturating towards a finite,
non-zero intensity, contrary to the assumption above of zero emissivity.
If this latter condition is not met, the profile modeling becomes extremely
more complex, the contributions from different layers not being writable
as a linear sum in $\ln r_{\lambda}$; also the order of layers along the
line of sight becomes important (the emission of one layer can be absorbed
only by the layers closer to us), which is not the case when pure
absorption is modelled. We have therefore not modelled those few cases in
quantitative detail. Qualitatively, it is suggested that the layers at
velocities $\sim -30$ to $-20$~km/s are sometimes hotter than those at
lower and higher velocities, their source function being much higher
than zero compared to the photospheres of the UVES targets. Temperatures
of several thousands K are therefore likely for those particular sodium
layers. We consider it likely that these layers are spatially adjacent
to those where recombination has recently taken place, as also supported
by their similar velocities.
The higher temperatures in the sodium layers at
$\sim -30$ to $-20$~km/s are also supported by their increased line
widths $\sigma$, as shown by Figure~\ref{na-rv-sigma}.
We remark that this figure shows intrinsic line widths $\sigma_i$, not the 
observed values, degraded by the instrumental resolution (as in
Eq.~\ref{eq4}). For this reason, their lower bound was set to zero in
the fitting; however, values of $\sigma_i$ lower than $\sim 3$~km/s should
be regarded as upper limits.

The wide variety of absorption profiles found within small angular
scales on the sky suggests strongly that the associated neutral gas lies
in the immediate vicinity of the Lagoon nebula, and not all along the
line of sight.
Also the observed large radial velocities of sodium absorption are not expected
for the general ISM gas in nearly circular orbits and observed almost in the
direction of Galactic center, and further support this argument.
The same cannot be said for the strong, saturated
component near $\sim -5$~km/s, which although not strictly identical in
all profiles is so widespread (and consistently narrow)
that we cannot rule out a line-of-sight
origin. A puzzling feature from the sodium spectra is the clear
existence of components at much larger negative velocities than found
from the ionic lines in Section~\ref{giraffe}: at face value, this would imply that the
expanding gas, after recombining at a typical approaching velocity of
$-5$ to $-10$~km/s (Fig.~\ref{ha-n2-s2}$a$), keeps accelerating towards us
to produce the neutral layers around $-30$ to $-40$~km/s.
Also the origin of the discrete velocity components (as opposed to a
continuous velocity distribution) in the sodium profiles is unclear,
i.e.\ whether it is related to a episodic energy input or to different
pre-existing layers of neutral gas, swept by the expanding envelope.
In the former case, a single star would be driving the phenomenon, or
otherwise the uncorrelated input from many stars would produce a
smoother velocity distribution. In the second, a continuously
(non-episodically) expanding shell would encounter discrete gas layers,
and decelerate as more and more mass is being pushed; this seems
contradictory with the above suggestion that neutral gas is accelerated
outwards, in the sodium-line formation region.

We have also computed equivalent widths for the D1 line (labeled D1
EW in Figure~\ref{sodium-1}), which fall in the range $0.4-0.6$~\AA\
for the bulk of stars. Using the relation given by Munari and Zwitter
(1997), this range would imply a $E(B-V)$ range of $0.2-0.5$, in good
agreement with that found by Sung \e (2000) for the NGC6530 massive stars.

\subsection{Spatial maps}
\label{maps}

\begin{figure*}
\includegraphics[width=17.5cm]{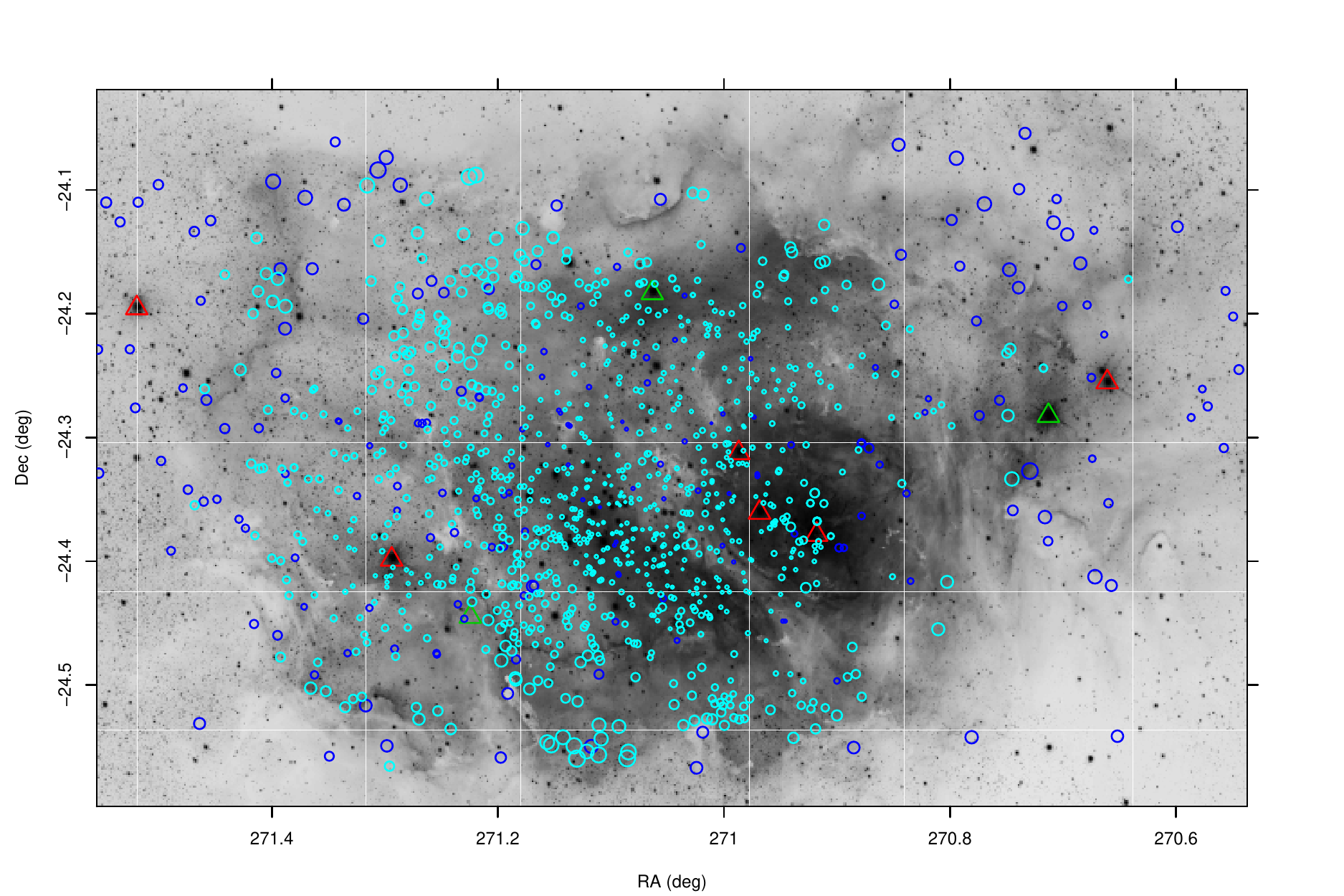} \\
\includegraphics[width=17.5cm]{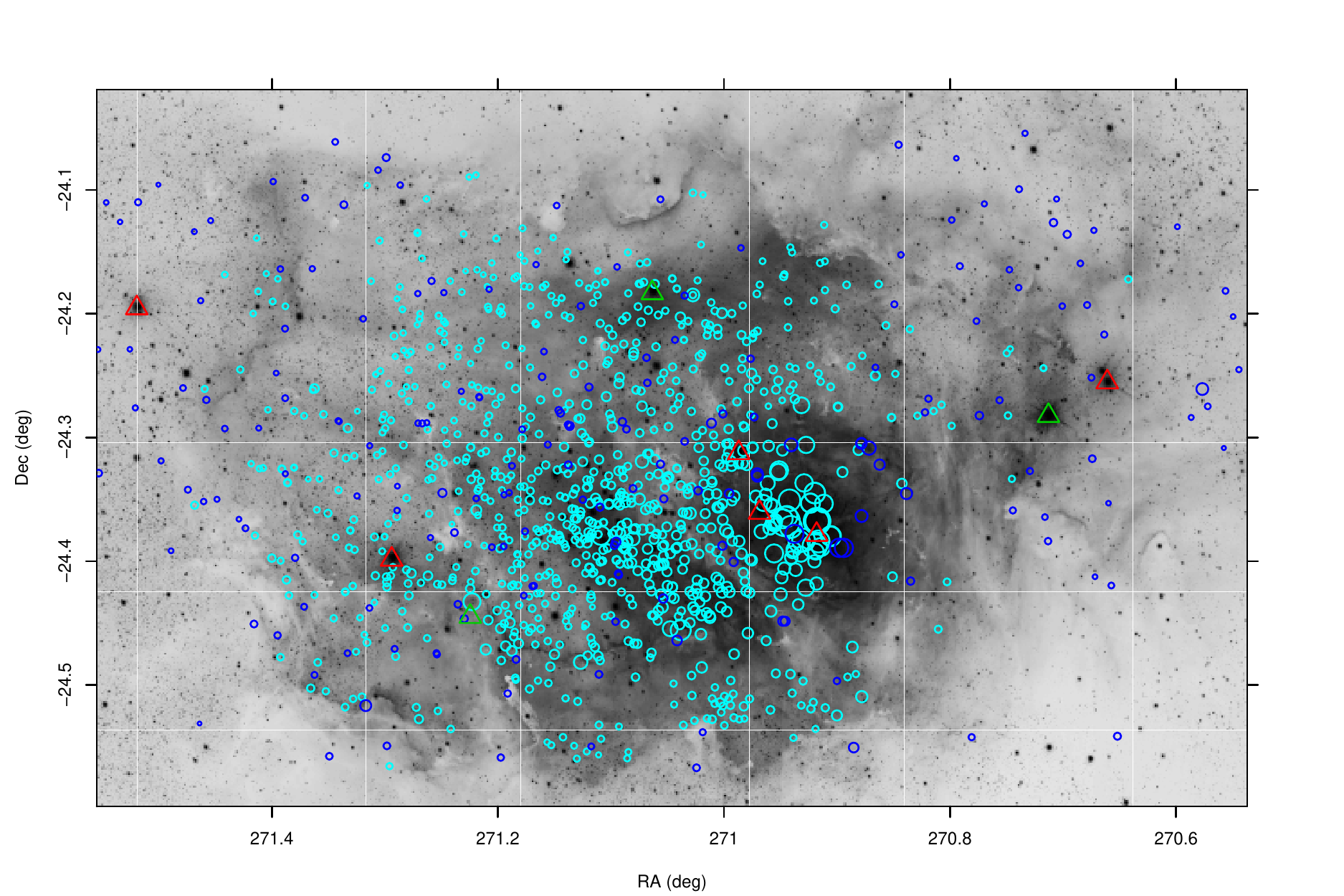}
\caption{
$a$ (upper panel): Map of intensity ratio between [N II] and \ha\ 
(proportional to circle size) from 1-g model fits,
superimposed to the same VPHAS$+$ image as in Fig.~\ref{vphas-fibers}.
Blue (cyan) circles refer to pure-sky (faint star) fibres.
Triangles have the same meaning as in Fig.~\ref{vphas-fibers}.
$b$ (lower panel): Map of [S II] 6731/6717 intensity ratio (proportional to
circle size, and increasing with density).
Symbols as in panel $a$.
\label{map-ha-n2-ratio}}
\end{figure*}


\begin{figure*}
\resizebox{\hsize}{!}{
\includegraphics[width=17.5cm]{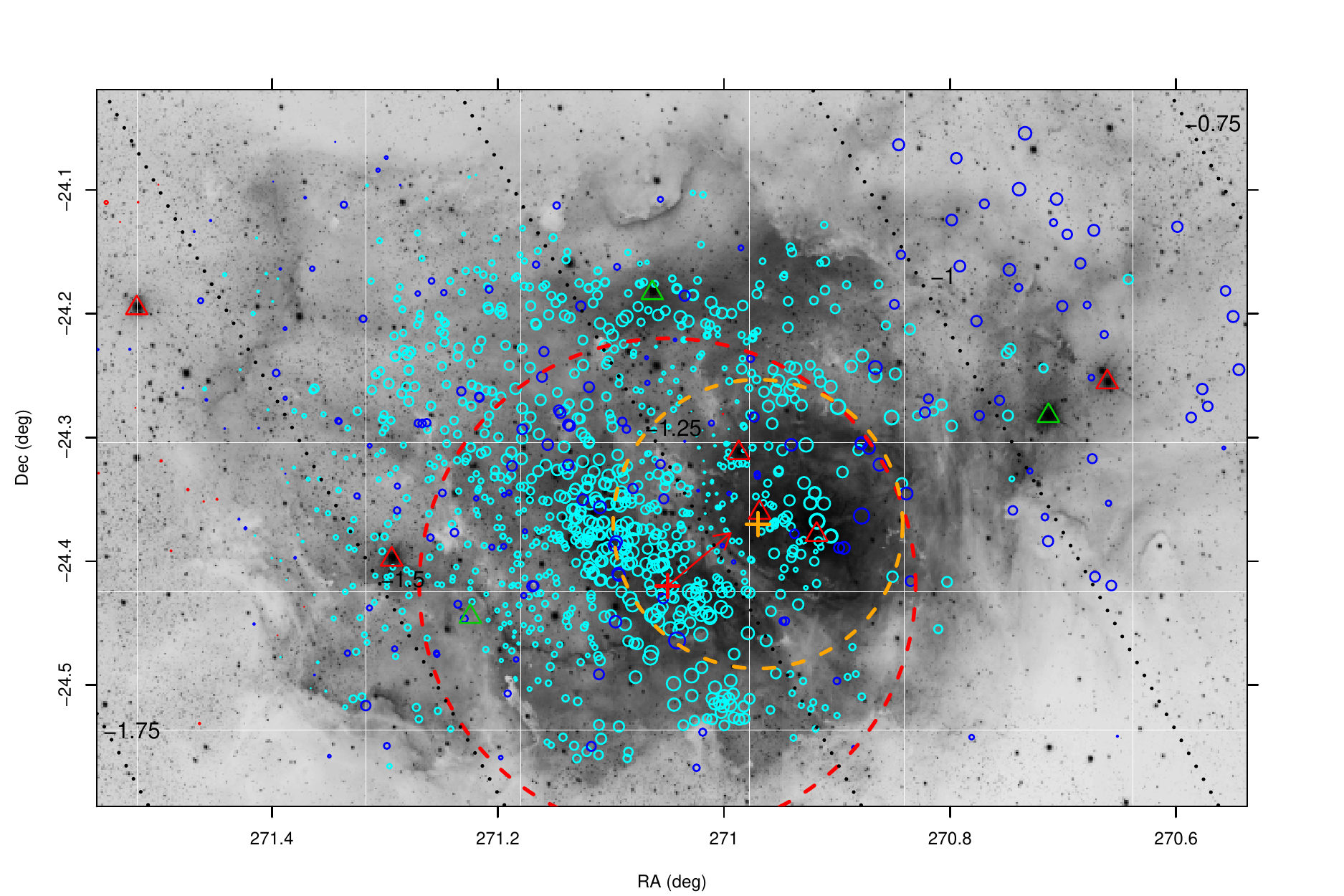}}
\caption{Map of \ha\ RV (absolute value proportional to circle size)
from 1-g model fits. Blue (cyan) circles: negative-velocity values from
sky (faint-star) fibres;
{
Red circles: positive-velocity values (only a dozen datapoints, all
near zero velocity and lying close to the eastern edge).
}
Triangles have the same meaning as in Fig.~\ref{vphas-fibers}.
Oblique dotted lines indicate Galactic latitudes $b=-1.75$ (left) to
$b=-0.75$ (right), in steps of $\Delta b=0.25$.
Plus signs indicate reference positions for the NGC6530 core
(red) and Hourglass region (orange). Centered on these
positions, two dashed circles are shown, of radii 12' (red) and 7'
(orange), respectively. The red arrow indicates the direction of
steepest RV gradient around the cluster core, nearly orthogonal to the
galactic plane.
\label{map-ha-rv}}
\end{figure*}

Most of the properties characterizing the nebular emission, as derived
from the line modelization discussed in Section~\ref{giraffe}, show distinct spatial
patterns, which we examine in detail here.
Figure~\ref{map-ha-n2-ratio}$a$ shows a map of the [N II]/\ha\ intensity
ratio (proportional to symbol size), superimposed on a VPHAS$+$ image of
the nebula as in Fig.~\ref{vphas-fibers}. As mentioned in
Section~\ref{giraffe}, a smaller
intensity ratio corresponds to higher ionization parameter. While one
would expect the latter to increase in the vicinity of the hottest
stars (red triangles in the Figure), this is not always found in our data.
Not considering the two O stars near the East (HD~165246) and West (HD~164536)
edges of the nebula, where our spatial coverage is sparse, we observe a
clear ionization increase around HD~165052 (in South-East region), while
the bright nebula surrounding the Hourglass shows no clear ionization
pattern around its three O stars (North to South: HD~164816, 9~Sgr, and
Herschel~36). The darker nebula parts (the ``Great Rift" in Lada \e
1976) around this brightest region shows distinctly higher ionization
than near the Hourglass. The relatively low ionization near 9~Sgr supports
further the suggestion by Lada \e (1976) that this star should lie
several pc in front of the nebular material, not in its immediate
vicinity. The irregular ionization near Herschel~36, instead, can be
attributed to the patchy distribution of dense gas and dust all around
this star, as suggested by the HST image of the Hourglass (Tothill \e
2008). Near the central NGC6530 cluster, to the S-E of the Great Rift,
ionization is often higher than in the Hourglass region; however, there
seems to be a gradient across the cluster, not a peak near its center,
so that the source of ionization is probably not internal to the
cluster, with the best candidate remaining the O4 star 9~Sgr, despite
being a few parsecs away. The issue will be examined in more detail in
Section~\ref{9sgr} below.
The bright blue supergiant HD~164865 (B9Iab) probably contributes to
ionization locally, but not predominantly since there is no associated
ionization peak near it.

\begin{figure*}
\resizebox{\hsize}{!}{
\includegraphics[width=17.5cm]{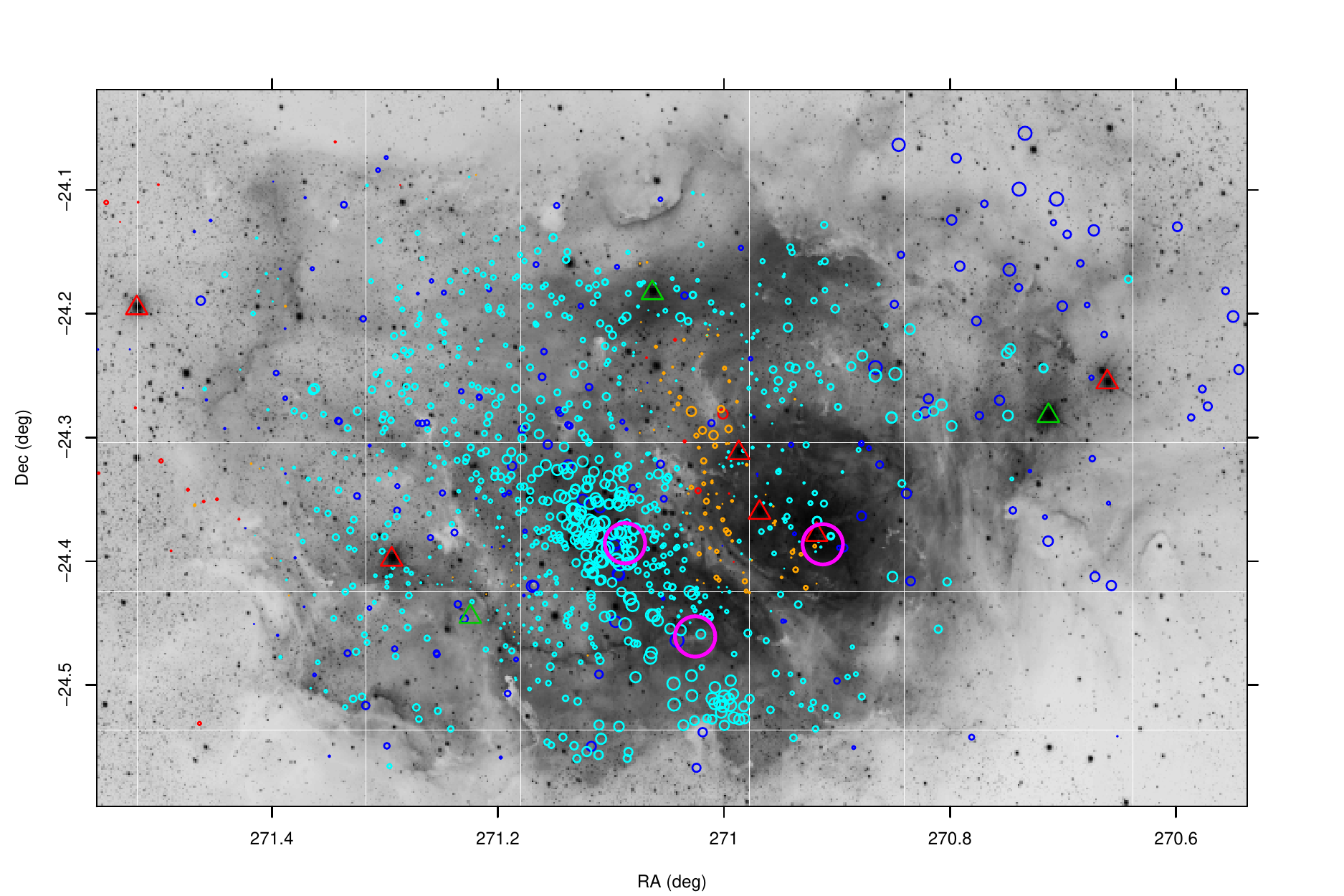}}
\caption{Map of [N II] RV from 1-g model fits, analogous to
Fig.~\ref{map-ha-rv}, with same meaning of symbols,
{
plus orange circles representing positive-velocity datapoints from
faint-star fibres.
}
The three large magenta circles indicate the positions of the CO bright
spots found by Lada \e (1976).
\label{map-n2-rv}}
\end{figure*}

We next consider the nebular electron density, as measured from the
ratio of [S II] 6717, 6731\AA\ lines (Figure~\ref{map-ha-n2-ratio}$b$).
The spatial pattern is here very different than the ionization pattern:
a strong and distinct increase in density ($\sim 3000 \times
\sqrt{10^4/T}$ cm$^{-3}$, in agreement with Bohuski 1973) is found throughout
the vicinity of the Hourglass, with the peak coinciding with the Hourglass
proper. The agreement between line ratios from pure-sky fibres and
faint-star spectra is very good, which is especially crucial here
because of the near coincidence between the [S II] 6717 line and a Ca~I
photospheric line. A very localized enhancement in density is also found
in the immediate vicinity of the bright rim near M8E-IR (green triangle
S-E of NGC6530). Near the stellar cluster core, instead, the density is
not particularly high ($\sim 800 \times
\sqrt{10^4/T}$ cm$^{-3}$), not higher than in the neighboring Great Rift,
despite the large difference in nebular brightness. The density
decreases very smoothly towards the nebula edges, to $\sim 50 \times
\sqrt{10^4/T}$ cm$^{-3}$. The density was derived from the doublet ratio
using analytic expressions in Weedman (1968) and Saraph and Seaton
(1970).

The radial velocity map for \ha, shown in Figure~\ref{map-ha-rv} provides a
vast amount of information. Velocities (proportional to circle sizes)
are here those derived from 1-g fits.
Since $RV_{cm} \sim 0$, absolute velocities are nearly the same as
$|RV - RV_{cm}|$, that is referred to the NGC6530 center of mass.
There is no central symmetry in
the velocity field. The cluster core (red plus sign) nearly coincides
with a (negative) velocity maximum: absolute velocities decrease towards
both S-E and N-W from this position. The location of these
low-absolute-velocity datapoints defines a reference direction,
indicated with a red arrow in the Figure; this is almost coincident with
the normal to the galactic plane.
The galactic plane itself is just off the figure region
to the right. The Hourglass region is also characterized by large
negative velocities, but no velocity minimum is detected West of it,
probably also because of the incomplete spatial coverage. In the
outermost nebula regions to the East the velocity smoothly decreases
towards $RV_{cm}$.
However, this does not happen in the western edge, where
velocities remain at large negative values, a surprising fact which will
be discussed in more depth in Section~\ref{pv-outer}.

The velocity map obtained from the 1-g [N II] line fits is instead shown in
Figure~\ref{map-n2-rv}: although it presents many similarities to the
analogous map for \ha\ of Fig.~\ref{map-ha-rv}, there are also important
differences: in the Hourglass region the negative velocity maximum is
much less pronounced; on the contrary, in the Great Rift even positive
velocities are found (orange in the Figure). That is, the \ha\ 
and the [N II] lines in this region indicate gas moving in opposite
directions with respect to the cluster center-of-mass velocity
$RV_{cm}=0.5$~km/s. In the N-W part, the puzzling large negative
velocities are again found.

\begin{figure*}
\includegraphics[width=17.5cm]{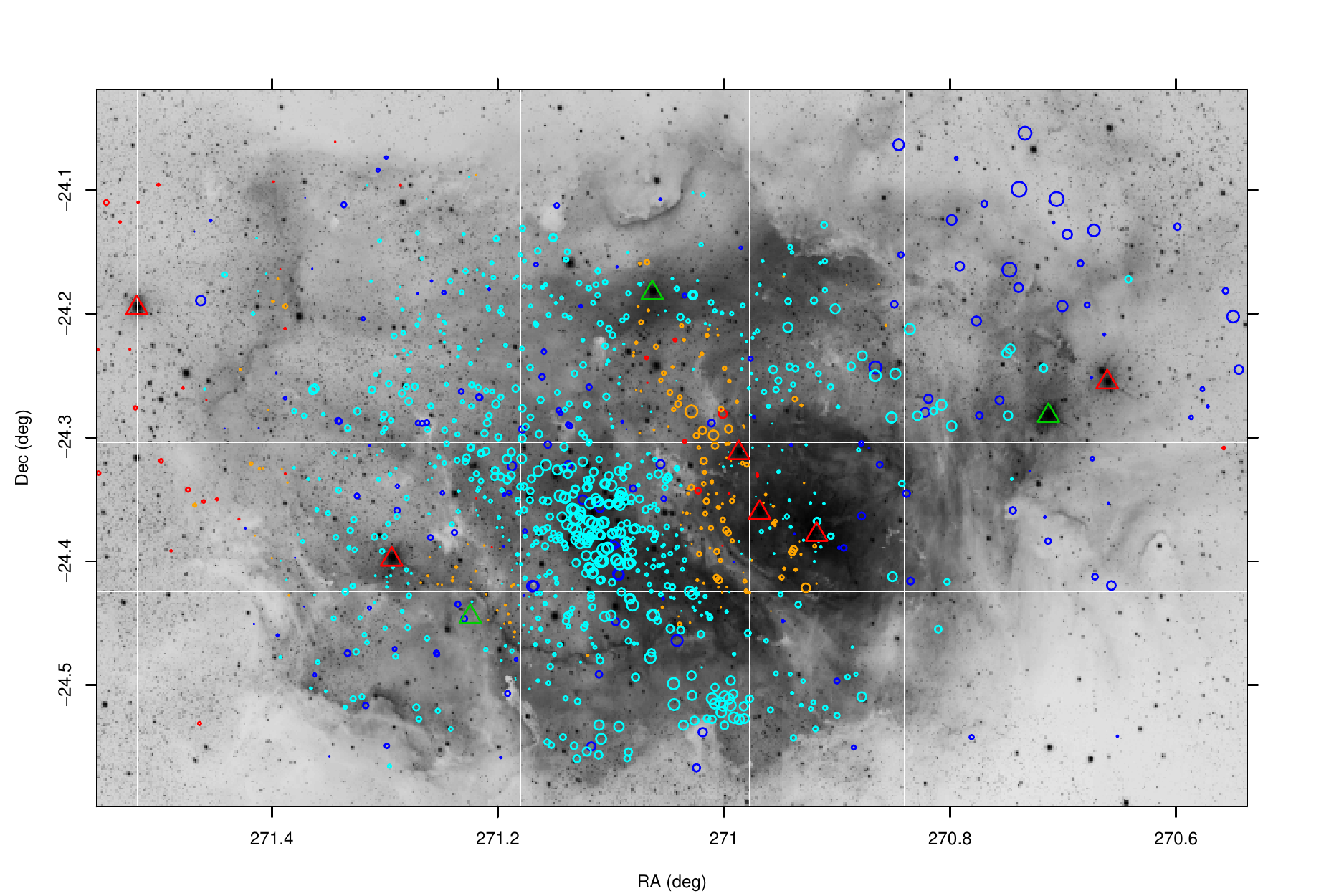} \\
\includegraphics[width=17.5cm]{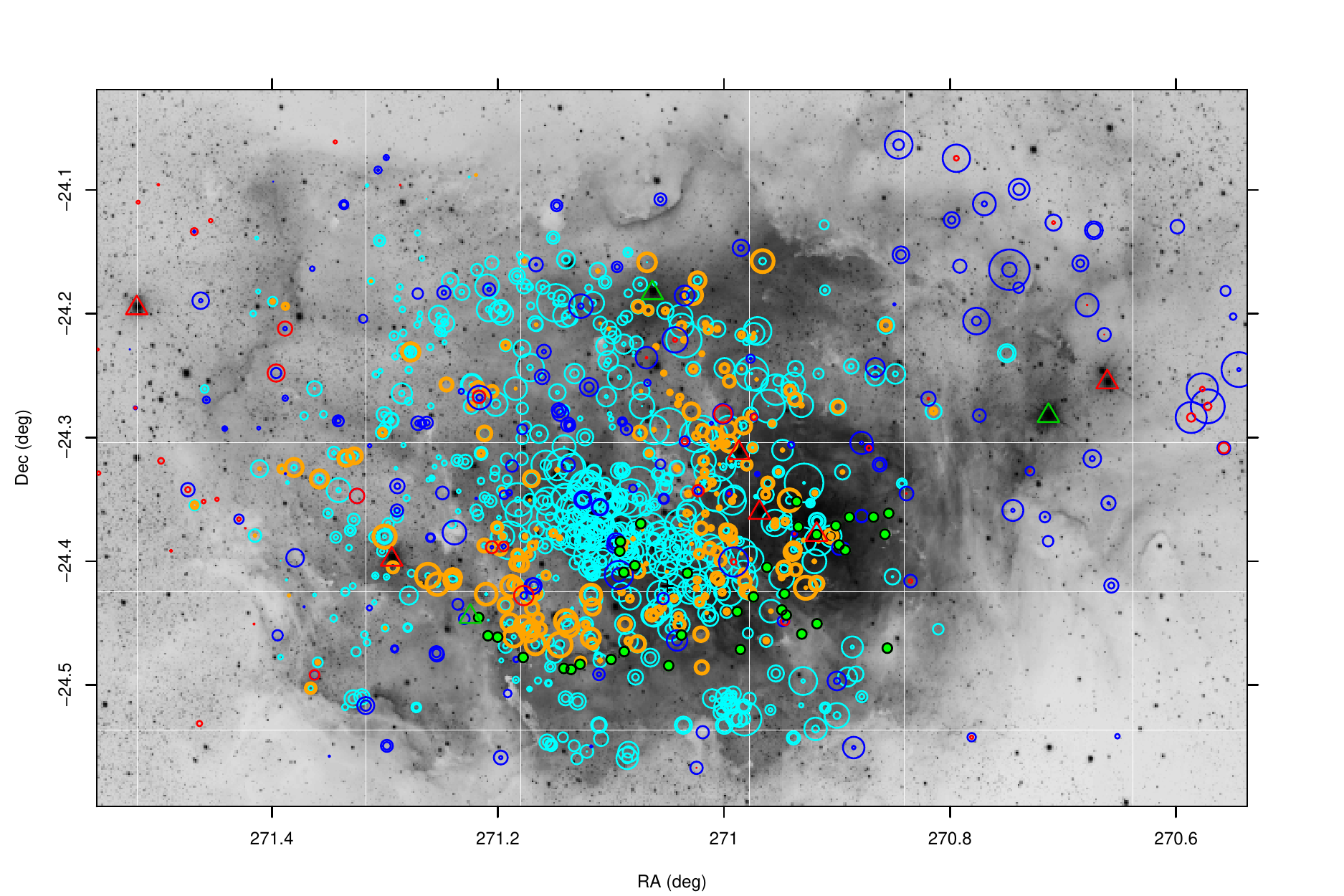}
\caption{
$a$ (upper panel): Map of [S II] RV from 1-g model fits, analogous to
Fig.~\ref{map-ha-rv}, with same meaning of symbols.
$b$ (lower panel): Map of [N II] RV (absolute value proportional to circle size)
from 2-g model fits. Symbols as in Fig.~\ref{map-ha-rv}, with the
addition of green filled dots indicating the submillimeter clumps found
by Tothill \e (2002).
Concentric circles of same color indicate that both RV components have
the same sign.
\label{map-s2-rv}}
\end{figure*}

\begin{figure}
\resizebox{\hsize}{!}{
\includegraphics{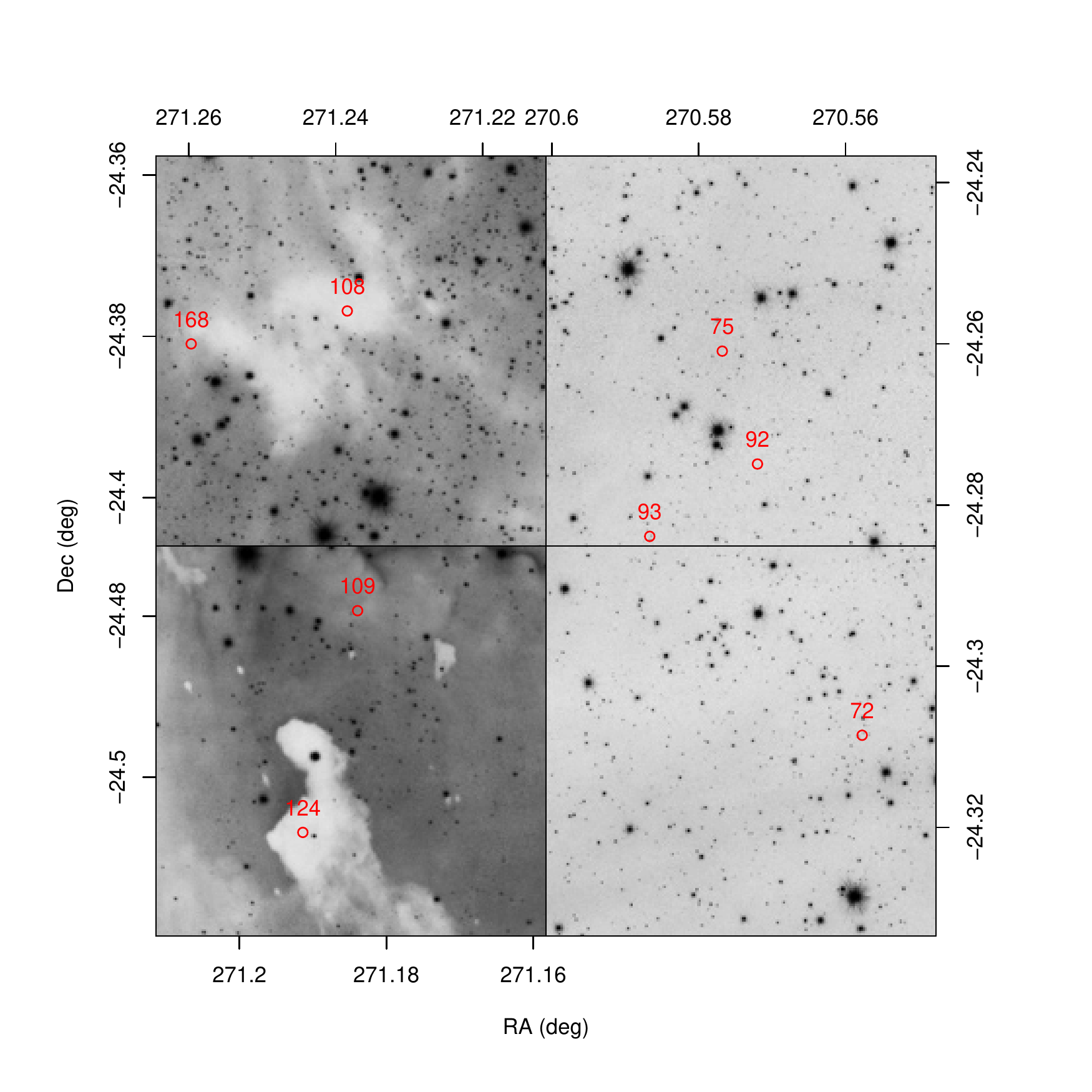}}
\caption{Examples of dark globules projected against bright background
(left), and the extreme West of the nebula. Sky fibre positions (only)
are indicated, for some of which spectra are shown in
Fig.~\ref{atlas-fits-3} and discussed in the text.
\label{dark-neb}}
\end{figure}

The [S II] velocity map of Figure~\ref{map-s2-rv}$a$ confirms these trends even
more, with slightly positive velocities being found also to S-E of
NGC6530 cluster core, almost parallel to the Great Rift. The
positive-velocity datapoints in this region follow closely the inner
border of
the bright-rimmed dark cloud hosting the massive protostar M8E-IR (green
triangle), while velocities just outside the bright rim (i.e., projected
against the most obscured part) become suddenly negative. Again,
the sharpest velocity gradients occur along a line joining M8E-IR with
the Hourglass region (arrow in Fig.~\ref{map-ha-rv}), so that a more
detailed understanding can be achieved from considering
position-velocity diagrams along this direction. Before doing that,
however, we consider the spatial maps obtained from results of the 2-g
fits.

A 2-g velocity map for [N II] is shown in
Figure~\ref{map-s2-rv}$b$. We prefer to study this line with respect to
the brighter \ha\ because the narrower line widths permit a better
determination of individual component velocities.
We have omitted the weakest components, which contribute more to the
noise than to show a clear pattern.
The most important features
shown are the crowdings of positive-velocity red/orange datapoints near
M8E-IR and along a vertical strip passing through 9~Sgr/Hourglass. With big
green dots are indicated the positions of the sub-mm knots found by
Tothill \e (2002; they did not explore the northern half of the nebula):
the orange circles fill almost exactly the arc-shaped region delimited
to the South by those knots, and ending with M8E-IR to the East. Also
the orange circles in the Hourglass region tend to fill the interior of
a region delimited by the sub-mm knots to the South. Almost everywhere
else, no receding component (with respect to $RV_{cm}$) is observed in
the ionized gas, even considering the 2-g line profile models.


\subsubsection{Peculiar locations}
\label{dark}

\begin{figure}
\resizebox{\hsize}{!}{
\includegraphics[angle=0]{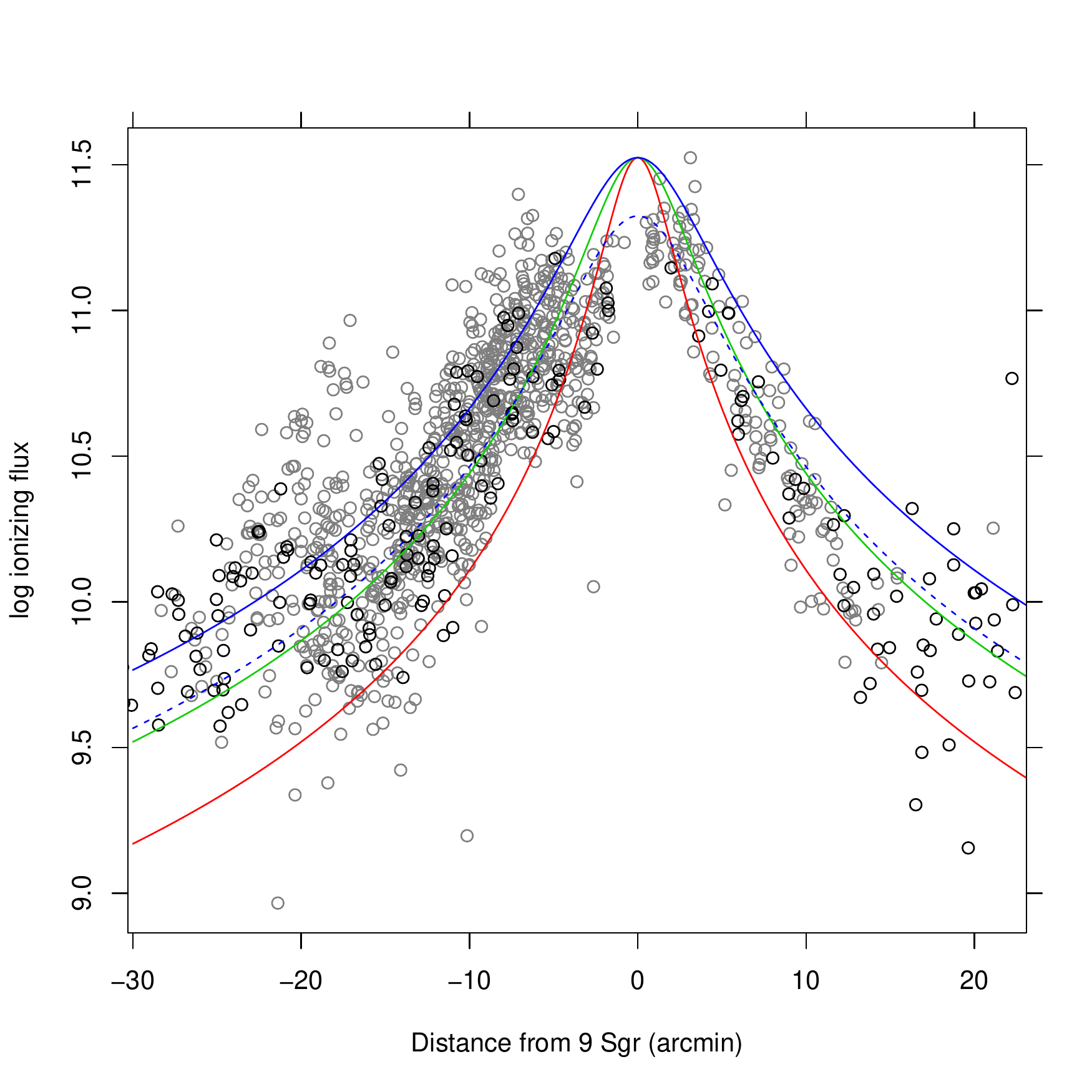}}
\caption{Ionizing flux vs.\ radial distance from 9~Sgr, shown separately
for the eastward and westward directions (negative and positive
distances, respectively). Black (gray) dots refer to
pure-sky (faint star) fibres. The blue, green, and red solid lines
refer to normal distances of 1.46, 1.09, and 0.73~pc, respectively. The
blue dashed line refers to a distance of 1.46~pc, adopting a different
maximum flux.
\label{radial-ioniz}}
\end{figure}

The image of the nebula shows several dark globules or ``elephant
trunks" projected against the bright nebular background. For some of
these dark nebulae we have fibre spectra, enabling us to discriminate
the properties of the foreground gas against that of the brighter background.
Two such examples are shown in Figure~\ref{dark-neb}, left panels.
The one in the lower left panel (called the ``Dragon" by Brand and Zealey 1978
and Tothill \e 2008) is one of the most evident, and its spectrum was
shown in Figure~\ref{atlas-fits-3}, labeled as nr.~124. This latter
reveals that the nebular emission, although attenuated by a factor $\sim
10$ with respect to adjacent unobscured positions, is still substantial,
and images show it as dark only because of the sharp contrast with the
surrounding bright emission; nevertheless, this spectrum shows a
peculiar component, as a significant emission residual at velocity $\sim
-50$~km/s. This is not seen, at least not as clearly as here, in any
other spectrum, including those of other obscured patches like that in
the upper left panel of Fig.~\ref{dark-neb}. Therefore, at least in the
direction of the ``Dragon", we see that ionized gas with velocity $\sim
-50$~km/s exists above a distance from the nebula enclosing $\sim$90\% of 
the bulk emission, from simple aperture photometry with respect to
nearby unobscured positions. The existence of faster gas component at large
distances agrees with the results from the sodium lines presented in
Section~\ref{uves}, and with those of Meaburn (1971) on [O III] lines.

The right panels of Fig.~\ref{dark-neb} show instead the locations, in
the extreme West of the entire nebula, where the most asymmetric or even
splitted line profiles are found in our dataset (spectra labeled as
nr.~72, 92, 93 both here and in Fig.~\ref{atlas-fits-3}); here again the
nebular emission is so weak as to appear nonexistent in the narrow-band
image, but is enough to be detected and studied in our spectra.
Figures~\ref{atlas-fits-3} and~\ref{dark-neb} together show that the
nebular line profiles vary smoothly with spatial position, which
confirms that their peculiarities do not arise from some random effect,
or isolated cloudlet. We will discuss therefore the implied large-scale motion
of this part of the nebula in Section~\ref{pv-outer}.

\subsubsection{The position of 9~Sgr}
\label{9sgr}

Another peculiar location is that of the most massive star, 9~Sgr, which
as mentioned is suspected to lie at some distance (5-10 pc according to
Lada \e 1976) in front of the cloud. Assuming like these authors that it
is the dominant ionizing source for the entire nebula (i.e., except in
the vicinity of Herschel~36 or HD~165052), its line-of-sight distance from the
nebula may actually be estimated from the decay of measured ionization
with sky-projected distance. We are able to estimate the ionization
parameter $q$ from the \ha/[N II] ratio, using e.g.\ the curves shown by
Viironen \e (2007), and the electron density $N_e$ from the [S II] doublet
ratio. The product $q N_e$ is proportional to ionizing flux $F$; this
will follow a spatial decay like $F = I_0/(r^2+d_{rad}^2)$, where $r$ is
the sky projected distance from 9~Sgr, and $d_{rad}$ is the
line-of-sight distance of 9~Sgr from the nebula, assumed flat.
Figure~\ref{radial-ioniz} shows the result of this experiment: a well
defined peak is indeed found, with additional, local enhancements in ionizing
flux near $+3$ and $-20$~arcmin due to Herschel~36 and HD~165052,
respectively; a minor enhancement near $-9$~arcmin corresponds instead
to the B stars in the NGC6530 core.
The proposed functional form for $F$ is shown by the
lines, with the green one corresponding to 3~arcmin, or $d_{rad}=1.09$~pc at the
nebula distance. If the adopted curve maximum is lowered, to account for
the fact that the actual maximum in the datapoints is due to
Herschel~36, a 4-arcmin curve (blue dashed) is also satisfactory,
corresponding to $d_{rad}=1.46$~pc. These normal distances are much
smaller than the estimates by Lada \e (1976); one possible explanation
is the assumed planar geometry for the illuminated nebula: a slightly
concave geometry would decrease center-to-edge differences, and require larger
$d_{rad}$ to produce the same observed effect.
However, lacking these detailed geometrical informations on
the nebula itself, we cannot derive better estimates, and may only
consider $d_{rad}=1.09-1.46$~pc as lower limits to the 9~Sgr distance
from the part of the nebula immediately behind it.

{
This distance is much larger than the distance between the Orion-nebula
ionizing star $\theta^1$~Ori~C and its molecular cloud, of
$\sim 0.25$~pc (Wen and O' Dell 1995, O' Dell 2001). This difference is
undoubtedly an important factor and may explain many of the differences
we find between the properties of M8 and the Orion nebula (see
Section~\ref{pv-6530}). It also suggests that 9~Sgr has excavated a
larger cavity in its parent cloud compared to $\theta^1$~Ori~C, and in
turn than the M8 \hii\ region as a whole is probably in a later evolutionary
stage than the Orion nebula.
}

\begin{figure}
\resizebox{\hsize}{!}{
\includegraphics[angle=0]{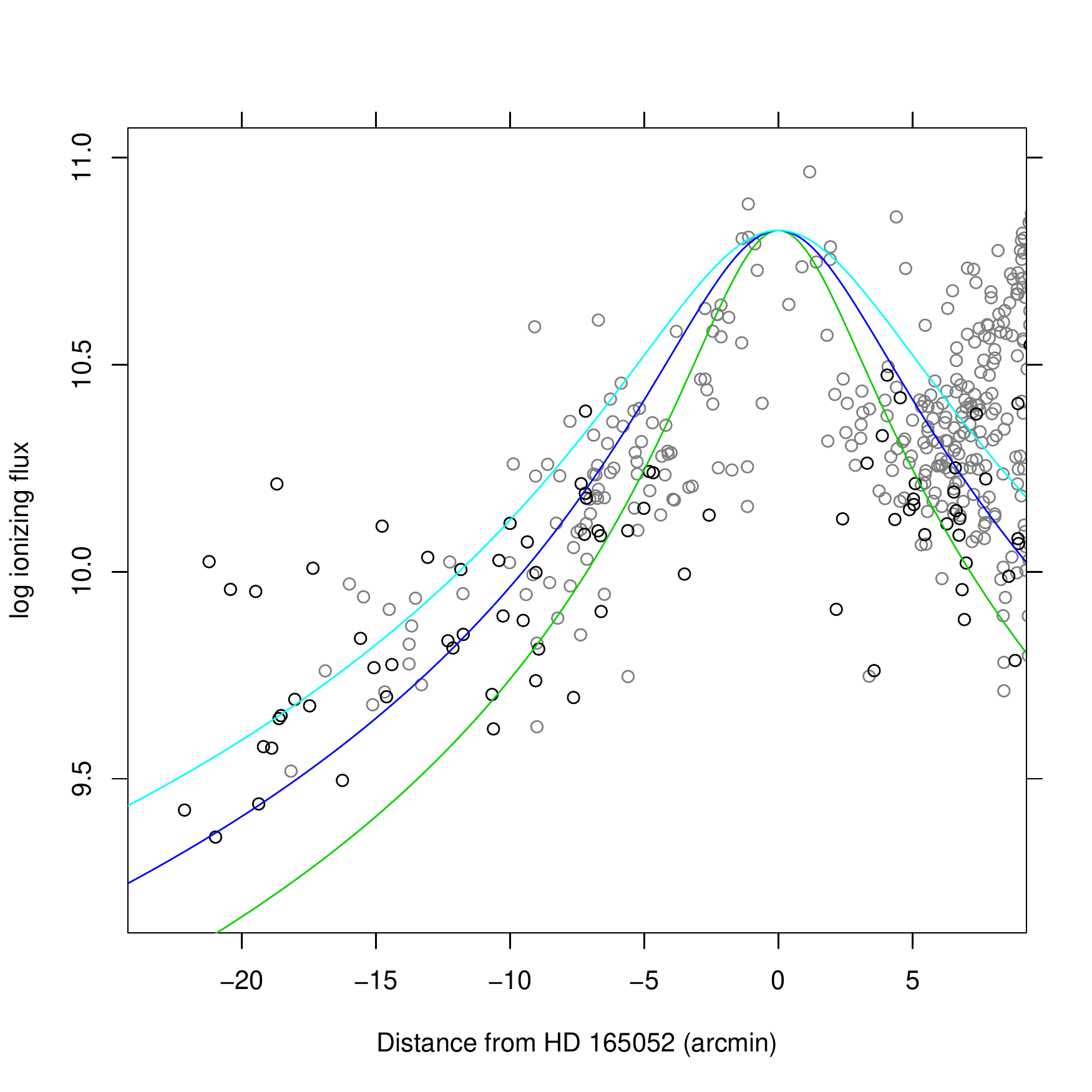}}
\caption{Ionizing flux vs.\ radial distance from HD~165052, shown separately
for the eastward and westward directions.
Symbols as in Fig.~\ref{radial-ioniz}.
Cyan, blue, and green, solid lines
refer to normal distances of 1.82, 1.46, and 1.09~pc, respectively.
\label{radial-ioniz-165052}}
\end{figure}

Both Figs.~\ref{radial-ioniz} and~\ref{map-ha-n2-ratio}$a$ suggest that
HD~165052 (O7Vz+O7.5Vz binary, Arias \e 2002) is instead the dominant
ionizing source in
its neighborhood, despite being almost irrelevant to the nebular dynamics
(Figs.~\ref{map-ha-rv} and~\ref{map-n2-rv}).
This is confirmed by the dependence of ionizing flux from
distance to this star, shown in Figure~\ref{radial-ioniz-165052},
analogous of Fig.~\ref{radial-ioniz}; five arcmin to the West of
HD~165052, the flux from 9~Sgr still dominates, but nearer to HD~165052
the 9~Sgr contribution becomes unimportant. We can therefore fit
(although with higher uncertainties) a profile depending on the normal
distance from HD~165052 to the cloud, as above: it turns out that also
this star, like 9~Sgr, is likely to be found distinctly above the cloud,
at a distance in the range 1.5-1.8~pc.


\subsection{Position-velocity diagrams}
\label{pos-vel}

\subsubsection{The NGC6530 region}
\label{pv-6530}

\begin{figure}
\resizebox{\hsize}{!}{
\includegraphics{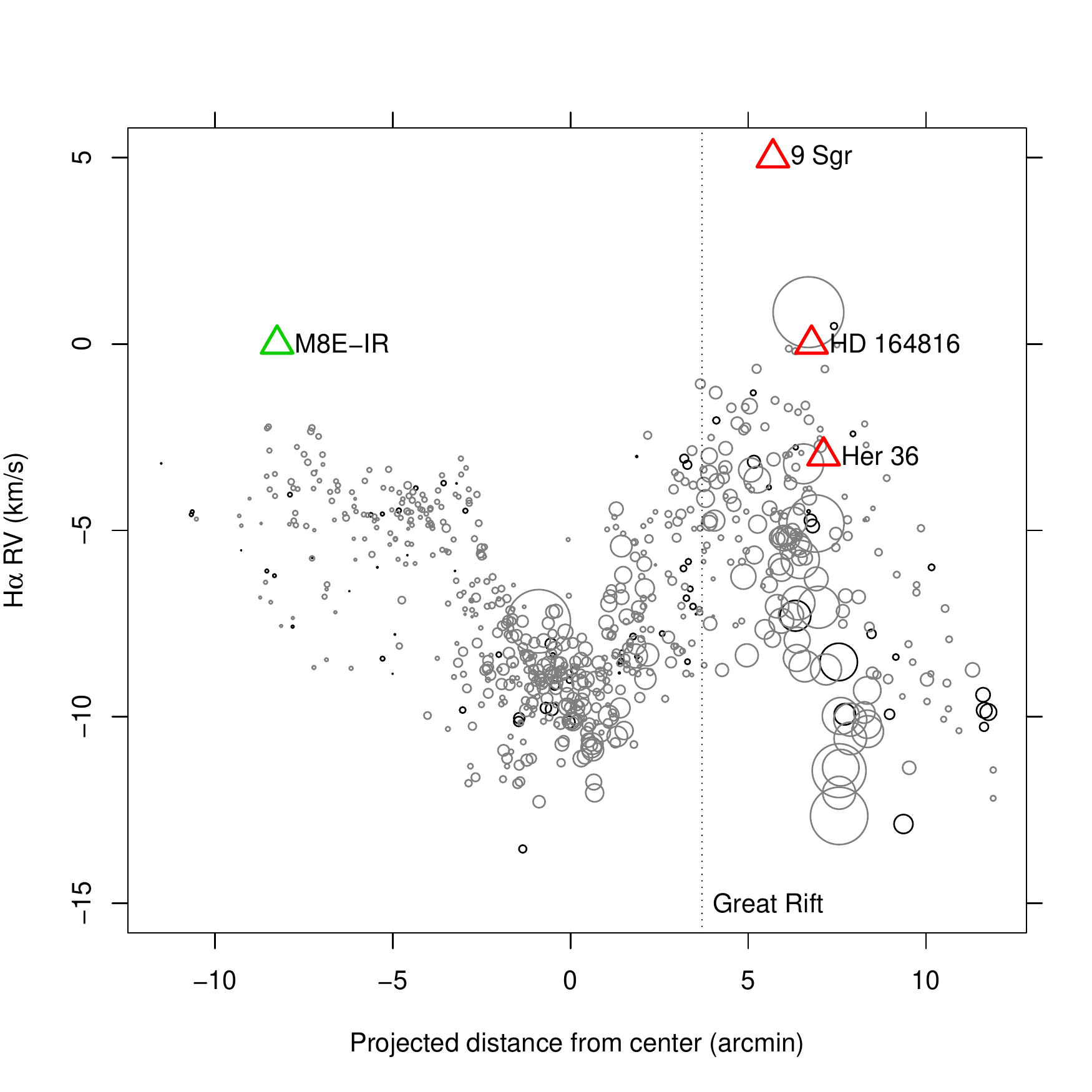}}
\caption{Position-velocity diagram for \ha, inside the cluster core
region (red dashed circle in Fig.~\ref{map-ha-rv}). The projected
distances in the abscissae are computed along the direction of the arrow
also shown in Fig.~\ref{map-ha-rv}. RVs from 1-g fits.
Symbol size is proportional to line intensity from 1-g fits.
Black (gray) circles refer to pure-sky (faint-star) fibres.
The O star HD~164816 and the massive object M8E-IR (triangles), for
which no RV measurements are available, are
plotted at RV=0, i.e.\ the NGC6530 cluster velocity, while the other
triangles indicate massive-star RVs as reported in Table~\ref{ostars}.
\label{proj-ha}}
\end{figure}


\begin{figure}
\resizebox{\hsize}{!}{
\includegraphics{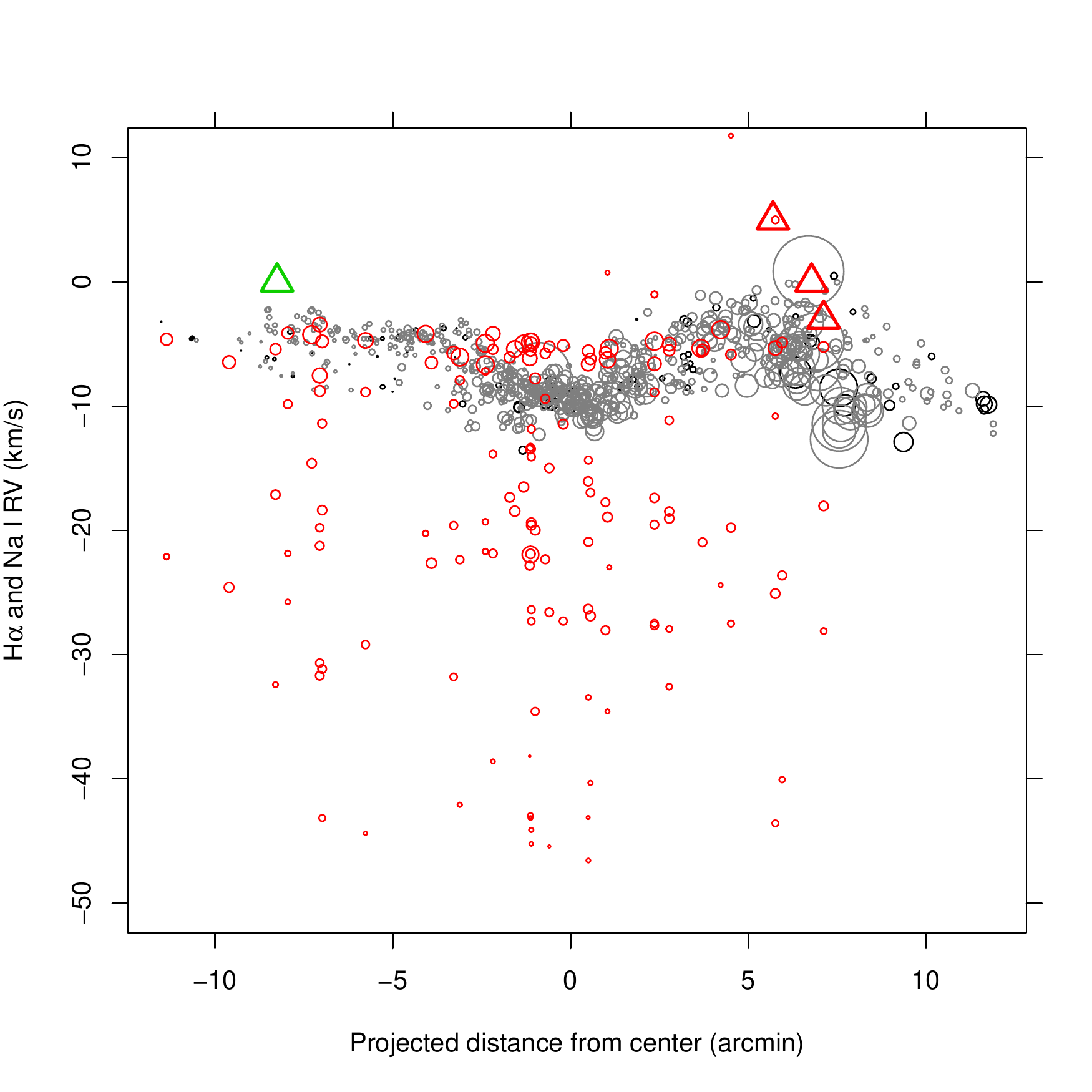}}
\caption{The same diagram as in Fig.~\ref{proj-ha}, but on a wider RV
range to show also the Na~I D2 absorption velocities (red circles).
\label{proj-ha-na}}
\end{figure}

Having discussed in Section~\ref{maps} the existence of a reference direction for the
nebular velocity field (red arrow in Fig.~\ref{map-ha-rv}), we study
here position-velocity diagrams along this direction. This symmetry
properties pertaining only to the region around the NGC6530 cluster
core, we consider here only the region (24~arcmin in diameter)
within the red circle in
Fig.~\ref{map-ha-rv}. The relevant position-velocity diagram for \ha\
(with velocities from 1-g fits) is shown in Figure~\ref{proj-ha}.
Positive projected distances are towards the arrow head of
Fig.~\ref{map-ha-rv}, i.e.\ towards the galactic plane (to N-W). The
origin of distances is at the position of the red plus sign in
Fig.~\ref{map-ha-rv}. As mentioned, the low-mass cluster stars have a
well-defined peak in their RV distribution, at $RV_{cm}=0.5 \pm 0.2$~km/s
(Prisinzano \e 2007); it is reasonable to assume that also the B stars
in the same spatial region have the same mean RV. The most massive star
in the cluster core region is HD~164906 (MWC280; type B0Ve, Levenhagen
and Leister 2006).
The cluster center coincides also with a CO bright spot (nr.~3 in Lada \e
1976), at velocity $RV \sim +6$~km/s (heliocentric, corresponding to
$v_{LSR} \sim 16$~km/s as reported in Lada \e), whose position is also
shown in Fig.~\ref{map-n2-rv}. As Fig.~\ref{proj-ha} shows, the cluster
core corresponds to the expansion center, in both position and velocity,
of a shell-like structure in the ionized gas, reaching maximum negative
speeds of $\sim -12$~km/s; no gas is found at $RV \sim RV_{cm}$ in the
vicinity of the cluster center (distance origin). Interestingly, the
colder CO molecular gas moves in the opposite direction with respect
to $RV_{cm}$, and Fig.4 of Lada \e (1976) also shows that it possesses a
velocity gradient along the N-S direction, suggesting a shell-like
geometry as well. The resulting picture is that of a localized expanding
gaseous bubble, pushed by the cumulative effect of winds from massive B
stars in the NGC6530 cluster core (there are no O stars inside it).
Therefore, in this part of the Lagoon nebula the nebular emission arises
in front of the star cluster, not in its background.
As the velocity map of Fig.~\ref{map-ha-rv} shows, such expansion is not
spherical, with near-zero radial velocity reached at small distances
only along of the direction of the arrow.
While dust extinction in front of the cluster stars is relatively low,
it rises considerably behind them (e.g.\ Damiani \e 2006), implying
large amounts of dust just behind the cluster. Interestingly, there is
no trace in Fig.~\ref{proj-ha} of any receding ionized shell, which we
interpret with the ionizing radiation being absorbed by the dust on the
rear side.
{
For comparison,
in the Orion nebula a rather regular sequence of velocities is found,
with some of the ionized gas layers having speeds within a few km/s
relative to the background CO ($\Delta RV \sim 3$~km/s for [S II],
$\sim 7$~km/s for [O III], $\sim 10$~km/s for \ha; O' Dell
\e 1993, O' Dell 2001).
Here instead (and see also Figure~\ref{proj2-n2-o3} below) we see little or
no ionized gas at velocities so close to that of background CO,
which points to important differences between the structure of the ionized
regions in M8 and in the Orion nebula.
Also the emission from the PDR on the molecular cloud surface is not
clearly recognizable in the position-velocity diagrams.
}

The stellar wind push
{
of the massive stars in the NGC~6530 core,
}
on the other hand, may be responsible
for the peculiar positive-velocity displacement of the CO emitting gas
(note that the other two CO spots found by Lada \e 1976, the brightest one
coincident with the Hourglass nebula, have velocities $\sim 0$~km/s,
heliocentric). The current view that the star cluster did form on the
near-side surface of the molecular cloud is in good agreement with the
proposed interpretation: the far side of the star cluster faces regions
with more dust, molecular gas, and higher-density gas in general, than
the near side, facing the outer, more rarefied parts of the cloud.
Therefore, also the absolute speed reached by the near-side ionized
diffuse gas ($\sim -12$~km/s in \ha) is larger than than of the far-side
denser and colder gas ($\sim +6$~km/s in CO).

At projected distances larger than $+5$~arcmin, Fig.~\ref{proj-ha} shows
a wide scatter in the RV distribution, with no clearly defined geometry;
this region corresponds to the Hourglass nebula, as also clear from the
size of the symbols in the Figure (proportional to intensity).
In the Figure are also shown the massive stars M8E-IR (green triangle) and
HD~164816, 9~Sgr, and Herschel~36 (red triangles, left to right).
Interpreting such a complex velocity field in term of expansion driven
by one or more massive stars requires to know the radial velocity of
these stars with sufficient accuracy. This information is not always
available for massive stars, which have few lines broadened by very fast
rotation, are found very often in binary systems, and are studied mostly with
single-epoch observations. For convenience we compile in
Table~\ref{ostars} the literature radial-velocity data on the most
massive stars discussed here, from the SIMBAD database, except for the 9~Sgr
velocity taken from Williams \e (2011); note however that this star is
a long-period SB2 binary, and the systemic velocity is subject to large
errors (Rauw \e 2012).
No Gaia-ESO velocity measurements are available for these massive stars.
Literature velocities are missing for four stars, including the massive
young object M8E-IR. These are plotted in our position-velocity diagrams
at a velocity $RV_{cm}$ for reference. The two stars in the western
regions have similar velocities ($\sim -10$~km/s), but strongly discrepant with $RV_{cm}$;
yet, they agree more with the approaching velocity of the \ha\ emission
in the same region, discussed in Section~\ref{pv-outer} below.
The positive velocity of 9~Sgr is
{
surprising at first sight, being so different from that of the CO
clouds; however, after considering that this star lies $\sim 1.5$~pc
above a massive molecular cloud ($2-6 \cdot 10^4 M_{\odot}$ for each of the CO
clouds in the region according to Takeuchi \e 2010,
with the cloud associated with the Hourglass being one of the most
massive), it becomes plausible that this star has gained a considerable
speed toward the cloud during the last few Myrs.
}

\begin{figure}
\resizebox{\hsize}{!}{
\includegraphics{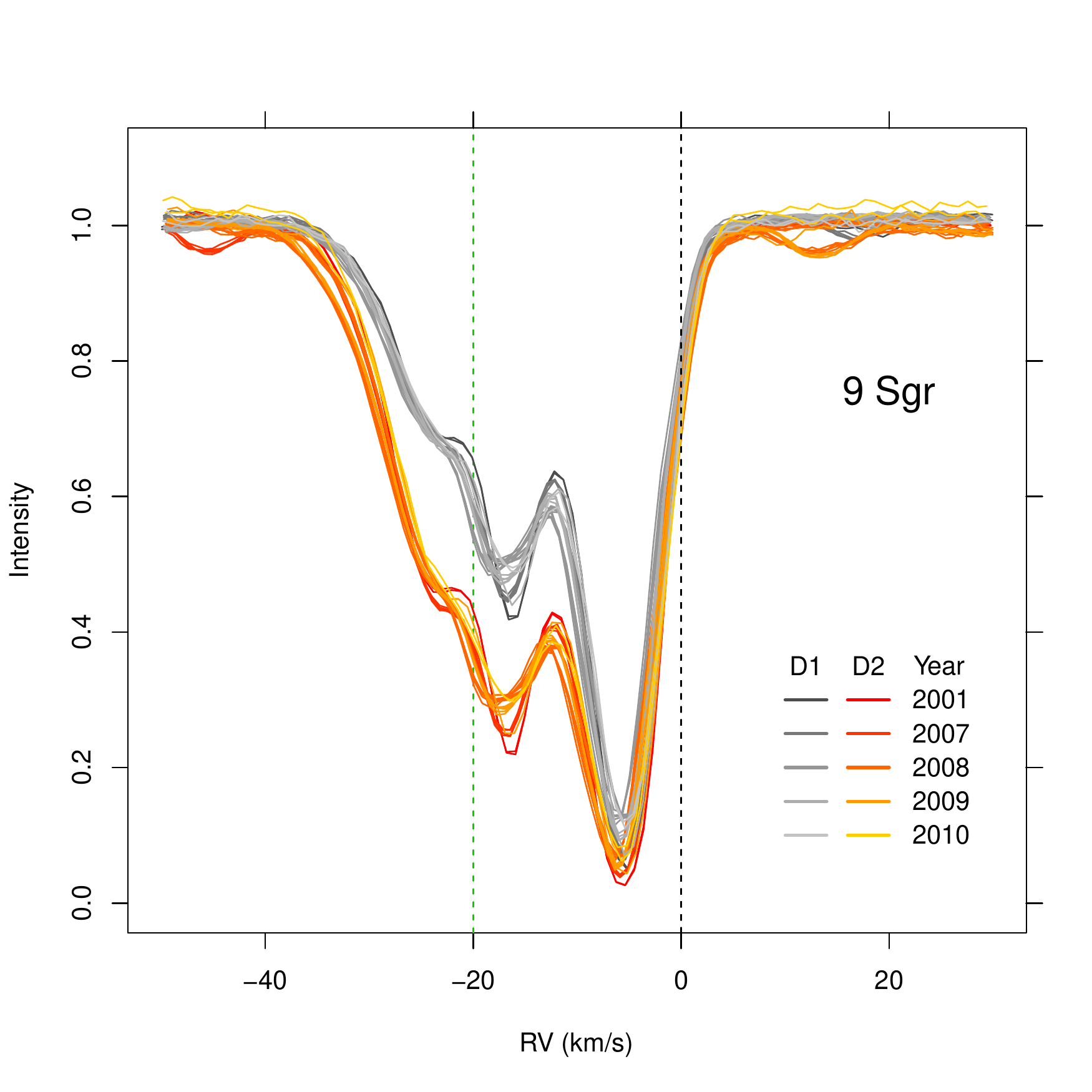}}
\caption{Sodium doublet absorption towards 9~Sgr
from UVES spectra between 2001-2010.
Velocity $RV=-20$~km/s, where significant line-profile changes are found,
is marked with a green dashed line.
\label{sodium-9sgr}}
\end{figure}

\begin{table}
\caption{Literature data for massive stars in the Lagoon nebula}
\label{ostars}
\centering
\begin{tabular}{lcclc}
\hline\hline
Name & RA & Dec & Spectral & RV \\
     & J2000 & J2000 & type & km/s \\
\hline
HD 164536   & 270.6609 & -24.2554 & O7.5V       &-10.5 \\
7 Sgr       & 270.7129 & -24.2825 & F2/F3II/III &-11.1 \\
Herschel 36 & 270.9180 & -24.3785 & O7:V        &-3.0 \\
9 Sgr       & 270.9685 & -24.3607 & O4V((f))z   & 5.0 \\
HD 164816   & 270.9869 & -24.3126 & O9.5V+B0V   &  \\
HD 164865   & 271.0634 & -24.1834 & B9Iab       &  \\
M8E-IR      & 271.2244 & -24.4448 &             &  \\
HD 165052   & 271.2940 & -24.3986 & O7V+O7.5V   & 1.2 \\
HD 165246   & 271.5195 & -24.1955 & O8V         &  \\
\hline
\end{tabular}
\end{table}

From Fig.~\ref{proj-ha} there is no apparent connection between the
position and motion of 9~Sgr and the ionized gas, despite this star
being the most massive of the region. This supports further the arguments
of Lada \e (1976) on its lying at some distance from the cloud, in its
foreground. In those outermost nebular regions, the local gas density is
likely so low that the ionized front approaching us becomes
undetectable. Outside the ionized regions, there is nevertheless neutral
gas associated with the Lagoon nebula, as discussed in connection with
sodium absorption in Section~\ref{uves}. Therefore,
Figure~\ref{proj-ha-na} shows the same \ha\ velocities as in
Fig.~\ref{proj-ha}, but on an expanded velocity scale, with the addition
of the Na~I D velocities: except for the dominant component near
$-5$~km/s, the neutral gas moves at much larger negative velocities than
the \ha-emitting gas. No clear pattern is seen, indicating that the
geometry of the neutral-gas expansion is different from that of the
ionized gas. It is interesting to remark that despite 9~Sgr is distant from
the nebula, the sodium absorption is still closer to us, since several
absorption components at negative velocities up to $\sim -25$~km/s or
more are evident also in the 9~Sgr UVES
spectra shown in Figure~\ref{sodium-9sgr}.
In this Figure a definite time variability of the sodium absorption
components is seen, especially near velocities $\sim -20$~km/s, in both
velocity and line width; this agrees with our arguments of
Section~\ref{uves} that the sodium layer at $\sim -20$~km/s is the one
most subject to dynamical changes.

\begin{figure}
\resizebox{\hsize}{!}{
\includegraphics{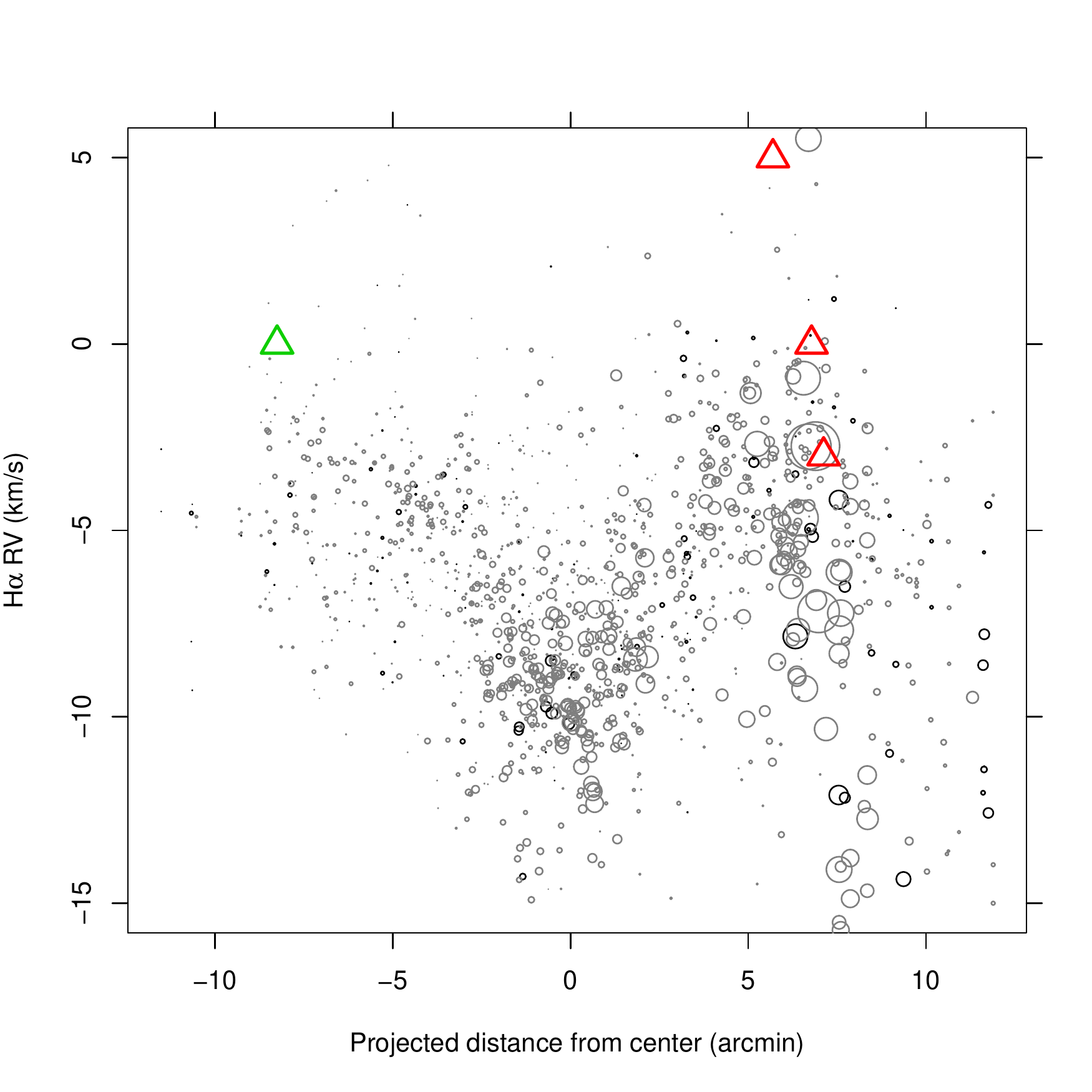}}
\caption{Position-velocity diagram for \ha, cluster core, with RVs from
2-g fits. Symbol meaning as in Fig.~\ref{proj-ha}.
\label{proj2-ha}}
\end{figure}

\begin{figure}
\resizebox{\hsize}{!}{
\includegraphics{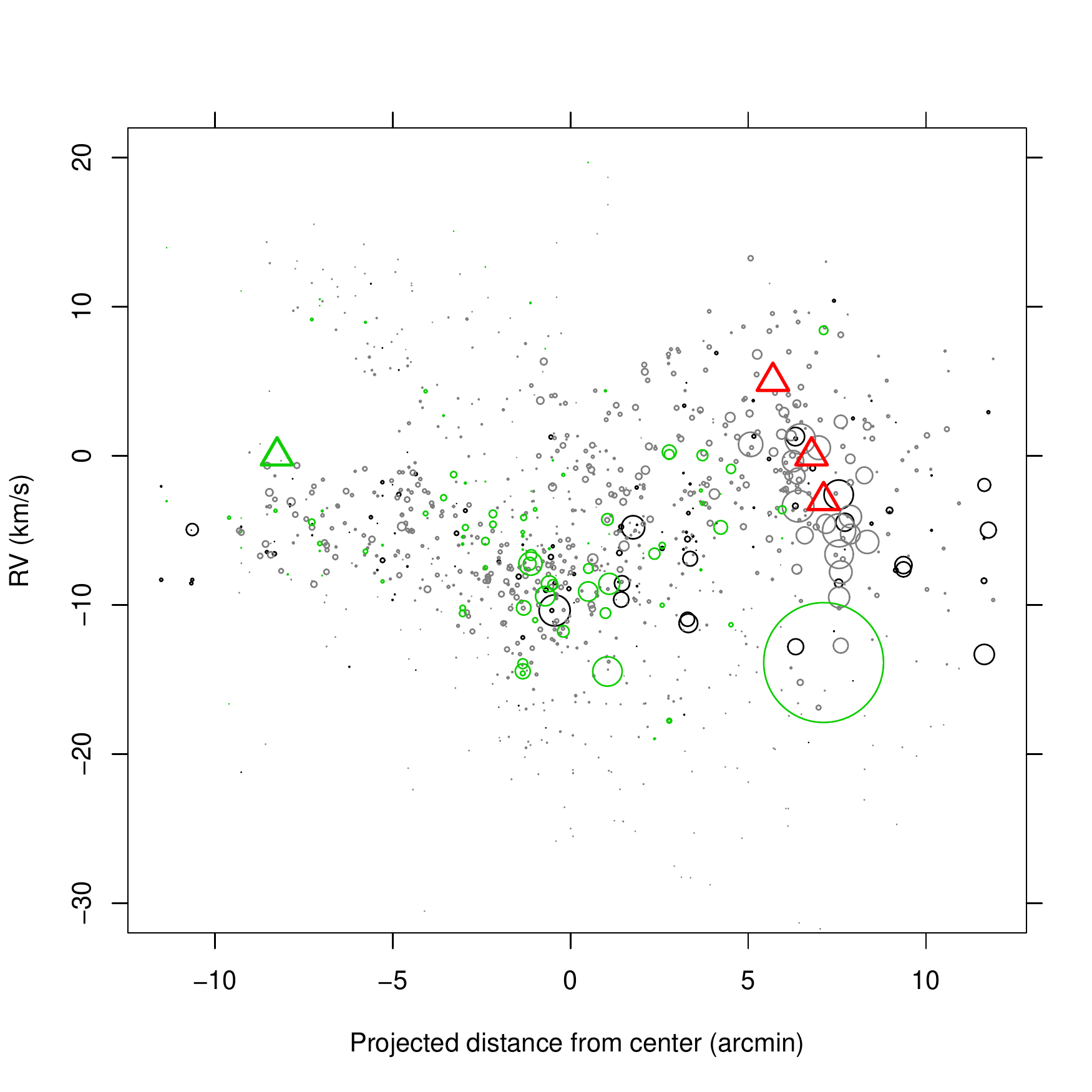}}
\caption{Position-velocity diagram for [N II], cluster core, from 2-g fits.
Also shown with green circles are the best-fit [O III] components.
\label{proj2-n2-o3}}
\end{figure}

Consideration of the 2-g model fits to the \ha\ line in the same cluster
core region provides us with only a marginally clearer picture
(Figure~\ref{proj2-ha}): the velocity splitting between blue and red
components is here mostly small compared to the absolute velocity
values. Again, near the shell center essentially no near-zero-velocity gas is
detected, while maximum negative velocities attain $\sim -15$~km/s.
In the Hourglass region the velocity spread is highest, again without
clear geometrical pattern; we recall that this latter property agrees
with the highly anisotropic brightness distribution of the Hourglass
nebula itself, whose oscuring material lets the radiation from
Herschel~36 leak only through irregularly-distributed ``windows".

More illuminating is the examination of the position-velocity diagram
involving 2-g fits to [N II] and [O III] lines, and shown in
Figure~\ref{proj2-n2-o3}.
While the emission at velocities between 0 to $-10$~km/s has not greatly
changed, new features are seen at both positive velocities (as in the map of
Fig.~\ref{map-s2-rv}$b$), and at velocities $<-10$~km/s. In the projected
distance range from $\sim -10$ to $\sim -3$~arcmin a weak but significant
positive-velocity component is found in both [N II] and [O III]. The
velocity is found to be largest near the projected position of M8E-IR
(green triangle), and to decrease gradually towards position $\sim
-3$~arcmin. There is no corresponding structure in the position-velocity
plane at negative velocity. This suggests strongly the existence of a
shell of ionized gas, expanding away from M8E-IR (or its
immediate vicinity), of which only the receding component is
visible to us, and limitedly to the part unobscured by the dense
bright-rimmed cloud.
This latter characteristics is easily explained assuming that any
approaching gas is blocked or hidden by the dark dusty structures seen
as bright-rimmed clouds, which also occult M8E-IR from our direct view.
While all the literature on this object (see the review in Tothill \e
2008) agrees that it must be very young and surrounded by thick layers
of dust, the existence of a emispheric shell ionized by this object
implies that the dust thickness between M8E-IR and the nebula behind it
is much less than the dust thickness in the direction towards us.
Alternatively, M8E-IR, a known outflow source (Mitchell, Maillard and
Hasegawa 1991), might be only the source of the mechanical push
exerted on the receding gas, which is instead ionized by another UV
source, maybe 9~Sgr further away. Even in this latter case, the total
column density of matter on our side of M8E-IR must be much larger than
on its rear side, in order to block any approaching gas expanding from
it.

Considering now the region around 9~Sgr (red triangle at $RV =+5$~km/s
in Fig.~\ref{proj2-n2-o3}), we observe that positive-velocity emission
is found near its position, up to $RV \sim +12$~km/s, and declining away
from the star position. Already in Fig.~\ref{map-s2-rv}$a$ the
positive-velocity datapoints were seen to follow an almost half-arch
around this star's position. While no (or very little) positive-velocity
gas is found in correspondence of the cluster core, blocked by the dense
molecular material behind, this blocking effect does not operate for the
gas receding from 9~Sgr, if this star lies at large distances in front of
the cluster as already discussed. Therefore, we obtain a coherent
picture by assuming that the positive-velocity gas is pushed by 9~Sgr
towards the nebula, counteracting its expansion locally.

The region around the Hourglass/Herschel~36 continues to show a rather
chaotic position-velocity pattern even using the 2-g model fits in
Fig.~\ref{proj2-n2-o3} as it was using 1-g models in Fig.~\ref{proj-ha}
above. The largest green circle in Fig.~\ref{proj2-n2-o3} represents the
strong negative-velocity [O III] emission in the Herschel~36 spectrum; a
corresponding, much weaker positive-velocity component is also found near $\sim
+8.5$~km/s, demonstrating again asymmetric expansion in the immediate
vicinity of this star, where the high-ionization [O III] line arises.

Finally, Fig.~\ref{proj2-n2-o3} shows, in correspondence to cluster
core, the largest negative velocities (up to $\sim -25$~km/s), which
overlap with the velocities of the sodium absorption where indications
of a hotter absorbing gas were found, as discussed in Section~\ref{uves}.

\begin{figure}
\resizebox{\hsize}{!}{
\includegraphics[angle=90]{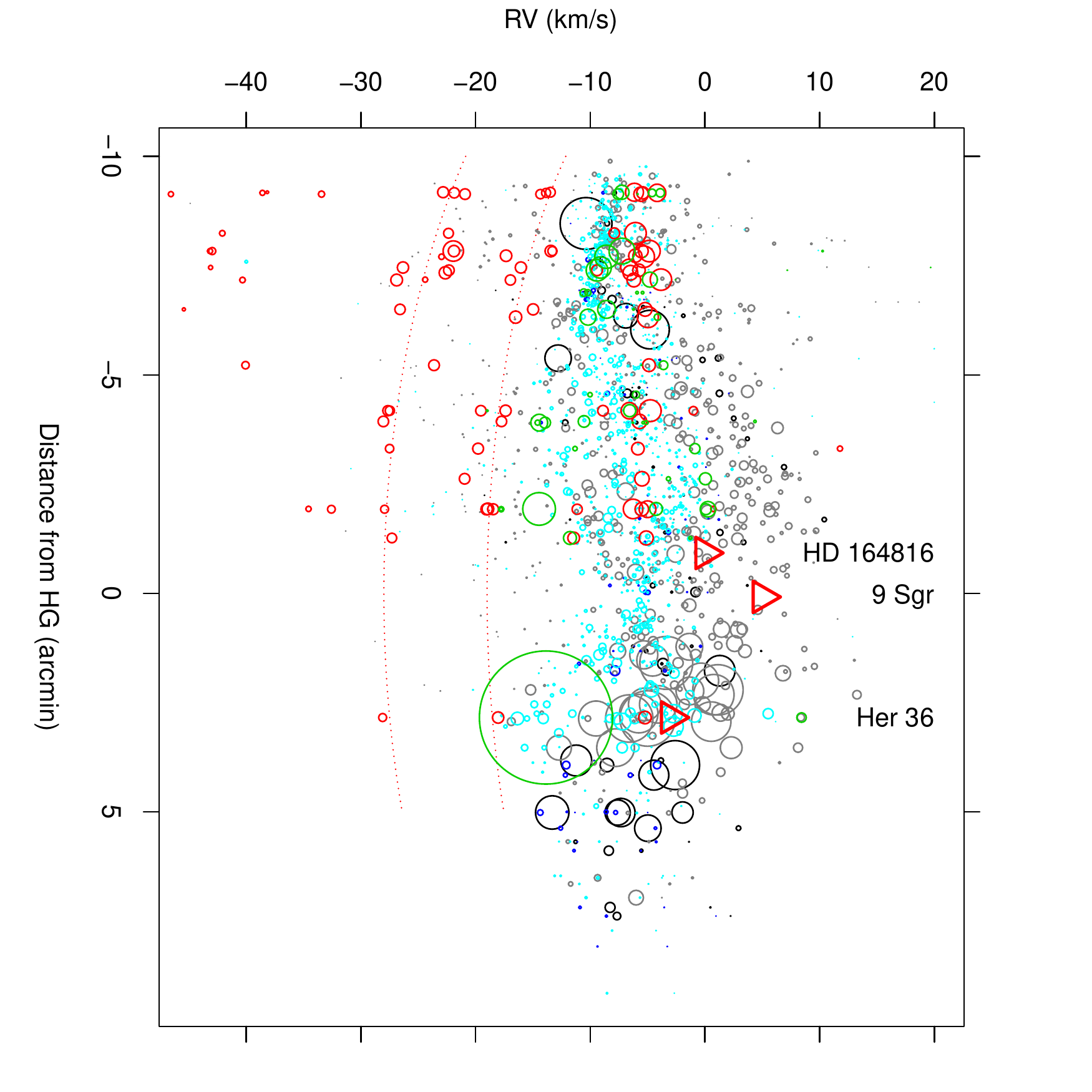}}
\caption{Position-velocity diagram, in the Hourglass region
(orange dashed circles in Fig.~\ref{map-ha-rv}), showing RVs from 2-g fits.
Positions in the abscissae are
along RA direction, from the reference point shown in Fig.~\ref{map-ha-rv}.
Black/gray circles refer to [N II] (sky fibres/faint stars as in
Fig.~\ref{proj-ha}),
Blue/cyan circles refer to \ha, green circles to [O III], and red
circles to Na~I absorption. The large dotted circle segments are
visually fitted to the sodium datapoints.
\label{proj2-hg-n2-na}}
\end{figure}

\subsubsection{The Hourglass nebula}
\label{pv-hourglass}

We next discuss position-velocity diagrams in the Hourglass region
(orange circle in Fig.~\ref{map-ha-rv}); since the positive-velocity
datapoints align along approximately the N-S direction in
Fig.~\ref{map-s2-rv}$a$, we take here the reference direction along RA for the
position axis. The position-velocity diagram of Figure~\ref{proj2-hg-n2-na}
shows together the 2-g fit results from four lines: [N II] (black/gray),
\ha\ (blue/cyan), [O III] (green), and Na I absorption (red), with circle
size proportional to intensity. The position origin is taken coincident
with the RA of 9~Sgr (triangle at $RV=+5$~km/s). Herschel~36 is the triangle
at $RV=-3$~km/s.
The diagram, although very complex, helps us to appreciate better several
effects. The [N II] emission (black/gray symbols), although found at both
positive and negative velocities in the neighborhood of 9~Sgr,
is on average stronger at positive RVs,
whereas \ha\ is largely absent at positive RVs.
This explains the discrepancy of bulk velocities between the two lines,
found in Figure~\ref{ha-n2-s2}$a$.
To the right of position origin, the velocity splitting in \ha\ increases
regularly until the position of Herschel~36: there, \ha\ shows both a
negative-velocity component (reaching $\sim -18$~km/s), and a slower
component, apparently at rest with respect to Herschel~36. On the other
hand, [N II] shows both a rest-frame component and another one with slightly
receding velocities with respect to Herschel~36. Also in this case as for
9~Sgr, the \ha\ to [N II] intensity ratio is different between the near and
far side of the massive star. We cannot make more quantitative studies
since the intensity ratios for the (unresolved) 2-g fit components are
affected by too large errors individually, as remarked in
Section~\ref{giraffe}.
The existence of positive-velocity [N II] emission from the envelope around
Herschel~36, unlike the cluster core region, suggests that this star like
9~Sgr lies at some distance above the dust-rich molecular cloud, otherwise
an inward-directed flow at positive velocities would have been blocked.

Much fainter than the positive-velocity [N II] emission, but still clearly
detected is negative-velocity [N II] at $\sim -20$~km/s, overlapping some
of the Na I absorption layers. As also discussed in
Section~\ref{uves}, the transition
between ionized and neutral gas should take place near this velocity range.
Although the spatial coverage of the UVES
data used for the study of the sodium line is much less dense than that of
the Giraffe data, we may tentatively identify in the Figure a regular
pattern in the sodium absorption, as indicated by the two dotted red
curves. These are centered to the 9~Sgr position, and suggest that the
large-scale expansion, at distances where the gas is neutral, is driven by
this star. The radii of the circle segments shown in Fig.~\ref{proj2-hg-n2-na}
are respectively 13 and 15 arcmin, corresponding to 4.74 and 5.47 pc.
The star 9~Sgr, showing sodium absorption at least around $-20$~km/s, must
be interior to at least the lower-velocity shell, in agreement with its
line-of-sight position derived in Section~\ref{9sgr}.
By dividing the inferred sodium-shell radii by their maximum velocities we
obtain timescales of order of $2-2.5 \cdot 10^5$~yr.
Note however that even stars farther away from 9~Sgr than 15 arcmin show
sodium absorption, so that the proposed geometry for the absorbing layers
must be only considered as tentative, and more complex neutral-gas structures
are certainly present across the entire face of the nebula.
{
These layers, well above the ionized gas, and at negative
velocities with respect to it, are very similar to the Orion Veil,
already mentioned in Section~\ref{giraffe}.
The existence of a blueshifted, neutral layer is also in very good
agreement with predictions of champagne-flow models of blister \hii\
regions (Tenorio-Tagle 1979), fully appropriate to a region like M8.
}

\begin{figure}
\resizebox{\hsize}{!}{
\includegraphics{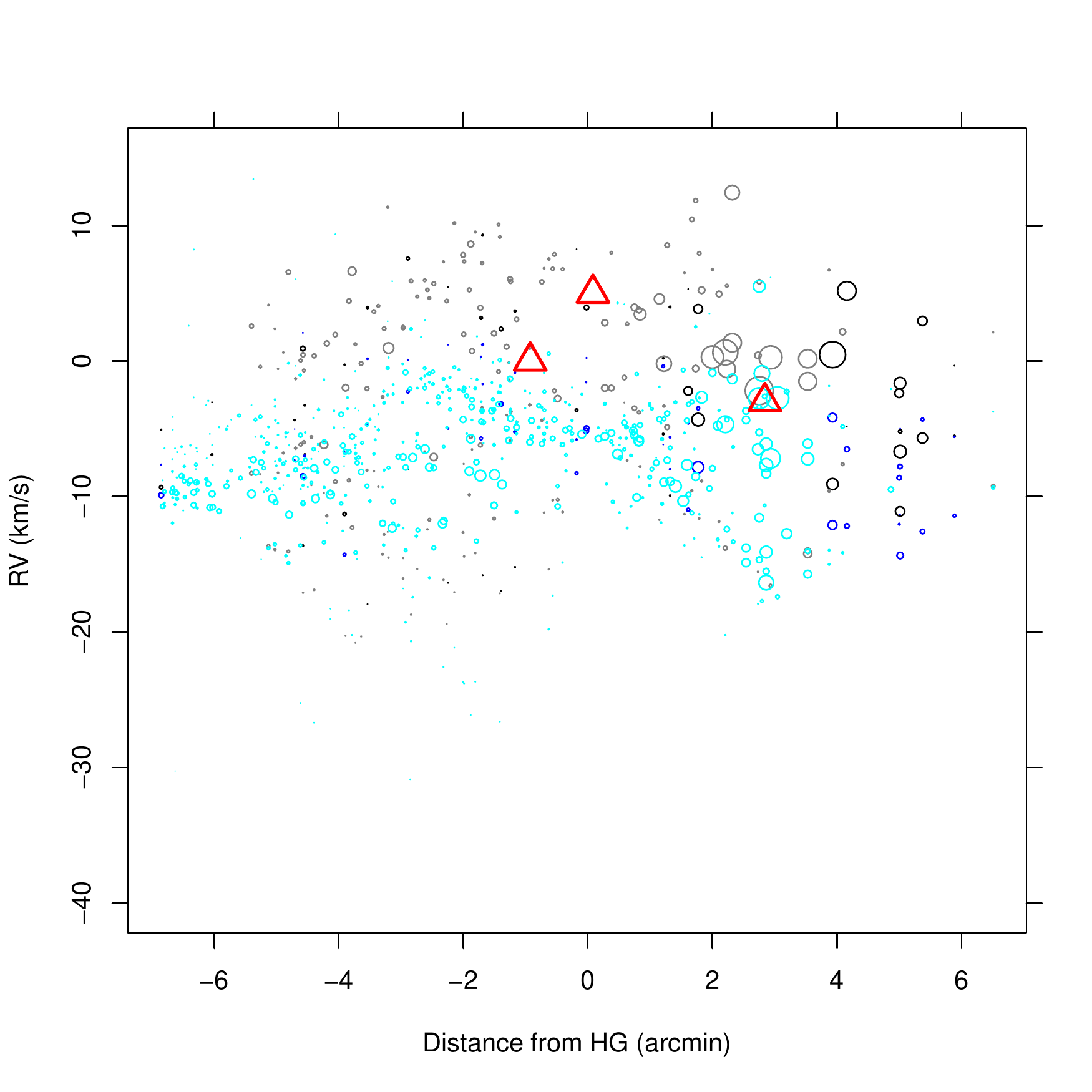}}
\caption{Analogous of Fig.~\ref{proj2-hg-n2-na}, but showing RVs from
2-g fits to [S II] (grey/black) and \ha\ (cyan/blue).
\label{proj2-hg-s2}}
\end{figure}

The results from the 2-g fits to [S II] lines, shown in
Figure~\ref{proj2-hg-s2},
provide independent confirmations of the findings just discussed. In
particular the brightest [S II] emission near Herschel~36 is found near the
stellar rest velocity (as for \ha), but clear [S II] emission near the
Hourglass is found also at positive velocities (while \ha\ shows large
negative velocities in the same region).

\begin{figure}
\resizebox{\hsize}{!}{
\includegraphics{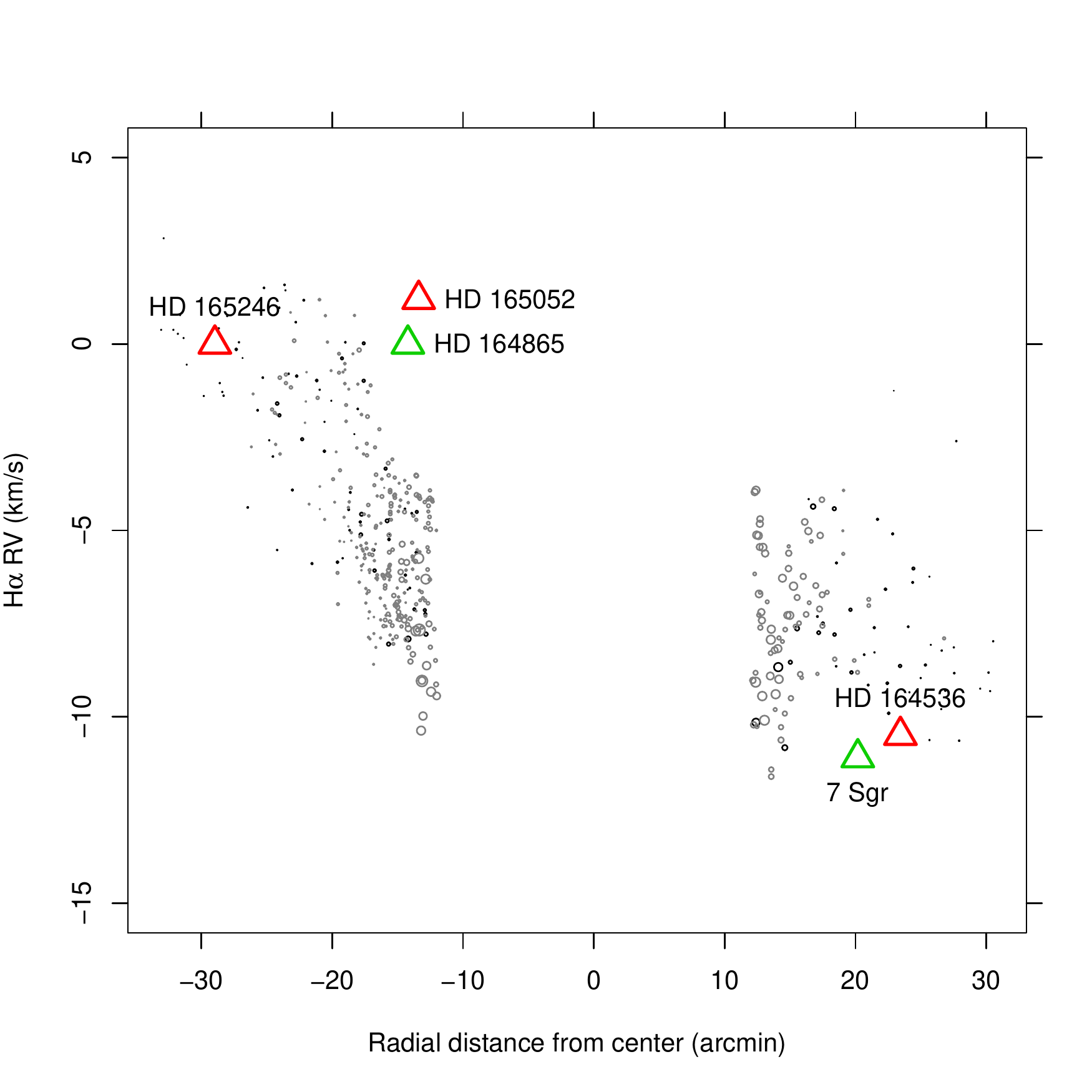}}
\caption{Large-scale RV dependence on radius, from \ha\ (1-g fits). Only
datapoints external to the cluster-core region (red dashed circle in
Fig.~\ref{map-ha-rv}) are shown. The abscissae show radial distances from
cluster center (red plus sign in Fig.~\ref{map-ha-rv}),
shown separately for the Eastern (left) and Western (right) parts.
Symbols as in Fig.~\ref{proj-ha}.
\label{radial-ha}}
\end{figure}



\begin{figure}
\resizebox{\hsize}{!}{
\includegraphics{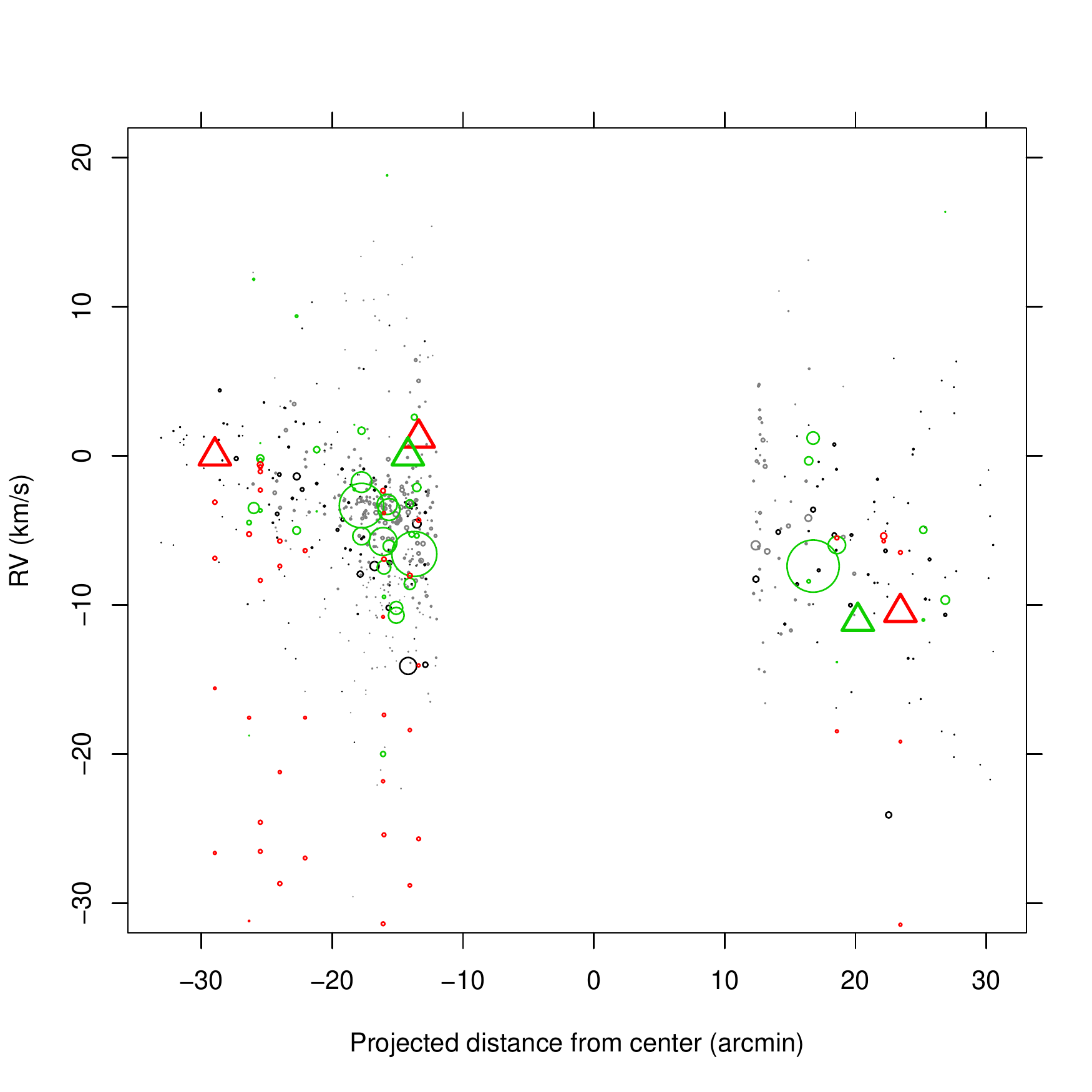}}
\caption{Same as Fig.~\ref{radial-ha}, but showing RVs from 2-g fits to
[N II] (black/gray) and [O III] (green).
Also shown are RVs of Na~I absorption (red).
\label{radial2-n2-o3}}
\end{figure}


\subsubsection{The outer parts of the Lagoon nebula}
\label{pv-outer}

Finally, we examine the large-scale velocity patterns of the nebula,
outside the central parts examined above. We have already remarked that the
Western regions do not show the same dynamics as the Eastern ones, and
therefore a radial coordinate does not prove useful. Instead, we consider
separately an East and a West radial coordinate, from the same NGC6530 cluster
center as in Section~\ref{pv-6530}. The dependence of 1-g \ha\ velocity
on these radial
distances is shown in Figure~\ref{radial-ha}. Towards East the gradual
decrease in absolute velocity (towards $RV_{cm}$) agrees well with a
global-expansion pattern, with velocity vectors becoming orthogonal to the
line of sight at the largest radii. To the West, there is no sign of this,
with velocities remaining at values from $-10$ to $-5$~km/s even where the
nebula becomes very faint. Our spatial sampling in the S-W outer regions is
almost nonexistent, so these result pertain essentially to the N-W parts,
i.e.\ those closest to the galactic plane. Adding complexity to the puzzle,
the two massive stars in the West (HD~164536 and 7~Sgr) both have negative
velocities, similar to the neighboring gas but contrasting with $RV_{cm}$.

One possible explanation of the East-West large-scale velocity gradient
in the Lagoon is rotation.  In order to keep in bound
keplerian rotation matter at a speed of 5~km/s at a radius of 10-11~pc,
and in the absence of internal pressure support,
a mass of $6 \cdot 10^4 M_{\odot}$ is required, which is not unreasonable
compared to the mass estimates for the CO clouds in M8 given by Takeuchi \e
(2010) using NANTEN, or the value of $10^4 M_{\odot}$ given for the
M8E region alone by Tothill \e (2008).
However, one obvious difficulty of this hypothesis is
that, if this was the case, then Fig.~\ref{radial-ha} would suggest for the
cloud center of mass a $RV \sim -5$~km/s, in strong disagreement with the
RVs of both the CO bright spots of Lada \e (1976) and the NGC6530 $RV_{cm}$.

\begin{figure*}
\resizebox{\hsize}{!}{
\centering
\includegraphics[width=17.5cm,height=14.6cm]{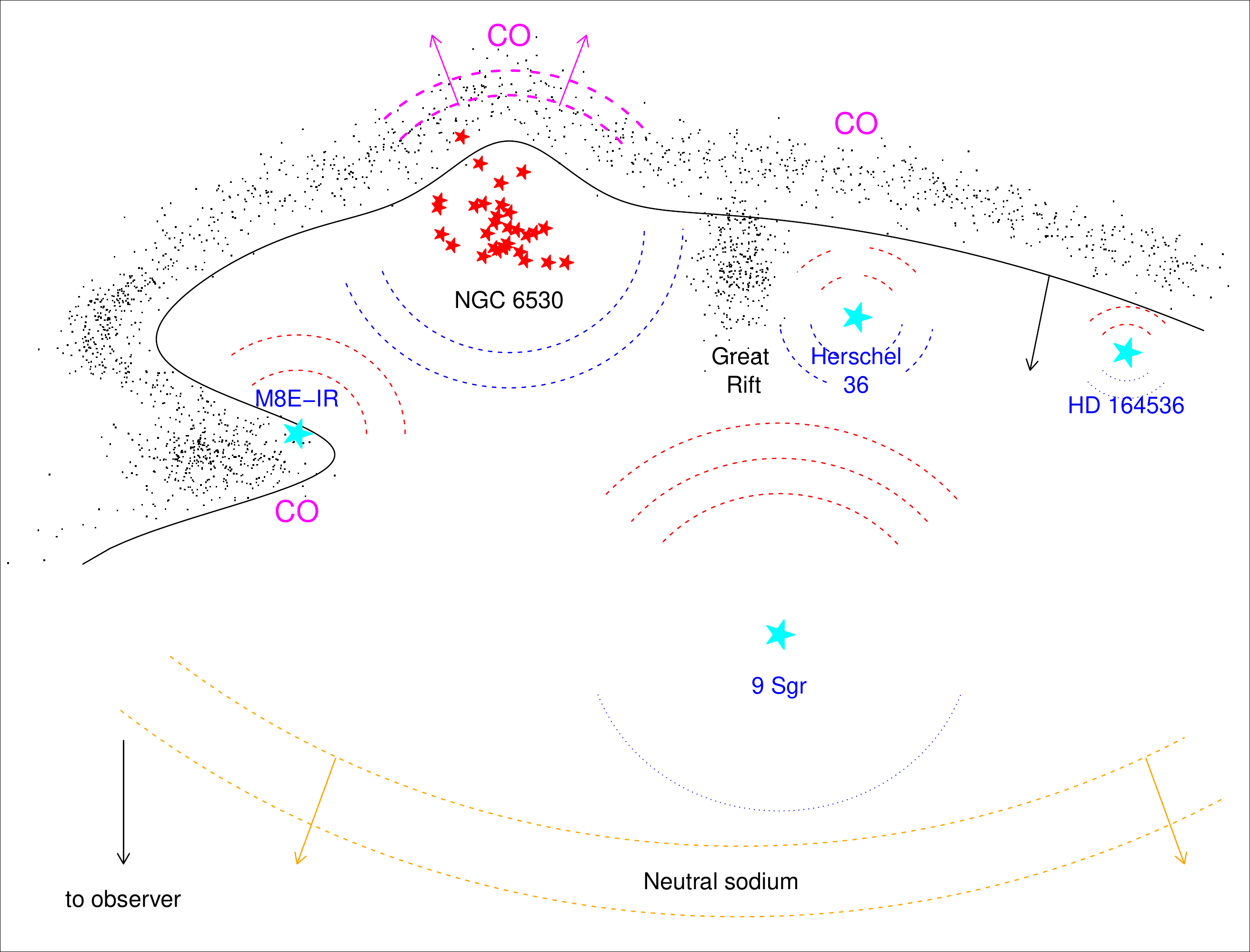}    }
\caption{Schematic representation of the Lagoon nebula geometry, along a
direction passing through M8E-IR and the Hourglass nebula.
The black line indicates the molecular cloud boundary. Dusty regions are
indicated with dots. Red (blue) dashed arcs indicate redshifted
(blueshifted) ionized gas. The position and motion of the known CO
clouds is indicated in magenta. The most massive stars and the NGC6530
cluster core are indicated with cyan and red stars respectively. The
large-scale sodium layers are also shown (yellow dashed lines),
probably not to scale.
\label{cartoon}}
\end{figure*}

The adoption of 2-g fits does not clarify the issue of large-scale
dynamics, even using [N II] and [O III] having smaller linewidths
(Figure~\ref{radial2-n2-o3}): not only the datapoint scatter is increased, but
the overall East-West velocity gradient becomes barely observable in
these lines, which therefore originate in layers well distinct from \ha.
For some locations, as discussed in Section~\ref{dark}, the [N II] lines
are split, with central RV very close to that of the nearby O star HD~164536:
these add to the scatter seen in the rightmost datapoints in
Fig.~\ref{radial2-n2-o3}, where local expansion adds to the average local
cloud velocity. The splitting center being so close to the O-star RV might
also be seen as a confirmation that the (poorly
studied) star HD~164536 lies within the Lagoon itself.
The figure also shows velocities of sodium absorption components, which may be
of some usefulness: if we consider the lowest-found (absolute) sodium
velocities at various radial distances, we may observe a regular
gradient from $RV \sim 0$~km/s at the East extreme, towards $RV \sim -5$~km/s
at the opposite one. As discussed in Section~\ref{uves}, this sodium layer, being the most
uniform of all found here, is probably the outermost one, and therefore
least affected by local phenomena. If it is really associated with the
Lagoon nebula, it might be considered as the best indicator of a global
rotation of the nebula. A much better spatial coverage in the sodium absorption
data would be needed, however, before accepting this possibility.

Alternatively, the observed velocity gradient might reflect a shear
motion, caused by interaction between the parts of M8 closest to the
galactic plane (to the Northwest) and other dense clouds.

\section{Discussion and summary}
\label{discuss}

The various pieces of evidence described in Section~\ref{results} enable
us to draw a complex picture of the ionized and neutral gas in the
Lagoon nebula, and of its physical connection to massive stars and
molecular material in the cloud. Perhaps the most effective way of
building a coherent and understandable picture is by means of a drawing.
We show therefore a graphical summary of most of the results obtained in
Figure~\ref{cartoon}, which refers to a section through the nebula along a
line joining M8E-IR with the Hourglass region, until the western nebula
parts. The Figure represents most of our findings with some level of detail.
The black solid line represents the approximate boundary of the
molecular cloud; dots indicate dusty regions.
O stars are shown as bigger cyan stars.

The red stars indicate the core of the NGC6530 cluster, where most of its B
stars are found; the entire cluster would fill most of the region shown.
The B stars drive a strong expanding shell towards us (blue dashed
half-circles) visible in the ionized-gas lines. On the opposite (far) side of
the cluster, no redshifted expansion is detected in the ionized lines,
suggesting that the cluster lies very close to the denser, dusty
molecular cloud, as shown. The CO clump in the same direction is
redshifted (Lada \e 1976, Takeuchi \e 2010), probably pushed by the same
cluster stars producing the blueshifted optical lines. We represent this
with the magenta dashed arcs. Note that the other two strong CO clumps
in this region (also labeled 'CO' in magenta) are found at rest with
respect to the NGC6530 $RV_{cm}$.

The Hourglass region around Herschel~36 lies probably at some distance
from the background molecular cloud, since some high-velocity redshifted
emission is seen around it, as well as blueshifted emission. The
geometry of any material expanding away from Herschel~36 is however very
irregular, because of the non-isotropic distribution of dense clouds all
around this star: we represent this as discontinuous shell fragments
(dashed), both redshifted and blueshifted.
We recall that the nebular densities found in this region are the
highest of the whole Lagoon nebula.

The M8-East region is also characterized by a partial-shell geometry,
since we detect the unobscured portion of
a redshifted expanding arc, centered on M8E-IR. This
star, or another sufficiently massive star near to it, must be heavily
obscured on our side (tens of magnitudes visually), but very little on
the ``inner" side, in order to be able to drive an inward flow. The
outward flow is instead blocked by the dense cloud, and its blueshifted
emission not detected accordingly.

A crucial role is played by the most massive star, 9~Sgr, which we find
to lie well separated (at least 1~pc, or more) from the cloud surface, in
agreement with previous works, but still inside the large blister concavity.
Being found in a low-density environment,
its radiative and mechanical push becomes detectable only against the
higher-density gas on the inward side of the star, as redshifted
emission in [N II] and [S II] (triple red dashed arcs); on its outward
side, very faint or undetectable blueshifted emission is present (dotted
blue arc). A combination of factors may therefore explain the
characteristics found in the ``Great Rift" region: it is denser than its
surroundings, being compressed by both sides (NGC6530 shell and
Herschel~36 shell), and pushed towards the cloud by the 9~Sgr wind and
radiation, hence the positive velocities on its ionized surface. It
would be interesting to examine whether such compression is able to
trigger new star formation.
There is no contradiction between the enhanced dust density in the
Great Rift, responsible for obscuration of background stars, and
the inconspicuous electron density found there from the [S II] doublet ratio
(Figure~\ref{map-ha-n2-ratio}$b$), since this latter only refers to its
ionized surface and not to its colder, inner parts.

Completing the picture, the western parts of the cloud are found to be
approaching us, as is the O star HD~164536, whose wind is probably
responsible of some line splitting in its vicinity, indicative of a
bidirectional expansion, as shown by the blue and red arcs. At large
distances in front of the whole cloud, discrete neutral layers are found, 
approaching us over a range of (negative) velocities and distances.
{
They might be named the ``Lagoon Veil'', by analogy with the Orion Veil.
}

In addition to these results, at least another one deserves some
discussion. The velocity profile around NGC6530, whatever the
diagnostic line and the modeling approach (1-g or 2-g) chosen, leaves
little doubt about an expanding shell being driven by stars in the cluster
core. At the same time, both the ionization parameter
(Fig.~\ref{map-ha-n2-ratio}$a$) and the ionizing flux (Fig.~\ref{radial-ioniz})
show a gradient across the cluster face, in the direction of 9~Sgr. The
latter Figure shows that, with respect to the 9~Sgr ionizing flux, only
a small excess (less than a factor of 2) is found at the NGC6530
position, attributable to the NGC6530 B stars themselves. Therefore
we reach the conclusion, on solid observational grounds, that the
NGC6530 shell is mechanically driven from inside, but ionized from
outside. This geometry is very unlike classical Str\"omgren spheres.
A deeper treatment of the problem is clearly outside our scopes here.
The biggest problem is, since we observe recombination in the \ha\ line,
where the recombined neutral gas lies: gas outside of the shell is
ionized by 9~Sgr, inside it is ionized by the NGC6530 B stars. Perhaps a
double shell develops, with an intermediate sheet of neutral gas:
this might account for the small velocity splitting indicated by our 2-g
fits to \ha\ lines. In the absence of a detailed modeling, we cannot
however derive any firm conclusion on this issue.

Another interesting issue is that related to rotation of the entire
cloud, as suggested by the blueshifted lines in the N-W parts. It is
worth noting in this respect that in the molecular CO lines the
appearance of the Lagoon nebula is very different than in the optical,
and splits in three main condensations, coincident with M8-East, NGC6530
core, and Hourglass regions respectively, with little in between
(Fig.~8 of Takeuchi \e 2010). The Lagoon N-W regions appear as an extension
of the Hourglass CO cloud, whose core is at rest with respect to
$RV_{cm}$. It is therefore possible that only the Hourglass molecular cloud
is rotating, not the entire Lagoon nebula; this motion does not involve
NGC6530 and is in better agreement with the other existing dynamical data.
Alternatively, the CO maps of Takeuchi \e (2010) make it clear that the
N-W region, the closest to the galactic plane, is also near other CO
clouds, and might be interacting with them. Therefore, while the bulk
motion of approach of the N-W Lagoon nebula region is an established
observational result, its interpretation
is not unambiguous.

\begin{acknowledgements}
We wish to thank an anonymous referee for his/her many interesting
comments and suggestions.
Based on data products from observations made with ESO Telescopes at the
La Silla Paranal Observatory under programme ID 188.B-3002. These data
products have been processed by the Cambridge Astronomy Survey Unit
(CASU) at the Institute of Astronomy, University of Cambridge, and by
the FLAMES/UVES reduction team at INAF/Osservatorio Astrofisico di
Arcetri. These data have been obtained from the Gaia-ESO Survey Data
Archive, prepared and hosted by the Wide Field Astronomy Unit, Institute
for Astronomy, University of Edinburgh, which is funded by the UK
Science and Technology Facilities Council.
This work was partly supported by the European Union FP7 programme
through ERC grant number 320360 and by the Leverhulme Trust through
grant RPG-2012-541. We acknowledge the support from INAF and Ministero
dell' Istruzione, dell' Universit\`a' e della Ricerca (MIUR) in the form
of the grant "Premiale VLT 2012".
T.~Z.\ acknowledges support from the Slovenian
Research Agency (research core funding No. P1-0188).
F.~J.-E.\ acknowledges financial support from the Spacetec-CM project
(S2013/ICE-2822).
M.~T.~C.\ acknowledges the financial support from the Spanish Ministerio de
Econom\'{\i}a y Competitividad, through grant AYA2016-75931-C2-1-P.
The results presented here benefit
from discussions held during the Gaia-ESO workshops and conferences
supported by the ESF (European Science Foundation) through the GREAT
Research Network Programme.
This work is also based on data products from observations made with
ESO Telescopes at the La Silla Paranal Observatory under programme
ID 177.D-3023, as part of the VST Photometric H$\alpha$ Survey of the
Southern Galactic Plane and Bulge (VPHAS$+$, www.vphas.eu).
This research has made use of the SIMBAD database,
operated at CDS, Strasbourg, France.
\end{acknowledgements}

\bibliographystyle{aa}

\begin{landscape}
\begin{table}
\centering
\caption{Fitting results (1-g models) for nebular emission lines.
Units of RV and $\sigma$ columns are km/s.
Units of Norm columns are ADU km/sec/minute.
Full table in electronic format only.} 
\label{table-sky}
\begin{tabular}{lcccccccccccccc}
  \hline
Id & RA & Dec & \multicolumn{3}{c}{H$\alpha$} & \multicolumn{3}{c}{[N II] 6584}
& \multicolumn{3}{c}{[S II] 6717} & \multicolumn{3}{c}{[S II] 6731} \\
 & \multicolumn{2}{c}{(J2000)} & RV & $\sigma$ & Norm & RV & $\sigma$ & Norm
& RV & $\sigma$ & Norm & RV & $\sigma$ & Norm \\ 
  \hline
SKY\_18041349-2406277 & 271.0562 & -24.10769 & -4.44 & 13.33 & 13604.47 & -3.32 & 9.82 & 5303.02 & -2.45 & 9.49 & 893.61 & -2.19 & 9.48 & 653.69 \\ 
  SKY\_18040836-2411076 & 271.0348 & -24.18544 & -8.67 & 13.35 & 71283.37 & -5.89 & 13.05 & 10866.23 & -3.70 & 12.88 & 2182.28 & -3.58 & 12.71 & 1608.08 \\ 
  SKY\_18041039-2413163 & 271.0433 & -24.22119 & -2.41 & 13.35 & 44833.48 & 2.03 & 10.90 & 6861.96 & 2.98 & 10.89 & 1360.72 & 2.84 & 10.59 & 997.84 \\ 
  SKY\_18043822-2413504 & 271.1592 & -24.23067 & -7.78 & 14.11 & 32206.53 & -4.93 & 11.19 & 6559.79 & -4.35 & 11.04 & 1178.57 & -4.06 & 10.83 & 856.31 \\ 
  SKY\_18043552-2406457 & 271.1480 & -24.11269 & -5.75 & 13.88 & 8302.00 & -4.48 & 10.82 & 3030.67 & -4.37 & 10.60 & 661.15 & -4.14 & 10.58 & 478.39 \\ 
  SKY\_18043039-2411382 & 271.1266 & -24.19394 & -7.91 & 13.73 & 45666.66 & -4.08 & 11.60 & 10108.42 & -2.68 & 11.34 & 1992.34 & -2.49 & 11.19 & 1479.74 \\ 
  SKY\_18044986-2410494 & 271.2077 & -24.18039 & -6.08 & 13.54 & 24539.49 & -3.82 & 11.38 & 8360.59 & -1.76 & 11.49 & 1839.73 & -1.83 & 11.10 & 1320.87 \\ 
  SKY\_18042865-2415336 & 271.1194 & -24.25933 & -8.52 & 13.18 & 44827.42 & -5.82 & 10.46 & 7341.40 & -4.70 & 10.00 & 1366.45 & -4.62 & 9.83 & 1034.42 \\ 
  SKY\_18043273-2417261 & 271.1364 & -24.29058 & -8.83 & 13.68 & 16638.05 & -4.96 & 11.79 & 2305.00 & -4.53 & 11.39 & 531.72 & -4.47 & 10.64 & 388.39 \\ 
  SKY\_18045577-2415459 & 271.2324 & -24.26275 & -7.12 & 14.10 & 21283.13 & -4.00 & 12.58 & 6772.99 & -4.41 & 12.38 & 1149.48 & -4.32 & 12.05 & 821.87 \\ 
  SKY\_18043471-2416492 & 271.1446 & -24.28033 & -8.51 & 13.62 & 46025.05 & -5.27 & 11.36 & 5759.30 & -3.64 & 11.00 & 1221.82 & -3.59 & 10.72 & 925.07 \\ 
  SKY\_18050209-2410250 & 271.2587 & -24.17361 & -5.00 & 14.01 & 11658.67 & -3.08 & 11.63 & 4058.60 & -2.10 & 11.18 & 937.26 & -1.71 & 11.51 & 690.53 \\ 
  SKY\_18050407-2417194 & 271.2670 & -24.28872 & -6.59 & 14.11 & 6440.37 & -5.05 & 11.84 & 1703.00 & -4.56 & 11.63 & 396.81 & -4.29 & 11.64 & 290.20 \\ 
  SKY\_18043741-2418209 & 271.1559 & -24.30581 & -9.02 & 14.09 & 53120.50 & -7.15 & 12.93 & 7183.70 & -5.76 & 12.30 & 1375.30 & -5.75 & 11.92 & 1008.91 \\ 
  SKY\_18042325-2423135 & 271.0969 & -24.38708 & -9.76 & 13.04 & 92195.82 & -9.26 & 12.77 & 10544.73 & -6.68 & 12.47 & 2340.38 & -6.23 & 12.22 & 1782.30 \\ 
  SKY\_18042233-2424399 & 271.0930 & -24.41108 & -10.13 & 13.35 & 81644.76 & -9.24 & 12.86 & 12434.13 & -8.79 & 12.66 & 2410.94 & -8.44 & 12.43 & 1806.04 \\ 
  SKY\_18035290-2419527 & 270.9704 & -24.33131 & -4.73 & 13.29 & 103114.41 & -0.16 & 11.45 & 15344.15 & 1.19 & 11.42 & 2948.04 & 1.59 & 11.08 & 2451.45 \\ 
  SKY\_18040022-2416517 & 271.0009 & -24.28103 & 0.48 & 14.22 & 58118.20 & 8.01 & 12.99 & 7824.35 & 6.78 & 14.20 & 1210.24 & 6.85 & 13.96 & 917.58 \\ 
  SKY\_18034573-2418216 & 270.9405 & -24.30600 & -9.93 & 13.65 & 84000.85 & -3.75 & 10.69 & 18035.57 & -2.44 & 10.97 & 3429.26 & -2.19 & 10.65 & 2984.15 \\ 
  SKY\_18030577-2416564 & 270.7740 & -24.28233 & -7.74 & 13.96 & 27976.36 & -6.61 & 11.41 & 8938.61 & -6.18 & 11.04 & 1992.20 & -5.86 & 11.16 & 1526.26 \\ 
   \hline
\end{tabular}
\end{table}
\end{landscape}

\begin{landscape}
\begin{table}
\centering
\caption{Fitting results for Na~I D2 lines.
Column Id is the Setup+star identifier used in Section~\ref{uves}.
Column Name is based on J2000 star coordinates. Units of $N_i$ are the
same as $\sigma_i$ by Eq.~\ref{eq3}.
Full table in electronic format only.} 
\label{table-sodium}
\begin{tabular}{lcccccccccccccccc}
  \hline
Id & Name & $v_1$ & $\sigma_1$ & $N_1$ & $v_2$ & $\sigma_2$ & $N_2$ & $v_3$
& $\sigma_3$ & $N_3$ & $v_4$ & $\sigma_4$ & $N_4$ & $v_5$ & $\sigma_5$ & $N_5$ \\ 
 & & (km/s) & (km/s) & (km/s) & (km/s) & (km/s) & (km/s) & (km/s) &
(km/s) & (km/s) & (km/s) & (km/s) & (km/s) & (km/s) & (km/s) & (km/s) \\
  \hline
580-1 & 18040126-2423474 & -5.01 & 1.48 &  208.52 & -11.14 & 10.10 &    6.69 & -18.47 & 2.20 &   10.29 & -27.94 & 5.63 &  2.13 &        &      &      \\ 
  580-2 & 18041116-2421452 & -1.01 & 8.68 &    2.66 &  -4.77 &  1.23 & 3071.49 & -17.38 & 3.36 &   11.62 & -27.48 & 3.75 &  4.28 &        &      &      \\ 
  580-3 & 18042056-2424556 & -4.98 & 1.13 & 2029.78 & -16.51 &  7.44 &   20.68 &        &      &         &        &      &       &        &      &      \\ 
  580-4 & 18042433-2415168 & -3.84 & 1.01 & 2523.49 & -24.40 &  2.57 &    0.55 &        &      &         &        &      &       &        &      &      \\ 
  580-5 & 18042502-2427453 & -6.50 & 3.00 &   80.45 & -22.64 &  4.72 &   21.36 &        &      &         &        &      &       &        &      &      \\ 
  580-6 & 18042663-2419321 & -5.34 & 1.12 & 2811.36 & -22.97 &  1.52 &    0.66 &        &      &         &        &      &       &        &      &      \\ 
  580-7 & 18042720-2422497 & -4.84 & 1.06 & 2667.80 & -13.44 &  7.88 &   14.49 & -21.95 & 0.57 & 1435.09 & -43.16 & 1.04 &  0.55 &        &      &      \\ 
  580-8 & 18043893-2424142 & -4.19 & 1.27 & 1458.15 & -20.24 &  3.10 &    1.66 &        &      &         &        &      &       &        &      &      \\ 
  580-9 & 18044279-2418339 & -4.75 & 1.26 &  455.51 & -14.07 &  3.00 &    5.41 & -19.38 & 1.73 &   12.75 & -26.37 & 2.66 &  4.45 & -45.23 & 3.00 & 0.48 \\ 
  580-10 & 18044593-2427191 & -4.24 & 1.24 & 3100.19 & -14.59 &  8.59 &   13.68 &        &      &         &        &      &       &        &      &      \\ 
  580-11 & 18045062-2425419 & -3.43 & 1.46 &  338.62 &  -8.78 &  3.19 &   35.49 & -21.24 & 9.12 &    8.65 & -30.68 & 1.26 &  6.12 &        &      &      \\ 
  580-12 & 18045273-2417525 & -4.94 & 1.14 & 2850.68 & -19.30 &  3.54 &    1.74 &        &      &         &        &      &       &        &      &      \\ 
  580-14 & 18053923-2407522 & -5.25 & 1.75 &  157.38 & -17.57 &  3.92 &    4.92 & -31.20 & 2.53 &    0.52 &        &      &       &        &      &      \\ 
  580-16 & 18055648-2416004 & -0.59 & 2.72 &  965.69 &  -0.74 &  7.46 &   27.64 & -24.58 & 2.41 &   35.00 & -41.33 & 0.21 &  0.14 &        &      &      \\ 
  520-1 & 18023863-2415195 & -6.48 & 2.33 &   27.77 & -19.17 &  2.70 &    2.83 & -31.45 & 3.79 &    6.19 &        &      &       &        &      &      \\ 
  520-2 & 18024192-2433360 & -5.38 & 1.03 & 2573.53 &  -5.72 &  3.00 &   12.44 &        &      &         &        &      &       &        &      &      \\ 
  520-3 & 18025328-2420170 & -5.50 & 2.11 &   38.28 & -18.47 &  4.37 &    5.14 & -33.29 & 2.22 &    4.39 &        &      &       &        &      &      \\ 
  520-4 & 18033016-2430506 & -5.33 & 0.97 & 2438.88 &  -5.58 &  3.00 &   13.45 &        &      &         &        &      &       &        &      &      \\ 
  520-5 & 18034033-2422427 & -5.23 & 2.98 &   24.67 & -18.03 &  5.00 &   12.82 & -28.11 & 5.00 &    2.10 &        &      &       &        &      &      \\ 
  520-6 & 18035837-2429128 & -5.12 & 1.61 &   54.09 & -11.45 &  4.54 &   16.17 & -27.30 & 3.87 &    4.89 &        &      &       &        &      &      \\ 
   \hline
\end{tabular}
\end{table}
\end{landscape}


\end{document}